%% file: MainVibroTagTMC.tex
\documentclass[10pt,journal,compsoc]{IEEEtran}
\usepackage{multirow, rotating, amsmath, wasysym}
\usepackage{tabularx}
\usepackage{booktabs} 
\usepackage{epstopdf}
\usepackage[tight]{subfigure}
\usepackage{algpseudocode}
\usepackage{adjustbox}
\usepackage{graphicx}
\usepackage{caption}
\usepackage{algorithm}
\usepackage{soul,xcolor}
\usepackage{todonotes}
\usepackage{array}
\usepackage{multirow}
\usepackage{placeins}

\input{new-commands}

\newcommand{\presec}{\vspace{-0.03in}}
\newcommand{\postsec}{\vspace{-0.03in}}
\newcommand{\presub}{\vspace{-0.03in}}
\newcommand{\postsub}{\vspace{-0.03in}}

\begin{document}
\title{Fine-grained Vibration Based Sensing Using a Smartphone}

\author{Kamran~Ali and Alex~X.~Liu
	\IEEEcompsocitemizethanks{\IEEEcompsocthanksitem K. Ali and A. X. Liu are with the Department Computer Science and Engineering, Michigan State University, Lansing, MI, USA, 48823.\protect\\
		E-mail: {alikamr3, alexliu}@cse.msu.edu 
	}
	\thanks{Manuscript under review since October 20, 2019.}
}

\IEEEtitleabstractindextext{
	\input{KamranTMC/abstract}
	
}

\maketitle

\thispagestyle{empty}

\IEEEdisplaynontitleabstractindextext

\IEEEpeerreviewmaketitle


\sloppy{

\input{KamranTMC/introduction}
\input{KamranTMC/relatedwork}

\input{KamranTMC/workingprinciples}
\input{KamranTMC/feature}
\input{KamranTMC/classification}
\input{KamranTMC/implementionevaluation}
\input{KamranTMC/usabilitystudy}
\input{KamranTMC/conclusion}
\input{KamranTMC/acknowledgements}
}
    
{ 
	\bibliographystyle{unsrt}
    \bibliography{VibrationTech,OtherRelatedTech,IndoorLocalizationTaggingTech}
}

%
\vspace{-0.6in}

\end{document}

%% file: new-commands.tex
\mathchardef\Gamma="0100 \mathchardef\Delta="0101
\mathchardef\Theta="0102 \mathchardef\Lambda="0103
\mathchardef\Xi="0104 \mathchardef\Pi="0105
\mathchardef\Sigma="0106 \mathchardef\Upsilon="0107
\mathchardef\Phi="0108 \mathchardef\Psi="0109
\mathchardef\Omega="010A

\newcommand{\outline}[1]{}

\usepackage{xspace}
\usepackage{url}
\usepackage{graphicx}
\usepackage{latexsym}
\usepackage{amssymb}
\usepackage{amsfonts}
\usepackage{psfrag}
\usepackage{wrapfig}
\usepackage{comment}
\usepackage{alltt}
\usepackage{color}

\newcommand{\ie}{\emph{i.e.}\xspace}
\newcommand{\eg}{\emph{e.g.}\xspace}

\newcommand{\etal}{\frenchspacing{}\emph{et al{.}}\xspace}

%% file: KamranTMC/abstract.tex
\begin{abstract}
Recognizing surfaces based on their vibration signatures is useful as it can enable tagging of different locations without requiring any additional hardware such as Near Field Communication (NFC) tags.
However, previous vibration based surface recognition schemes either use custom hardware for creating and sensing vibration, which makes them difficult to adopt, or use inertial (IMU) sensors in commercial off-the-shelf (COTS) smartphones to sense movements produced due to vibrations, which makes them coarse-grained because of the low sampling rates of IMU sensors.
The mainstream COTS smartphones based schemes are also susceptible to inherent hardware based irregularities in vibration mechanism of the smartphones. 
%
Moreover, the existing schemes that use microphones to sense vibration are prone to short-term and constant background noises (\eg intermittent talking, exhaust fan, etc.) because microphones not only capture the sounds created by vibration but also other interfering sounds present in the environment.
In this paper, we propose VibroTag, a robust and practical vibration based sensing scheme that works with smartphones with different hardware, can extract fine-grained vibration signatures of different surfaces, and is robust to environmental noise and hardware based irregularities.
%
%
%
%
%
We implemented VibroTag on two different Android phones and evaluated in multiple different environments where we collected data from 4 individuals for 5 to 20 consecutive days. 
%
%
Our results show that VibroTag achieves an average accuracy of 86.55\% while recognizing 24 different locations/surfaces, even when some of those surfaces were made of similar material.
VibroTag's accuracy is 37\% higher than the average accuracy of 49.25\% achieved by one of the state-of-the-art IMUs based schemes, which we implemented for comparison with VibroTag.
%
%
%
\end{abstract}

%% file: KamranTMC/introduction.tex
\presec
\presec
\section{Introduction}
\postsec

\presub
\subsection{Motivation}
\postsub
Vibration based sensing has been shown to be a low-cost and effective approach to recognizing different surfaces \cite{griffin2008user, shafer2013learning, kunze2007symbolic, cho2012vibration}.
The key intuition is that different surfaces respond to the same vibration differently because the surfaces may be made of different materials, and even if they are made of the same material, they may have different shapes and sizes.
Even if two surfaces are made of the same material and have the same shape and size, they may have different objects placed on them, such that those surfaces still exhibit different vibration patterns because the objects placed on them may respond to the same vibration differently.
Recognizing surfaces based on their vibration signatures is useful as it can enable tagging of different locations without requiring any additional hardware such as Near Field Communication (NFC) tags.
Such tagging of locations can provide us with indirect information about the user activities and intentions without any dedicated infrastructure, based on which we can enable useful services such as context aware notifications/alarms \cite{ho2005using}. 
For example, a user can set their phone to automatically go into silent mode when it is placed on their bed side table.
Such vibration based tagging can also be used to reduce the search space for the user when they lose their phone. 
For example, the ``find my phone'' feature can report the GPS location of a lost phone as well as what material the phone is laying on.
%

A robust and practical vibration based sensing scheme should satisfy three key requirements. 
First, it should work with commercial off-the-shelf (COTS) smartphones with different hardware, so that it can be easily deployed and widely adopted.
Second, it should be able to extract fine-grained vibration signatures, so that it can accurately differentiate different surfaces.
Third, it should be robust to environmental noise and hardware based irregularities, so that its accuracy stays consistent across different environments and devices.

\begin{figure*}[htbp]
	\vspace{-0.085in}
	\centering
	\captionsetup{justification=centering}    
	\subfigure[Bed]{
		\includegraphics[width=0.3\columnwidth]{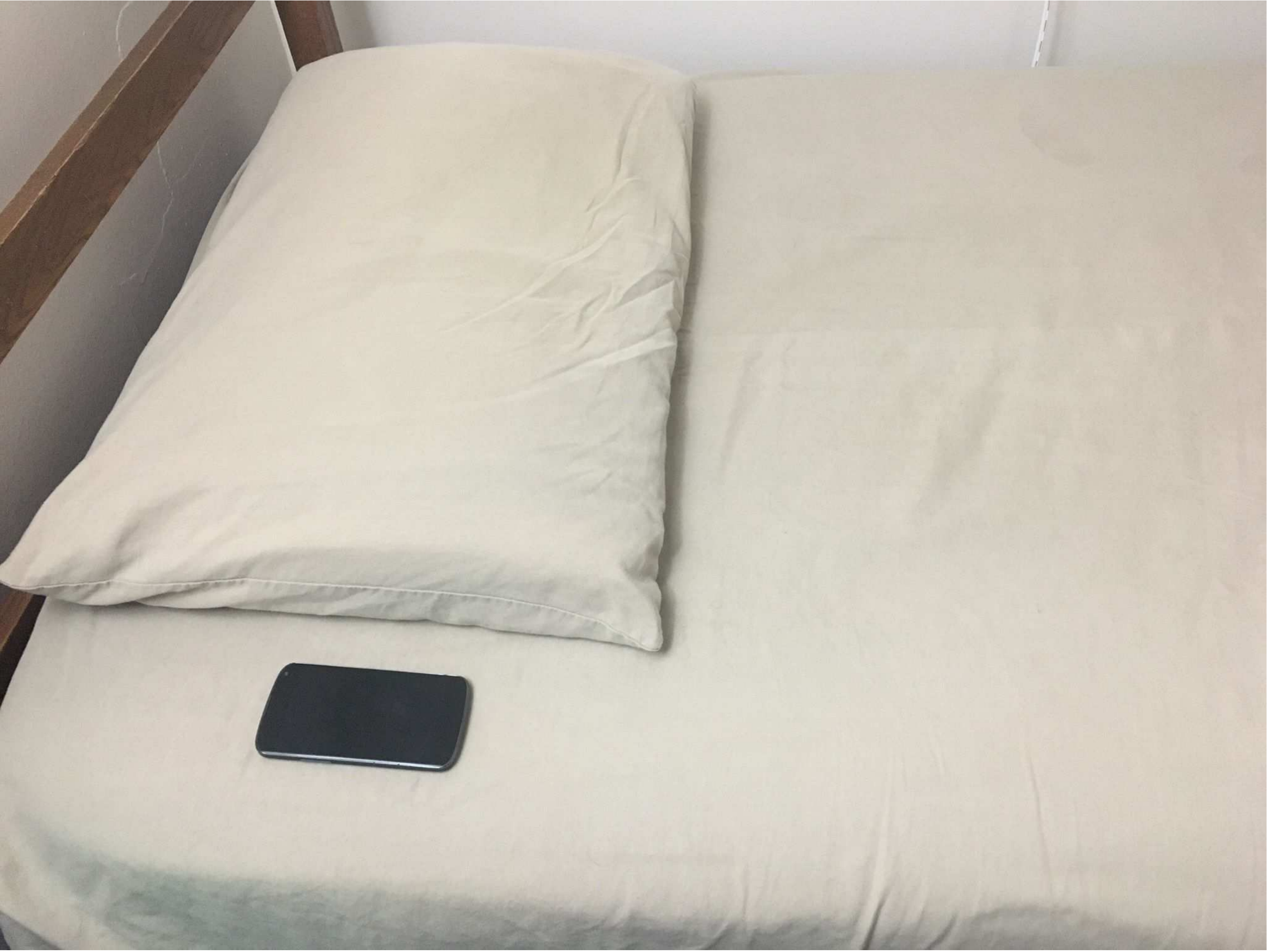}
		\label{fig:homebed}}
	\subfigure[Bed-Table]{
		{\includegraphics[width=0.3\columnwidth]{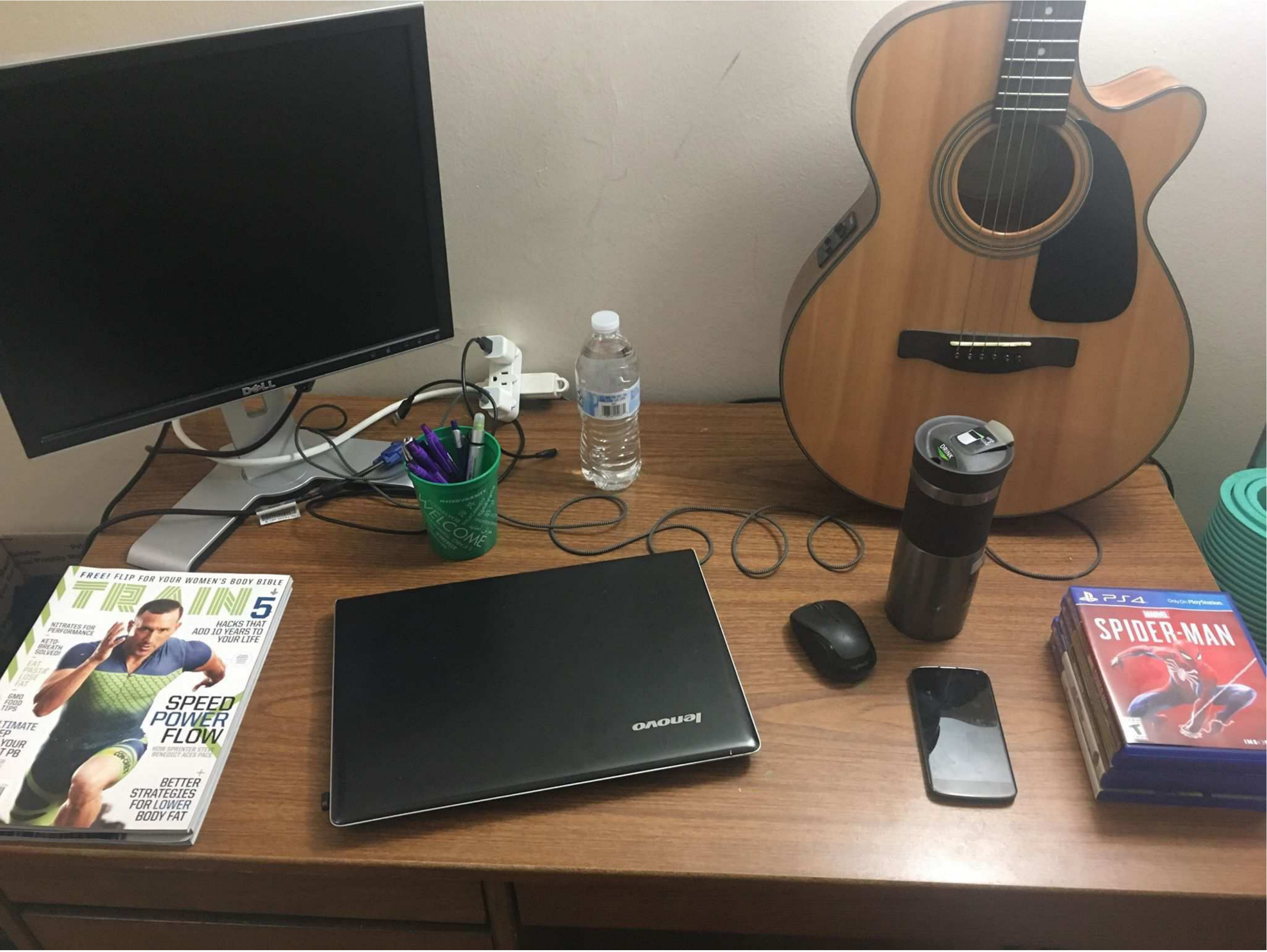}}
		\label{fig:homecontrolled}}
		\subfigure[Kitchen]{
			{\includegraphics[width=0.3\columnwidth]{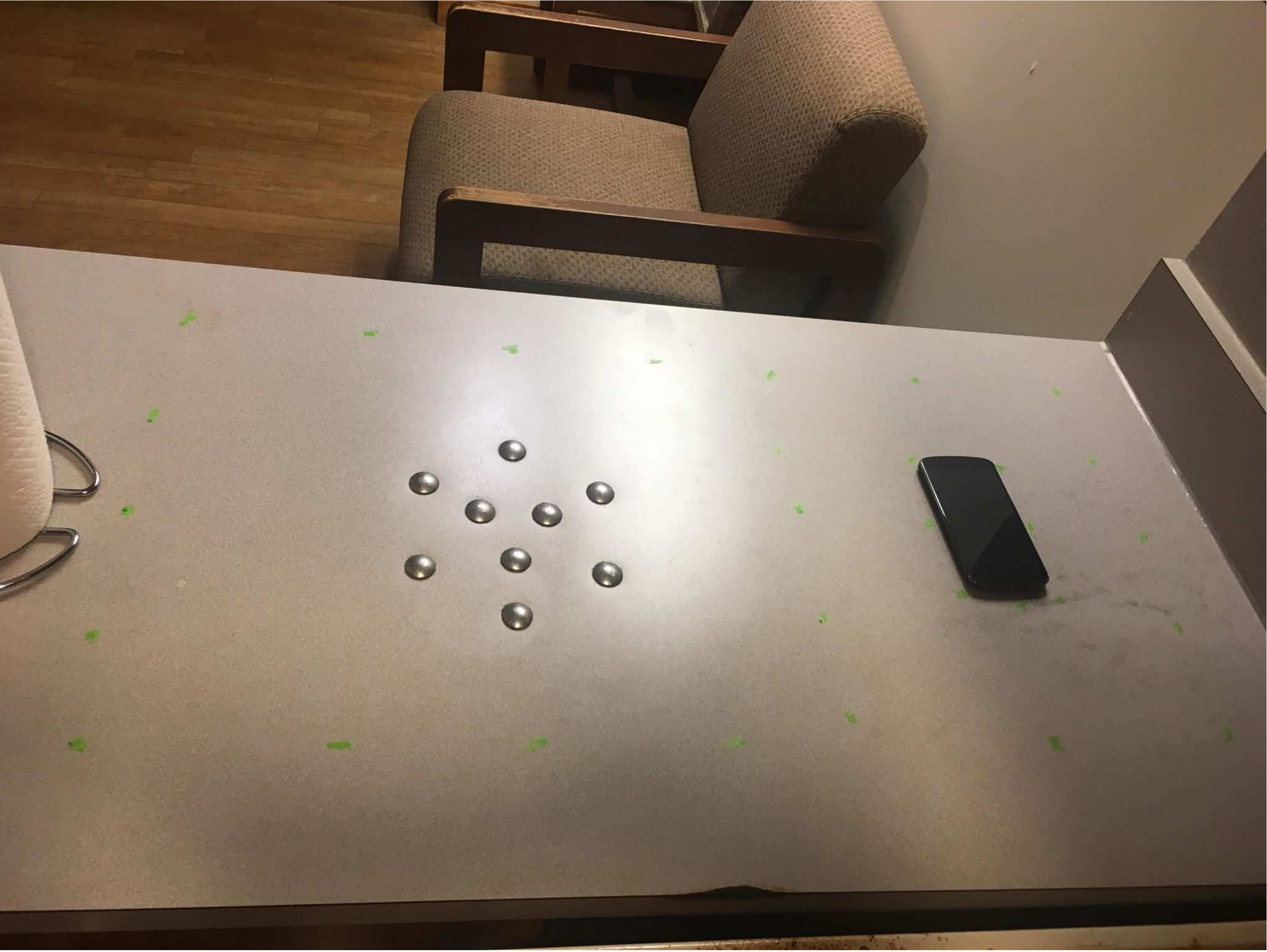}}
			\label{fig:kitchen}}
			\hspace{-0.001in}
	\subfigure[Sofa]{
		{\includegraphics[width=0.3\columnwidth]{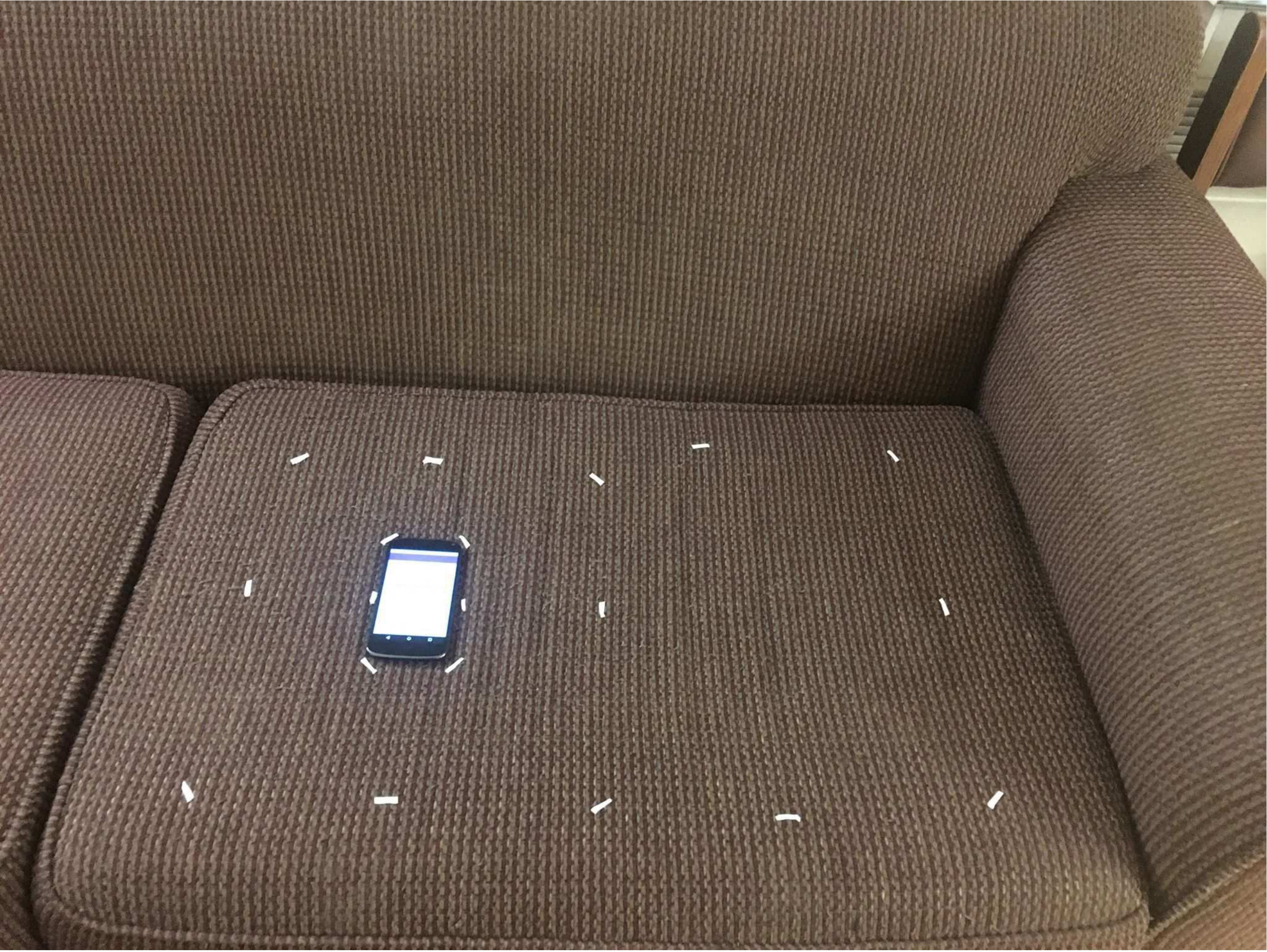}}
		\label{fig:sofa}}
	\hspace{-0.001in}
	\subfigure[Work-Table]{
		{\includegraphics[width=0.3\columnwidth]{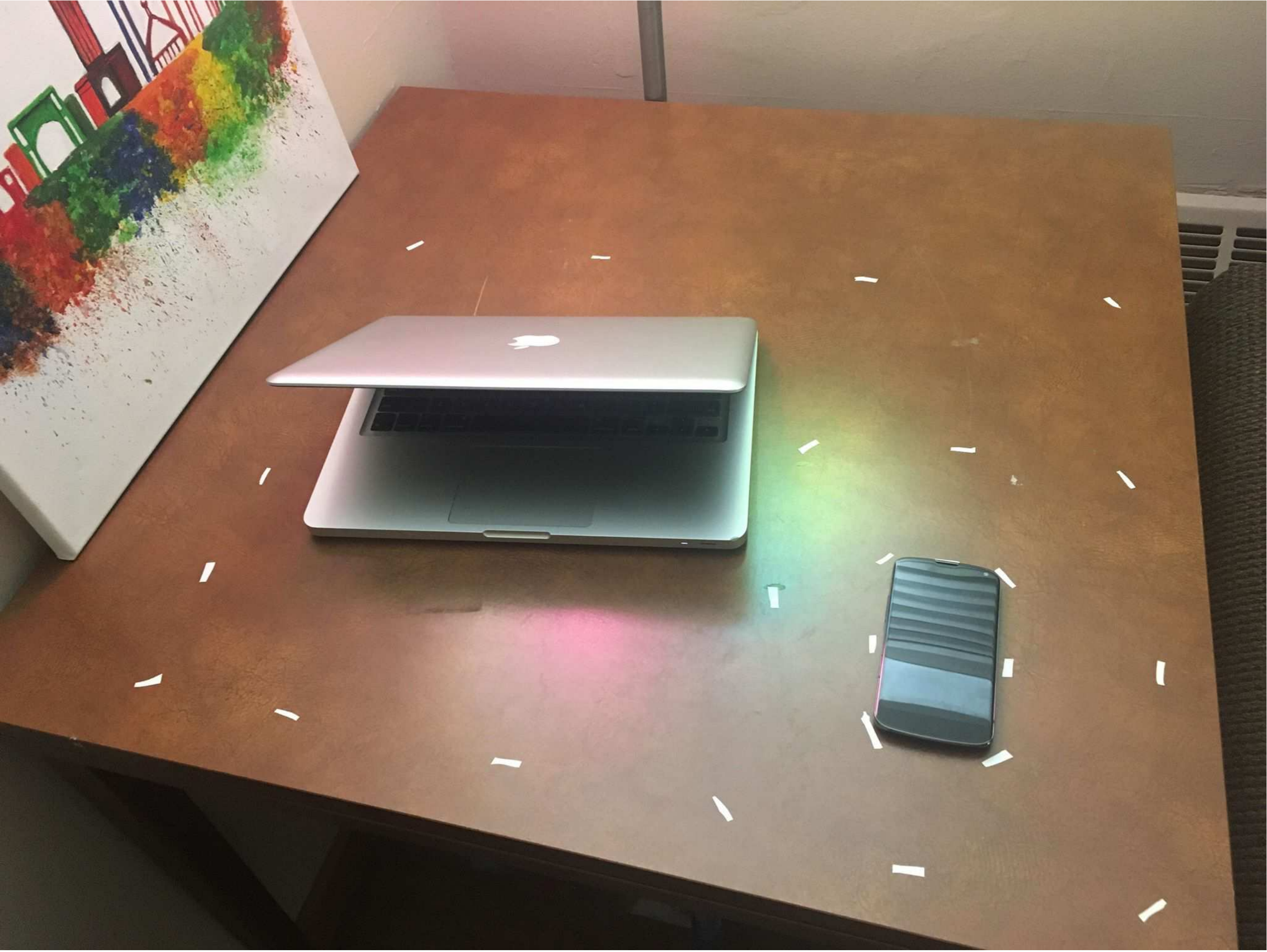}}
		\label{fig:homework}}
	\hspace{-0.001in}
	\subfigure[Restroom]{
		{\includegraphics[width=0.3\columnwidth]{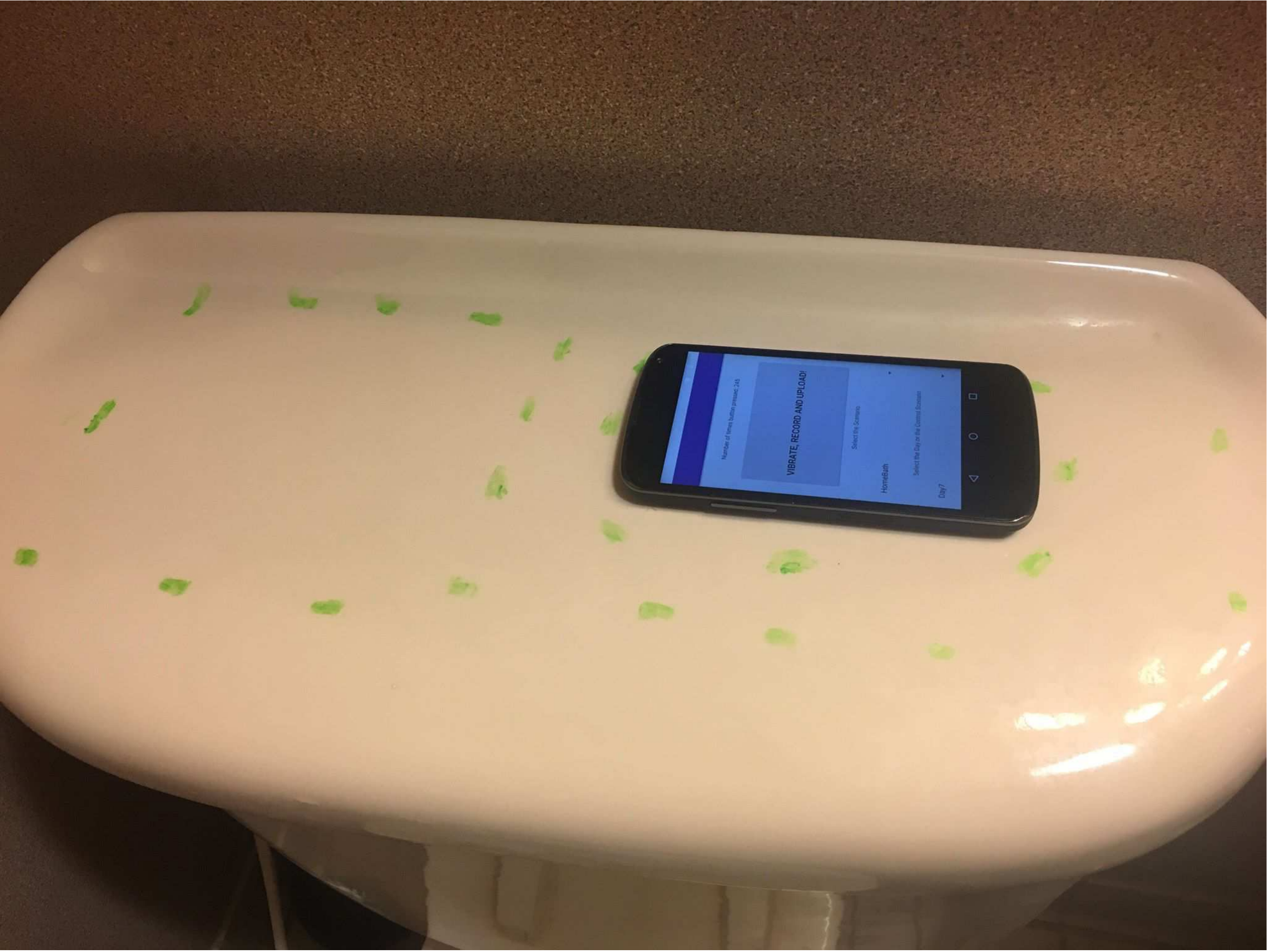}}
		\label{fig:restroom}}
	
	\vspace{-0.1in}
	
	\subfigure[Bed]{
		{\includegraphics[width=0.3\columnwidth]{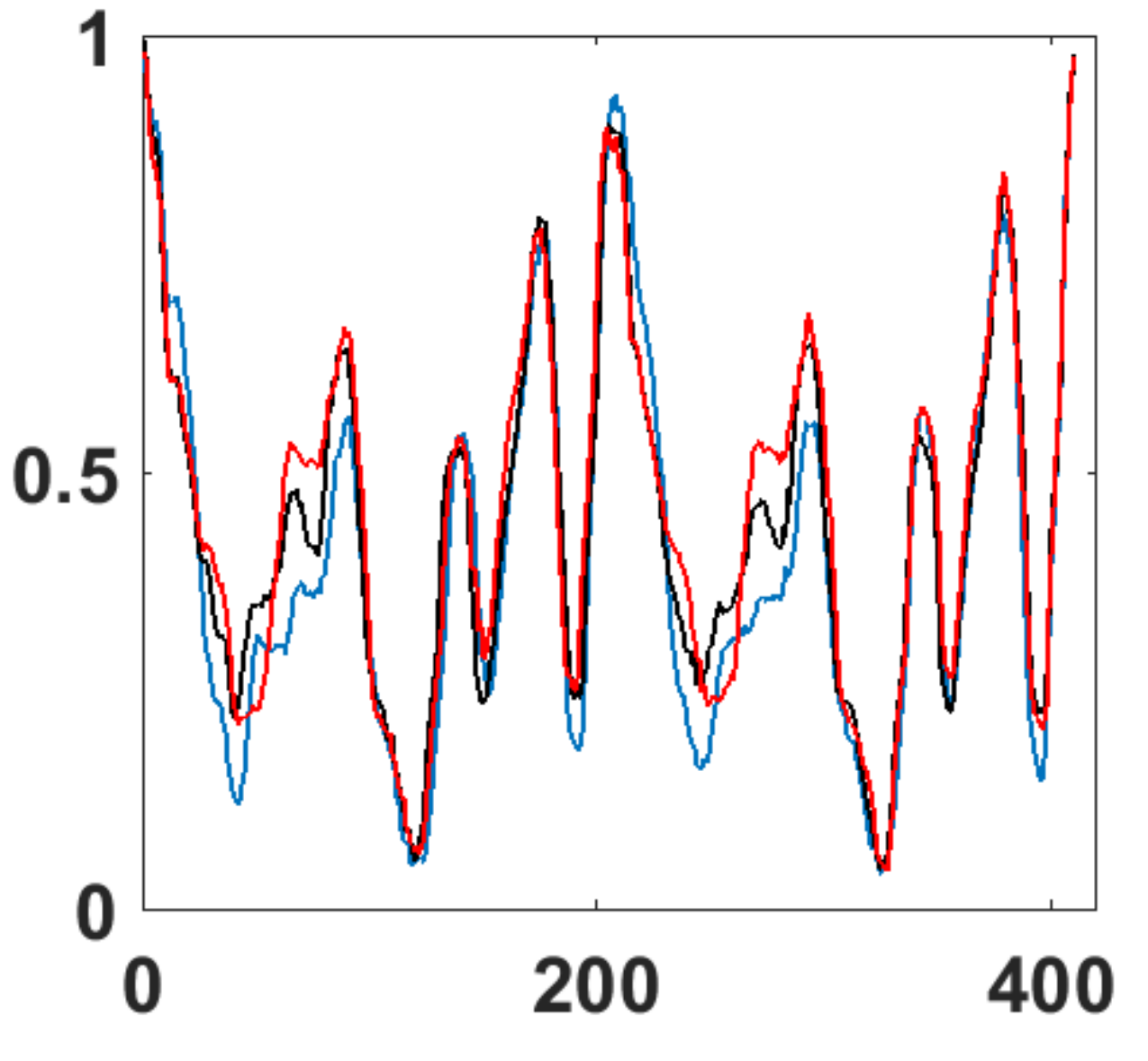}}
		\label{fig:bedfeature}}
	\hspace{-0.001in}
	\subfigure[Bed-Table]{
		{\includegraphics[width=0.3103\columnwidth]{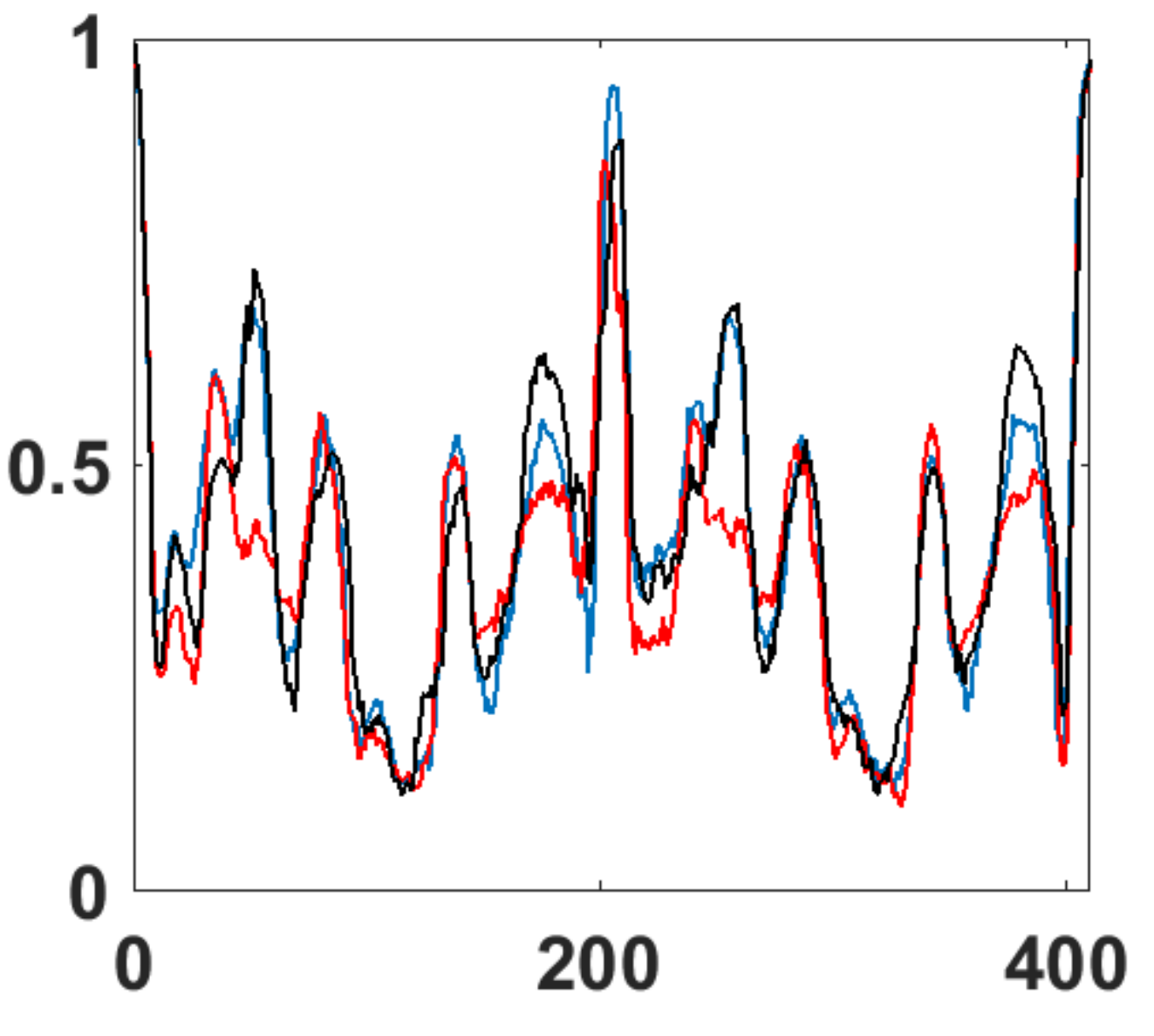}}
		\label{fig:controlledfeature}}
	\hspace{-0.05in}
		\subfigure[Kitchen]{
			{\includegraphics[width=0.3187\columnwidth]{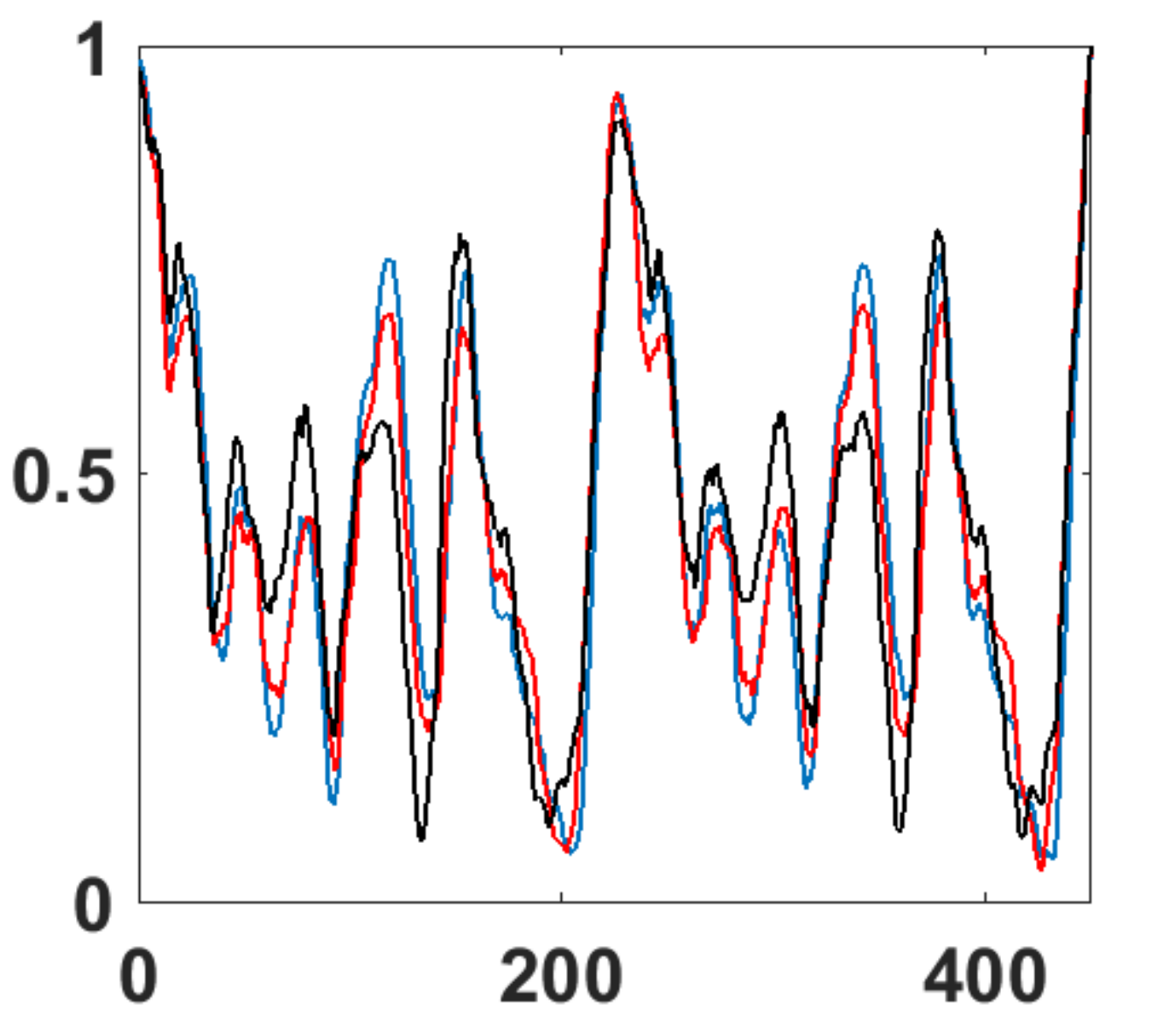}}
			\label{fig:kitchenfeature}}
		\hspace{-0.05in}
	\vspace{-0.01in}
	\subfigure[Sofa]{
		{\includegraphics[width=0.3\columnwidth]{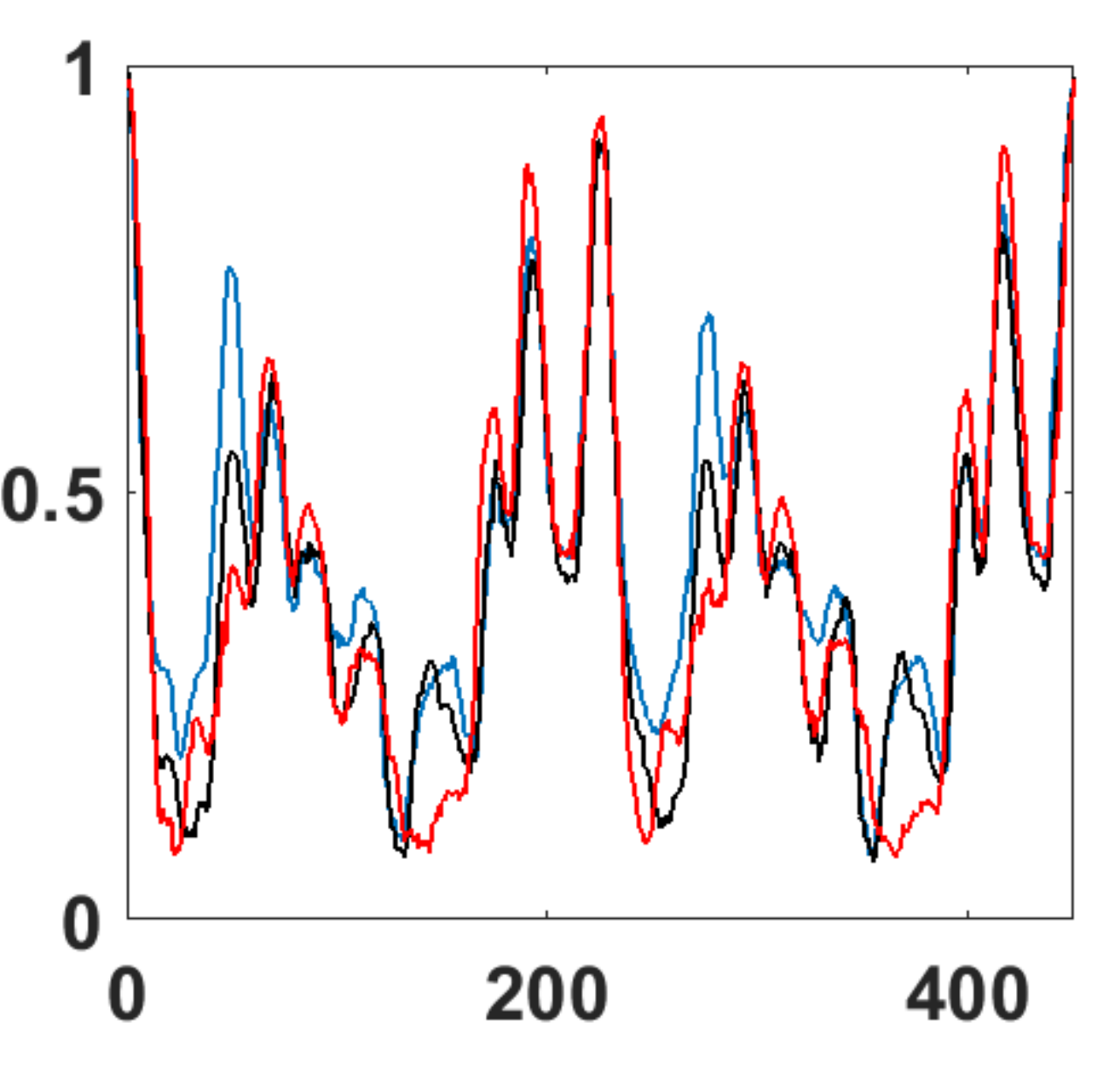}}
		\label{fig:sofafeature}}
	\hspace{-0.001in}
	\vspace{-0.01in}
	\subfigure[Work-Table]{
		{\includegraphics[width=0.3\columnwidth]{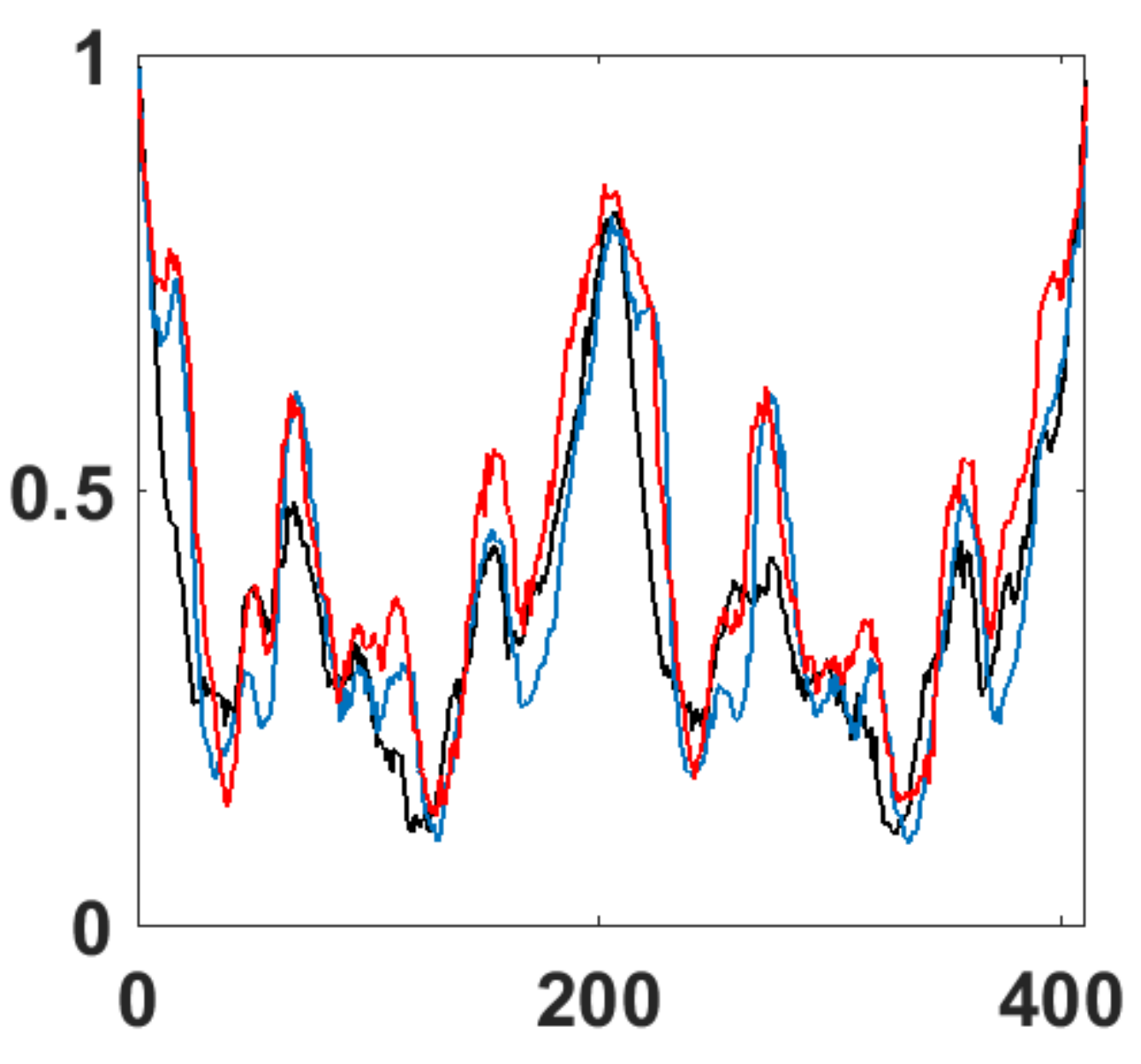}}
		\label{fig:workplacefeature}}
	\hspace{-0.001in}
	\subfigure[Restroom]{
		{\includegraphics[width=0.302\columnwidth]{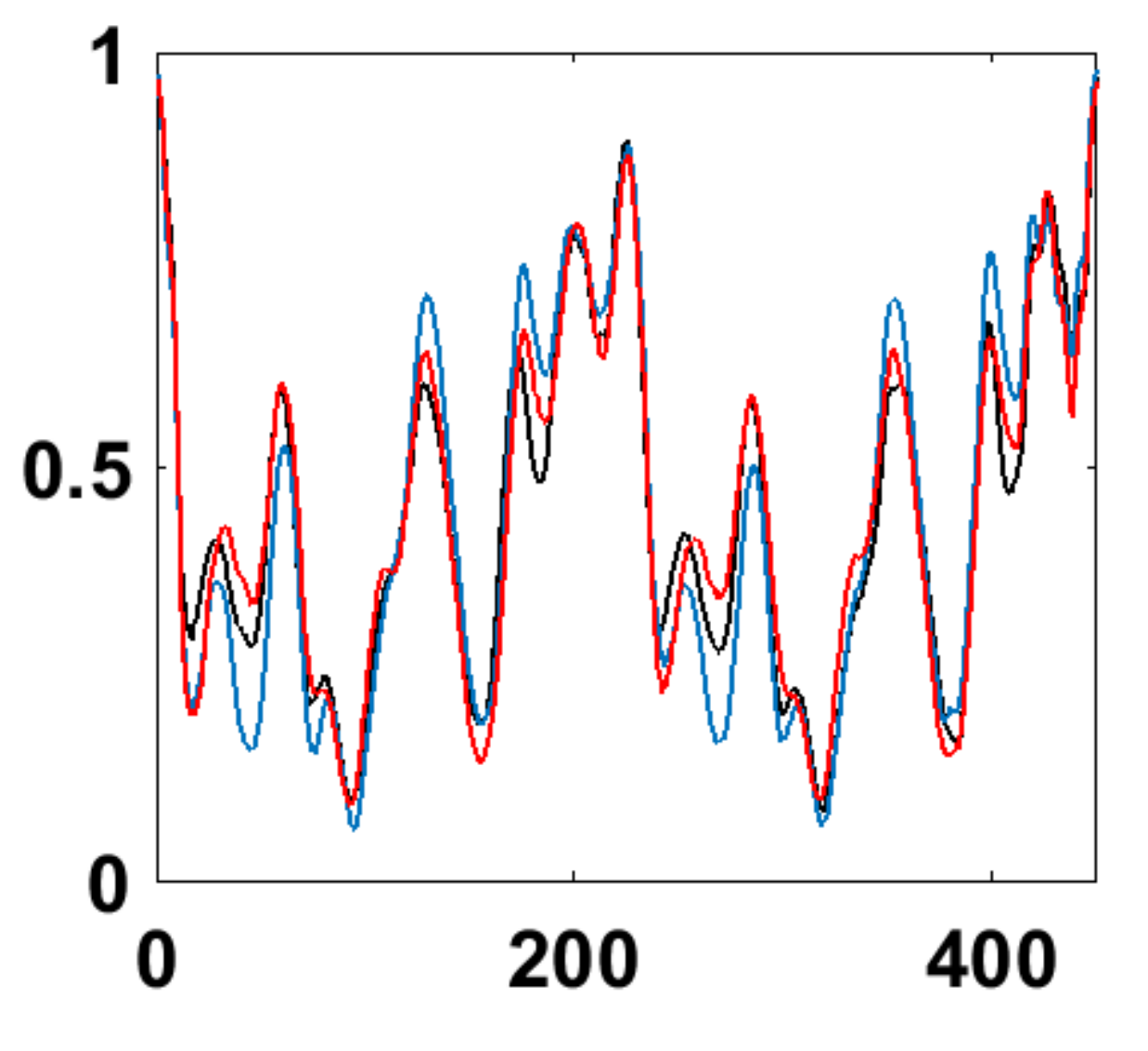}}
		\label{fig:restroomfeature}}
	\hspace{-0.001in}
	\vspace{-0.15in}
	\caption{Experimental scenarios and their corresponding extracted acoustic time-series based vibration signatures}
	\label{fig:examplesintro}
	\vspace{-0.2in}
\end{figure*}

\presub
\presub
\subsection{Limitations of Prior Art}
\postsub
Several vibration based sensing schemes have been proposed in the past to realize different kinds of applications \cite{goel2012gripsense, wiese2013phoneprioception, hwang2013vibrotactor, darbar2015surfacesense, cho2016vibephone, kunze2007symbolic, laput2016viband, liu2016vibkeyboard, liu2017vibsense, shafer2013learning, ono2013touch, fan2015soqr, goel2014surfacelink}.
However, none of them satisfies all the above three requirements.
Existing vibration based sensing schemes can be divided into two categories: custom hardware based and COTS smartphones based. 
The custom hardware based schemes use separate hardware including a micro-controller, a vibrator motor, and some piezoelectric (\eg a microphone) or IMU sensors \cite{kunze2007symbolic, laput2016viband, liu2016vibkeyboard, liu2017vibsense, shafer2013learning}, which gives them fine-grained low level control over different physical layer parameters of their underlying hardware.
However, these schemes are incompatible with COTS smartphones and are not easily generalizable to different hardware because most COTS smartphones have limited sensing capabilities and control over the hardware installed in them.
Most COTS smartphones based schemes rely on motion based features extracted from built-in IMU sensors \cite{griffin2008user, cho2012vibration, goel2012gripsense, wiese2013phoneprioception, cho2016vibephone}.
%
%
%
However, these features are coarse-grained because the sampling frequencies of IMU sensors on COTS smartphones are much lower (usually between 200-300Hz and less than 4kHz even after driver modifications \cite{laput2016viband}) compared to microphones (usually above 44kHz), which naturally leads to lower classification accuracies.
Because of this reason, such schemes can only broadly differentiate between different types of surfaces (\eg wood and plastic) and cannot differentiate between similar surfaces (\eg two different wooden tables in Figs. \ref{fig:homecontrolled} and \ref{fig:homework}).
Also, IMU readings get significantly affected by the smartphone's own motion in space (\eg when a user moves his hand while holding his smartphone or when the smartphone slides during vibration when placed on a smooth surface).
Moreover, the mainstream COTS smartphones based schemes (may they be IMU or microphone based) are susceptible to inherent hardware based irregularities in vibration mechanism of the smartphones. We discuss the impact of such irregularities in Section \ref{sec:robustnesshardware}.
%
Finally, the existing schemes that use microphones to sense vibration \cite{kunze2007symbolic, hwang2013vibrotactor, darbar2015surfacesense} are prone to short-term and constant background noises (\eg intermittent talking, clapping, exhaust fan, etc.) because microphones not only capture the sounds created by vibration but also other interfering sounds present in the environment. We discuss the impact of such noises in Section \ref{sec:robustnessnoise}.
\presub
\subsection{Proposed Approach}
\postsub
%
In this paper, we propose VibroTag, a vibration based sensing approach that can recognize different surfaces based on their unique vibration signatures.
Compared to previous work, VibroTag is robust and practical because it works with COTS smartphones, it can extract fine-grained features representative of different surfaces, and it is robust to hardware irregularities and background environmental noises.
The key intuition is that as the vibrating mass inside a smartphone's vibrator motor repeatedly moves to and fro, the vibrating mass causes the whole smartphone structure and the hardware inside it to vibrate in a peculiar pattern, which depends upon the vibration response (or absorption properties) of the surface that smartphone is placed on.
These vibrations produce peculiar sound waves that VibroTag detects using the smartphone's microphone.
Figure \ref{fig:examplesintro} shows the unique vibration signatures that VibroTag extracted for 6 different surfaces.
We observe that vibration signatures of even two similar surfaces, \ie Bed and Sofa, are quite different from each other.

To make VibroTag easily scalable and compatible with COTS smartphones, we design VibroTag's signal processing pipeline such that it relies only on built-in vibration motors and microphone for sensing, and it is applicable to different phones with different hardware.
To reliably extract fine-grained vibration signatures from the sound signals recorded during vibration, we propose a novel time-series based approach, which is robust to hardware irregularities and environmental noise.
The key idea behind our vibration signature extraction approach is that even if there are irregularities in vibration frequencies due to hardware imperfections, the time-series patterns created during different vibration cycles are very similar.
When a phone vibrates during a specific period of time, such as 3 seconds, multiple such patterns occur and get distributed all over the time-series of recorded sound signals. 
VibroTag finds multiple of these patterns in randomly selected intervals of the time-series, and then combines them into single time-series features that ensures consistency even if there are irregularities in the occurrence of those patterns and/or if the environment is slightly noisy.
Afterwards, it uses these features to differentiate between surfaces.
%



\presec
\subsection{Technical Challenges and Our Solutions}\label{sec:technicalchallenges}
\postsec
%
The first technical challenge is to reliably extract fine-grained vibration signatures.
Based on our experiments on two different phones, we observed that the frequency response of a surface to vibrations introduced by vibration motors installed in COTS smartphones exhibit repeated irregularities, which makes extraction of reliable features a challenging task.
This happens because the phone, its vibrator motor, and the rest of its hardware vibrate at irregular frequencies during every experiment, which occurs due to hardware imperfections.
This behavior is random and uncontrollable, and therefore, is bound to create significant variations within features extracted at the same location, which will lead to classification inaccuracy.
%
%
The existing techniques that use microphone to extract sound (sampled in the order of kHz) based straightforward frequency domain features (\eg vibration sound spectrum \cite{kunze2007symbolic}) are not only considerably susceptible to such hardware based irregularities, but also to short-term and constant environmental noises, where even intermittent talking or noise from a restroom's exhaust fan can significantly affect their performance.
To address this challenge, we take a time-series based vibration signature extraction approach.
First, we differentiate the recorded sound signals and take their root mean square (RMS) envelope, which removes most of the unrelated constant and higher frequency background noise.
Second, we develop a specialized peak-detection based algorithm to extract unique time-series patterns corresponding to vibrations from the RMS envelope, and then use them as \textit{vibration signatures} to represent different surfaces. 
Our extraction algorithm is based on the observation that even if there are irregularities in vibration frequencies due to hardware imperfections, the time-series patterns created during different vibration cycle are very similar.
When a phone vibrates during a specific period of time, multiple such patterns occur all over the time-series of recorded sound signals, which can be successfully extracted by our algorithm.
To make the vibration signatures robust to environmental noise, VibroTag extracts numerous such vibration patterns across time during an experiment and combines them by taking their median.
%
%

The second technical challenge is to compare vibration signatures of any two surfaces. 
The midpoints of extracted vibration signatures of the same surface rarely align with each other because the start and end points determined by extraction algorithm are never perfectly aligned.
Moreover, the lengths of different vibration signatures also differ slightly because the duration of vibration cycle can often be a little different due to hardware irregularities. 
Consequently, the midpoints and lengths of vibration signatures do not match either. 
Another issue is that the shape of different vibration signatures of the same surface are often distorted versions of each other, which occurs due to hardware based irregularities in the vibration mechanism. 
Therefore, two vibration signatures cannot be compared using
standard measures like \textit{correlation coefficient} or \textit{Euclidean
distance}. 
To address this challenge, we use the Dynamic Time Warping (DTW) to quantify the distance between any two vibration signatures. 
DTW can find the minimum distance alignment between two waveforms of different lengths.
For classification, we employ a \textit{Nearest-Neighbor} (NN) classifier with DTW distance as the comparison metric between different vibration signatures.


\presec
\subsection{Key Novelty and Advantages} 
\postsec
The key technical novelty of this paper is on proposing the first fine-grained vibration based sensing scheme that can recognize different surfaces using the vibration mechanism and microphone of a single COTS smartphone.
Furthermore, we propose a novel signal processing technique to extract fine-grained vibration signatures that are robust to hardware irregularities and background environmental noises.
The key insight is that even if there are irregularities in vibration frequencies due to hardware imperfections, the time-series patterns created during different vibration cycles are very similar.
VibroTag finds many such patterns in the sound signals recorded during vibration, and combines them into single consistent vibration signatures.
Compared to previous schemes, VibroTag works with COTS smartphones, it can extract fine-grained features representative of different surfaces, and it is robust to hardware irregularities and background environmental noises.
%

\presec
\subsection{Summary of Experimental Results}
\postsec
We implemented VibroTag on two Android based smartphones, \ie Nexus 4 and OnePlus 2, for which we developed an application for generating vibrations and to sample sound signals simultaneously.
We tested our system for 4 different individuals, from whom we collected data for 5 - 20 days.
We show that VibroTag achieves an average accuracy of 86.55\% while recognizing 24 different locations, even when some of those surfaces were made of similar material, with as few as 15 training samples per location.
%
%
Moreover, VibroTag maintains an average accuracy of up to 85\% without any re-training requirements after 3-4 days of training.
We also implement one of the state-of-the-art IMUs based vibration sensing schemes for single COTS smartphones proposed in \cite{cho2012vibration}, and compare its surface recognition accuracy with VibroTag.
We chose this scheme for comparison because this is the only existing scheme that is insusceptible to acoustic noises in the environment.
We show that VibroTag achieves more than 37\% higher accuracy when compared to the IMUs based scheme, while recognizing the 24 different locations.


%% file: KamranTMC/relatedwork.tex
\presec
\vspace{-0.03in}
\section{Related Work}
\postsec
Existing work related to our paper consists of some vibration based sensing schemes \cite{cho2012vibration, tung2016expansion, goel2012gripsense, wiese2013phoneprioception, hwang2013vibrotactor, darbar2015surfacesense, cho2016vibephone, kunze2007symbolic, laput2016viband, liu2016vibkeyboard, liu2017vibsense, shafer2013learning, ono2013touch, fan2015soqr, goel2014surfacelink, griffin2008user, fan2015soqr} and sound based symbolic localization schemes \cite{tung2015echotag, azizyan2009surroundsense}.

\smallskip \noindent\textbf{Vibration Based Sensing:}
Vibration based sensing schemes leverage the response of different surfaces to a specific vibration pattern to recognize those surfaces.
Existing vibration based sensing schemes can be divided into two categories, \ie custom hardware based, and COTS smartphones based. 
The custom hardware based schemes use separate customized hardware made using a set of micro-controller, vibrator motor, and piezoelectric (\eg microphones) or IMU sensors \cite{kunze2007symbolic, laput2016viband, liu2016vibkeyboard, liu2017vibsense, shafer2013learning}, so that they have fine-grained low level control over different physical layer parameters of their hardware.
ViBand uses variations introduced due to vibrations produced by different objects to identify those objects, \eg electric tooth brush \cite{laput2016viband}.
%
%
VibKeyboard \cite{liu2016vibkeyboard} and VibSense \cite{liu2017vibsense} develop a virtual keyboard based on the idea that the impact of a touch on a surface such as a table or door causes a shockwave to be transmitted through the material that can be passively detected with accelerometers or more sensitive piezo-vibration sensors.
Kunze \etal \cite{kunze2007symbolic} develop customized hardware to recognize surfaces through active sampling of acceleration and sound signatures.
However, the above schemes are incompatible with COTS smartphones and are not easily generalizable to different hardware
because most COTS smartphones have limited sensing capabilities and control over the hardware installed in them.
Most COTS smartphones schemes rely on motion based features extracted from built-in IMU sensors \cite{griffin2008user, cho2012vibration}.
Cho \etal \cite{cho2012vibration} and Shafer \etal \cite{shafer2013learning} use built-in vibrator and accelerometer of a COTS smartphone to recognize surfaces.
Griffin \etal use vibration detected by an acceleration signal to determine if a phone is in the user's hand \cite{griffin2008user}.
%
%
%
However, these features are coarse-grained because the sampling frequencies of IMU sensors on COTS smartphones are much lower (usually between 200-300Hz and less than 4kHz even after driver modifications \cite{laput2016viband}) compared to microphones (usually above 44kHz), which naturally leads to lower classification accuracies.
Because of this reason, such schemes can only broadly differentiate between different types of surfaces (\eg wood and plastic) and cannot differentiate between similar surfaces (\eg two different wooden tables in Figs. \ref{fig:homecontrolled} and \ref{fig:homework}).
Also, IMU readings get significantly affected by the smartphone's own motion in space (\eg when a user moves his hand while holding his smartphone or when the smartphone slides during vibration when placed on a smooth surface).
Moreover, the mainstream COTS smartphones based schemes (may they be IMU or microphone based) are susceptible to inherent hardware based irregularities in vibration mechanism of the smartphones. We discuss the impact of such irregularities in Section \ref{sec:robustnesshardware}.
%
Finally, the existing schemes that use microphones to sense vibration \cite{kunze2007symbolic, hwang2013vibrotactor, darbar2015surfacesense} are prone to short-term and constant background noises (\eg intermittent talking, clapping, exhaust fan, etc.) because microphones not only capture the sounds created by vibration but also other interfering sounds present in the environment. We discuss the impact of such noises in Section \ref{sec:robustnessnoise}.
%
%
%
%
%
%
Compared to all the above schemes, VibroTag is robust and easily deployable because it works with COTS smartphones, it can extract fine-grained features representative of different surfaces/locations, and it is robust to hardware irregularities and background environmental noises.


\smallskip \noindent\textbf{Sound Based Symbolic Localization:}
Sound based symbolic localization systems leverage the propagation of the sound generated using speakers of a device, such as a smartphone, to determine the symbolic location of that device (\eg whether the device is in the user's kitchen or at his bedroom table).
SurroundSense uses sensor data from a microphone, a light sensor, the wireless radio, and passive accelerometer data for localization \cite{azizyan2009surroundsense}.
However, their technique can only be used for very coarse-grained localization (\eg room level) and not for finer-grained localization (\eg whether the phone is on user's study table or his bedroom table).
EchoTag generates ultrasound signals and then uses the reflections from the environment to achieve centimeter level tagging \cite{tung2015echotag}.
However, their work requires strict millimeter level marking of the tagged locations because the ultrasound signals based signatures that they use are highly location dependent, where even small variations in the phone's position leads to significant localization errors.
This makes their scheme unsuitable for symbolic localization, and also puts significant calibration effort on the user end.
In contrast to above schemes, VibroTag uses vibration instead of speaker generated sound signals for such symbolic localization.
Moreover, VibroTag achieves finer-grained localization, and does not require strict marking of the tagged locations. 
%

%% file: KamranTMC/workingprinciples.tex
\presec \section{Understanding Vibrations} \label{sec:workingprinciples} \postsec
 \subsection{Vibrator Motors in Smartphones} 
\begin{figure}[htbp]
	\centering
	\captionsetup{justification=centering} 
	\includegraphics[width=1\columnwidth]{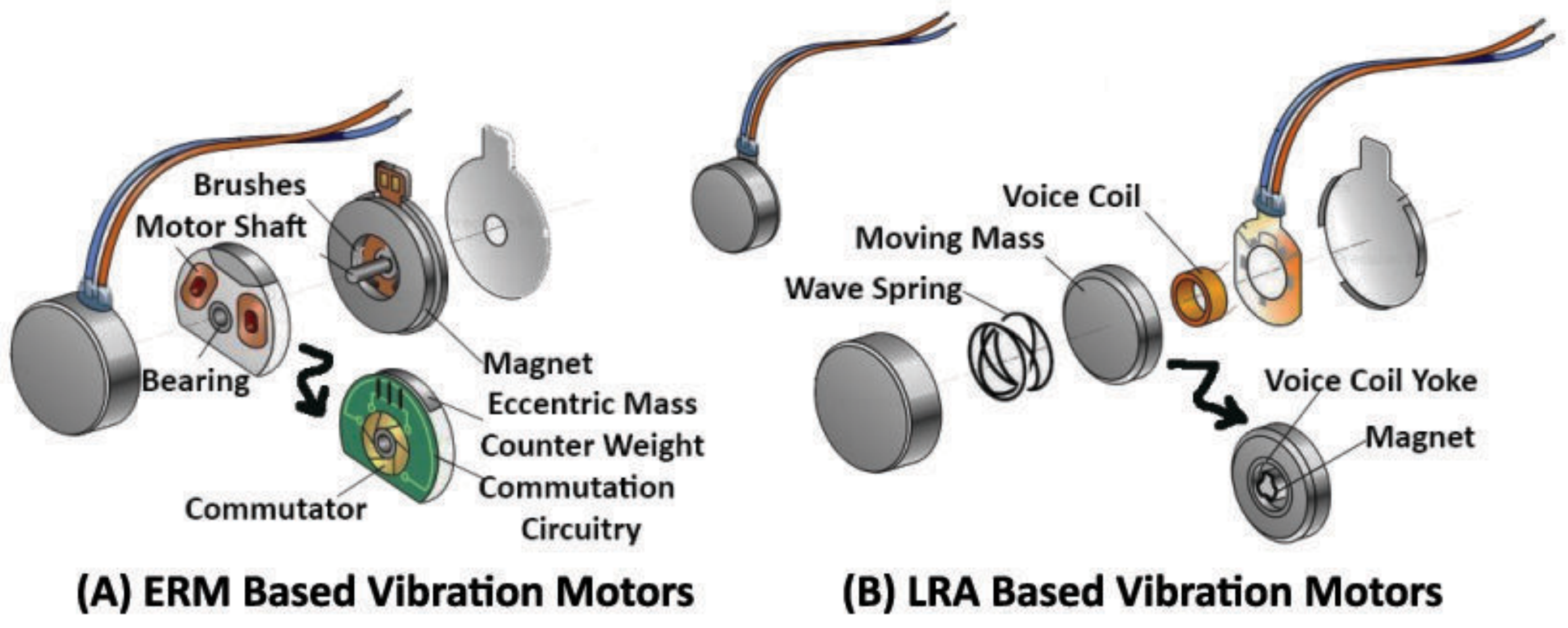}
	\vspace{-0.15in}
	\caption{ERM and LRA based vibration motors \cite{sparkfunguide}}
	\label{fig:motors}
	\vspace{-0.15in}
\end{figure}

\postsub
Electric vibrator motors generate vibrations by periodically moving an unbalanced mass around a center position using the principles of electromagnetic induction.
The vibrator motors used in today's smartphones are often known as \textit{coin-type vibration motors} due to their coin-like shapes and sizes.
There are two types of coin-type vibrator motors that are widely adopted in smartphones: (i) \textit{Linear Resonant Actuator} (LRA) based (\eg used in Nexus 4) and  (ii) \textit{Eccentric Rotating Mass} (ERM) based (\eg used in OnePlus 2).
Figure \ref{fig:motors} shows the internals of ERM and LRA based vibration motors.
ERM based motors use a DC motor to rotate an eccentric mass around an axis. 
As the mass is not symmetric with respect to its axis of rotation, it causes the device to vibrate during the motion. 
Both the amplitude and frequency of vibration depend on the rotational speed of the motor, which can in turn be controlled through an input DC voltage. 
With increasing input voltages, both amplitude and frequency increase almost linearly and can be measured by an accelerometer.
In LRA based motors, vibration is generated by the linear movement of a magnetic mass suspended near a coil, called the ``voice coil''. 
When an AC current  is applied to the motor, the coil behaves like a magnet (due to the generated electromagnetic field) and causes the mass to be attracted or repelled, depending on the direction of the current. 
This generates vibration at the same frequency as the input AC signal, while the amplitude of vibration is dictated by the signal's peak-to-peak voltage. 
Thus, LRAs offer control on both the magnitude and frequency of vibration. 
%

\presub \subsection{Physics of Surface Response to Vibrations} \postsub
Sound is essentially pressure waves created by vibrating matter. 
These waves are longitudinal, \ie they oscillate along the axis of travel, where the oscillation is composed of compression and rarefaction of molecules in the medium (\eg air).
For example, human speech is based on vibrations created inside our vocal chords, and audio speakers generate sound by translating an electrical signal into physical vibrations via mechanical excitation of a diaphragm using an electromagnet.

VibroTag is based on the intuition that different surfaces exhibit different response to vibrations introduced by smartphone.
When a smartphone vibrates, it mechanically excites not only it's own structure and hardware inside it, but also the surface on which it is placed.
On one hand, some surfaces tend to absorb most of the vibration energy (\eg Sofas), while on the other hand, some surfaces may exhibit a resonant response where they start vibrating in sync with the smartphone (\eg the smartphone's surface vibrates in sync with the vibrator motor inside).
Moreover, the effect of these vibrations can reach different objects placed nearby, which may get mechanically excited as well (especially the lighter objects); therefore, leading to more peculiar sounds.
As different surfaces respond to the vibrations differently (in terms of their absorption/dampening effect on smartphone's movements), and as different surfaces often have different objects placed on them, which also respond to those vibrations differently, pressure waves peculiar to those surfaces are created during the vibration, which we can sense using a piezoelectric device (\eg a built-in microphone) and then leverage to differentiate those surfaces.

%% file: KamranTMC/feature.tex
\presec\postsec \section{Feature Extraction} \postsec
To differentiate between different locations, we need to extract features that can uniquely and consistently represent those locations. 
In VibroTag, a smartphone is vibrated for about 3 seconds while the surface response to the vibration is recorded simultaneously via the phone's built-in microphone.
Sounds produced during vibration are sampled at fixed $F_s =$ 44.1 kHz.
The recorded sound is analyzed in both frequency and time domains to extract robust surface/location specific vibration signatures.
There are two key challenges in feature extraction for VibroTag to be robust.
The first challenge is on reducing impact of background noises (such as those created by fans and short-term human speech).
The second challenge is on accommodating smartphone hardware imperfections (\ie microphones and vibrator motors mainly) that degrades the quality of the signals collected when a smartphone vibrates.

\presub \subsection{Robustness to Background Noise} \label{sec:robustnessnoise} 
\begin{figure*}[htbp]
	\centering
	\captionsetup{justification=centering}    
	\captionsetup[subfigure]{aboveskip=20pt,belowskip=20pt}
	\subfigure[No noise]{
		\includegraphics[width=0.72\columnwidth]{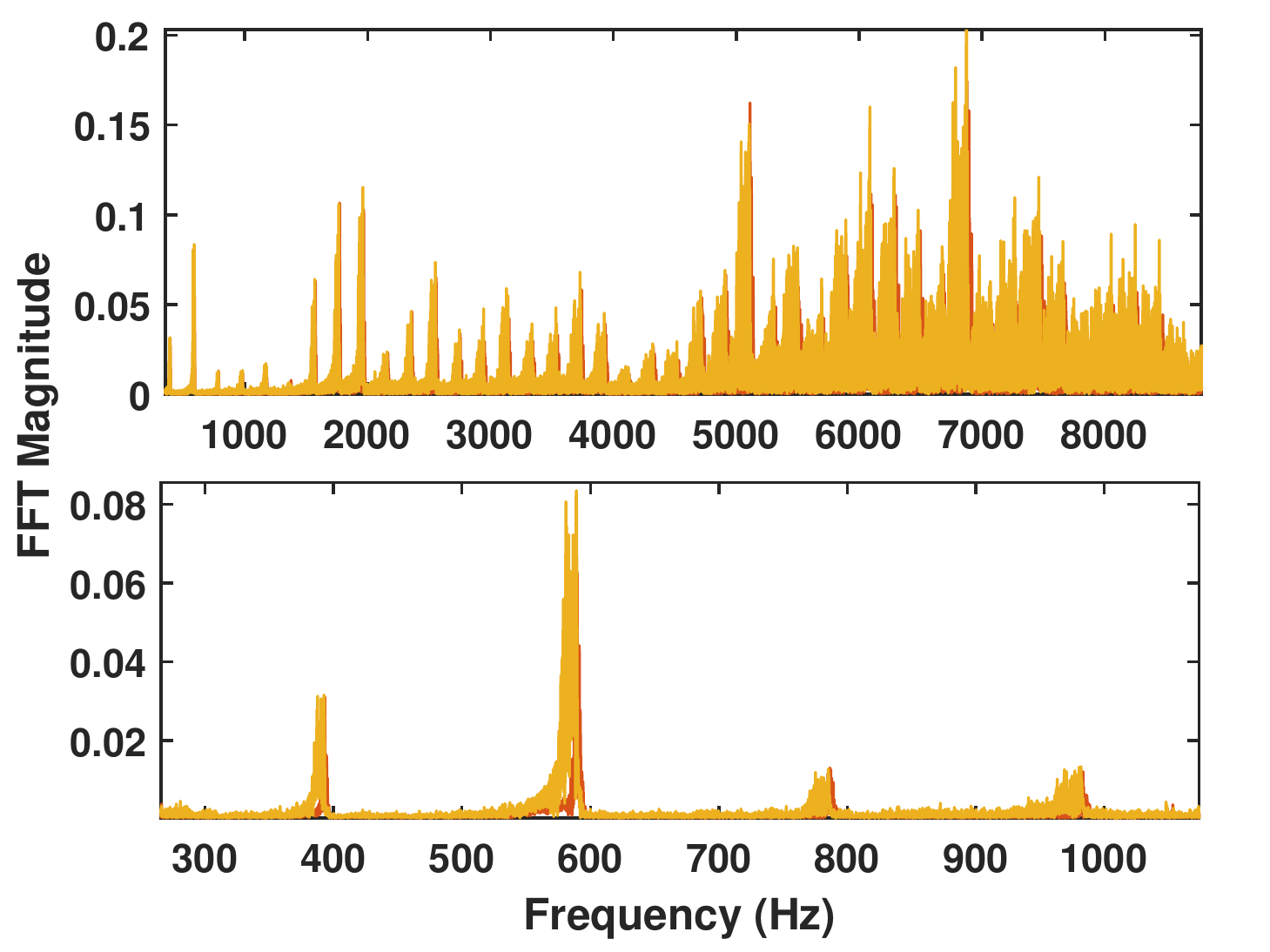}
		\label{fig:noise1}}
	\hspace{-0.13in}
	\subfigure[Intermittent Talking]{
		\includegraphics[width=0.72\columnwidth]{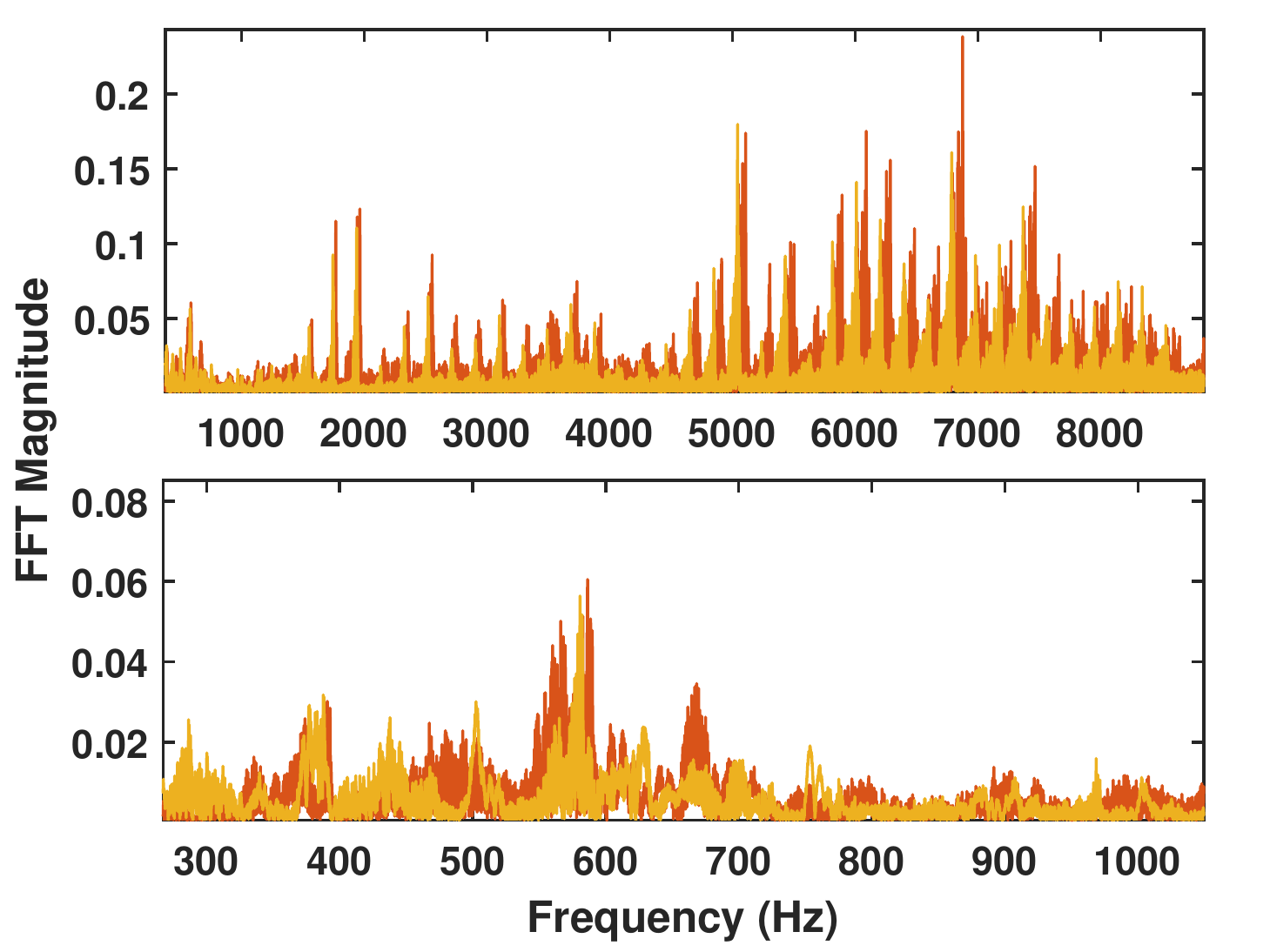}
		\label{fig:noise2}}
	\hspace{-0.12in}
	
	\vspace{-0.14in}
	
	\subfigure[Clapping]{
		{\includegraphics[width=0.72\columnwidth]{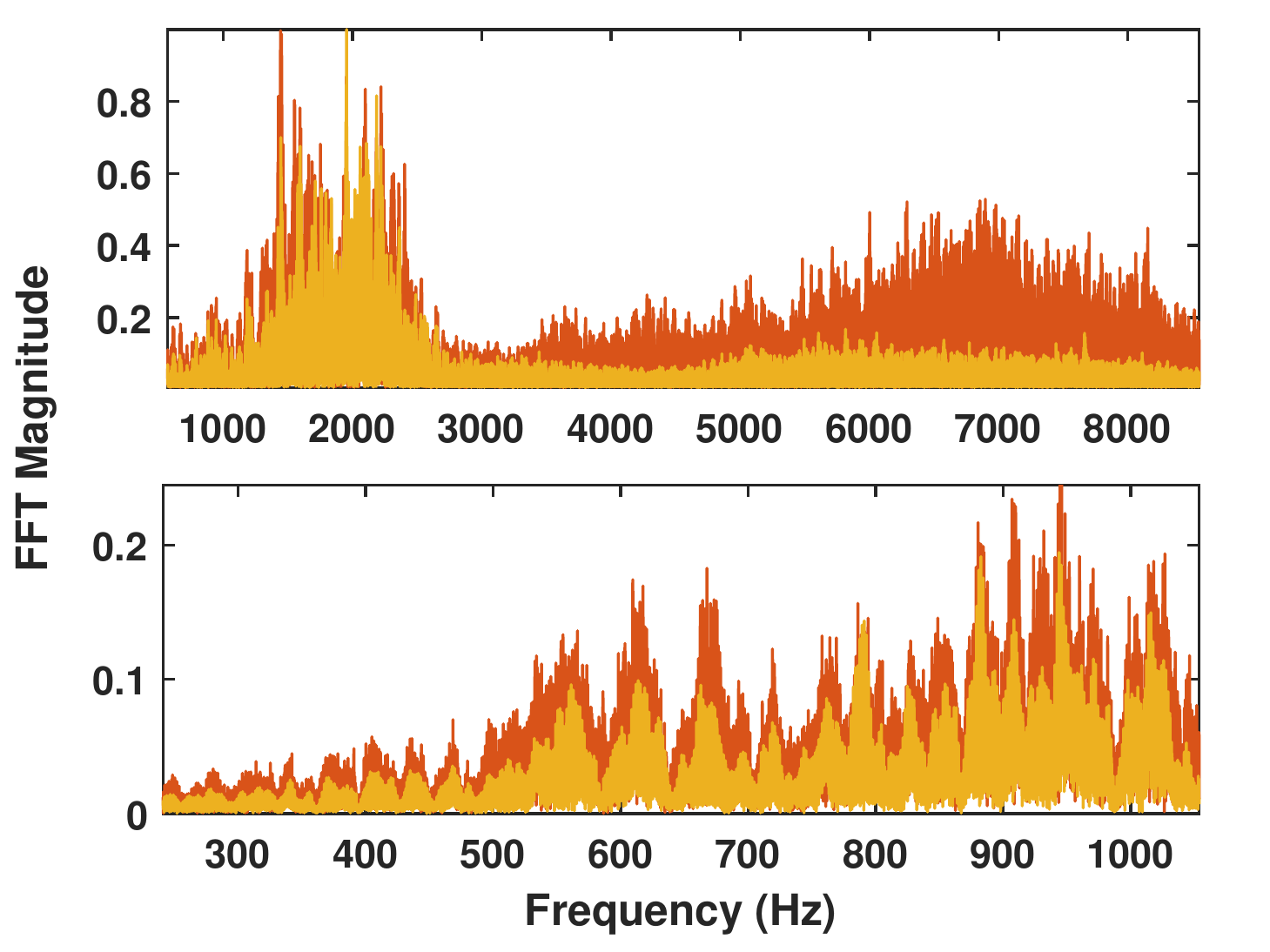}}
		\label{fig:noise3}}
	\hspace{-0.12in}
	\subfigure[VibroTag signatures (a)-(c)]{
		{\includegraphics[width=0.708\columnwidth]{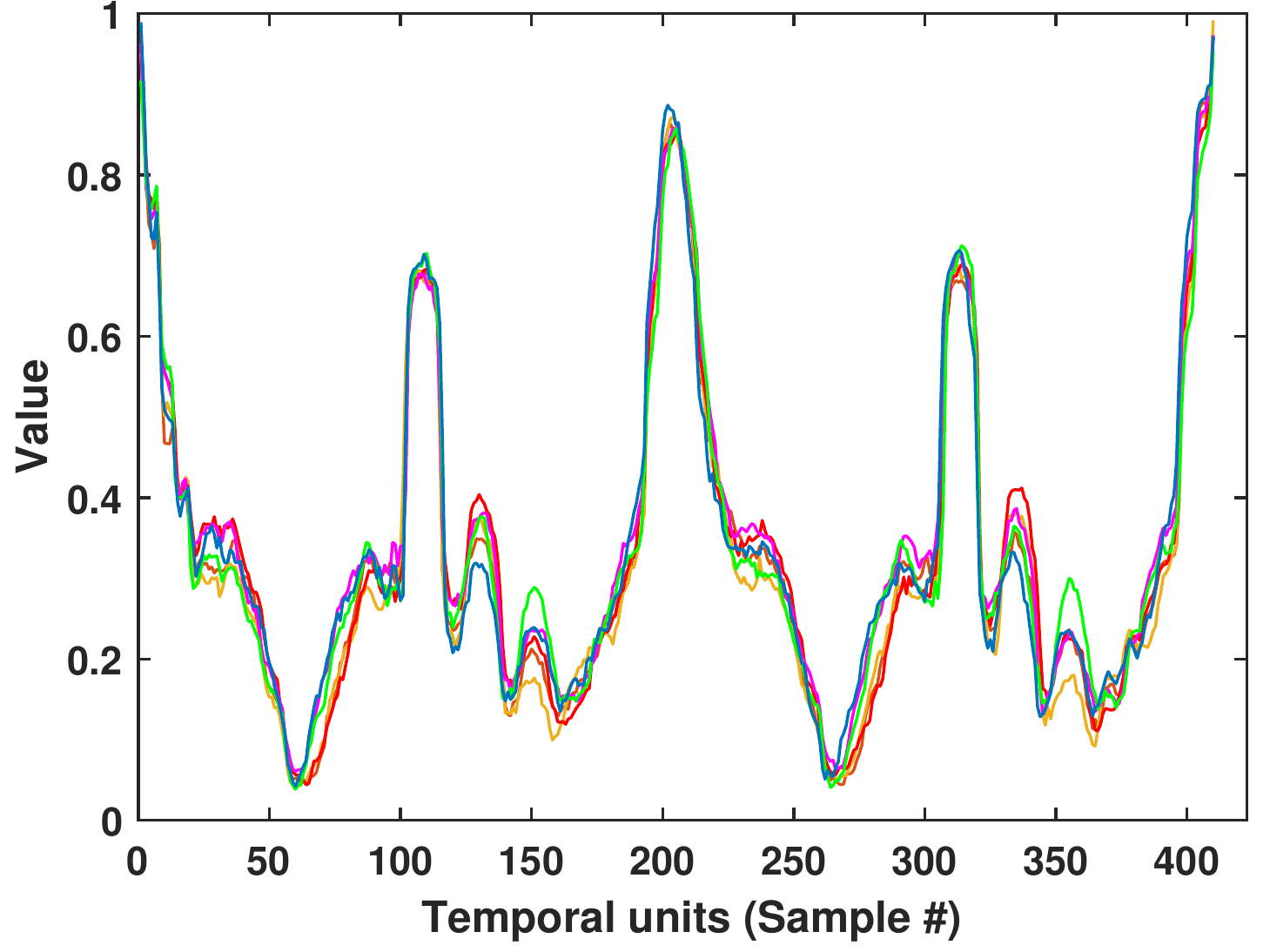}}
		\label{fig:noise4}}
	\label{fig:noise}
	\vspace{-0.12in}
	\caption{Impact of background noises on features extracted by traditional techniques and VibroTag}
	\vspace{-0.12in}
\end{figure*}

\postsub
To understand the challenge posed by background noise, we use Fast Fourier Transform (FFT) based Power Spectral Density (PSD), which is one of the mainstream frequency based feature extraction techniques for acoustic sensing.
%
Figure \ref{fig:noise1}, \ref{fig:noise2}, \ref{fig:noise3} show the FFT coefficients for both lower and higher frequency ranges corresponding to our experiments conducted at the same location on a wooden chair's cushion for three scenarios: (a) no noise, (b) intermittent human speech, and (c) clapping, respectively.
We can observe that the FFT features are significantly affected by the background noises because the frequencies produced by these noises directly interfere with the frequency bands for vibration based sensing.
It also shows that mainstream techniques such as FFT or PSD are unsuitable for vibration based sensing on COTS smartphones when there are background noise sources present in the environment.
In this paper, we propose two schemes to reduce the impact of constant and intermittent short-term background noises, respectively.

To reduce the impact of constant background noises in VibroTag, we take the \textit{first order difference}  of the recorded sound signals and then take their \textit{root mean squared} (RMS) envelope.
We choose RMS envelope for our analysis as it gives us a measure of the power of the vibration signals, while producing a waveform that is easy to analyze. 
Moreover, higher frequency noisy variations are averaged out in the envelope signal, while it still keeps most of the useful vibration response related information intact.
We take the RMS envelope of the signals over a sliding window of N samples, where N = 15 audio samples in our current implementation of VibroTag.
In the rest of this paper, when we mention sound signals, we mean the first order difference of the RMS envelope of those sound signals.

To reduce the impact of intermittent short-term noises (similar to Figures \ref{fig:noise2} and \ref{fig:noise3}), we vibrate the phone for at least 3 seconds, and extract multiple vibration patterns across time from the processed sound signals; then, we combine these virbration patterns to get a single consistent vibration signature.
We will discuss how we extract such signatures in Section \ref{sec:timeseries}.
Figure \ref{fig:noise4} shows the signatures extracted by VibroTag for the experiments corresponding to Figures \ref{fig:noise1}, \ref{fig:noise2}, \ref{fig:noise3}, from which we observe that our signatures are consistent and almost identical even though there are intermittent short-term noises.


\presub \subsection{Robustness to Hardware Imperfections} \label{sec:robustnesshardware} 
\begin{figure*}[htbp]
	\vspace{-0.1in}
	\centering
	\captionsetup{justification=centering}    
	\captionsetup[subfigure]{aboveskip=20pt,belowskip=20pt}
	\subfigure[Home Bed]{
		\includegraphics[width=0.74\columnwidth]{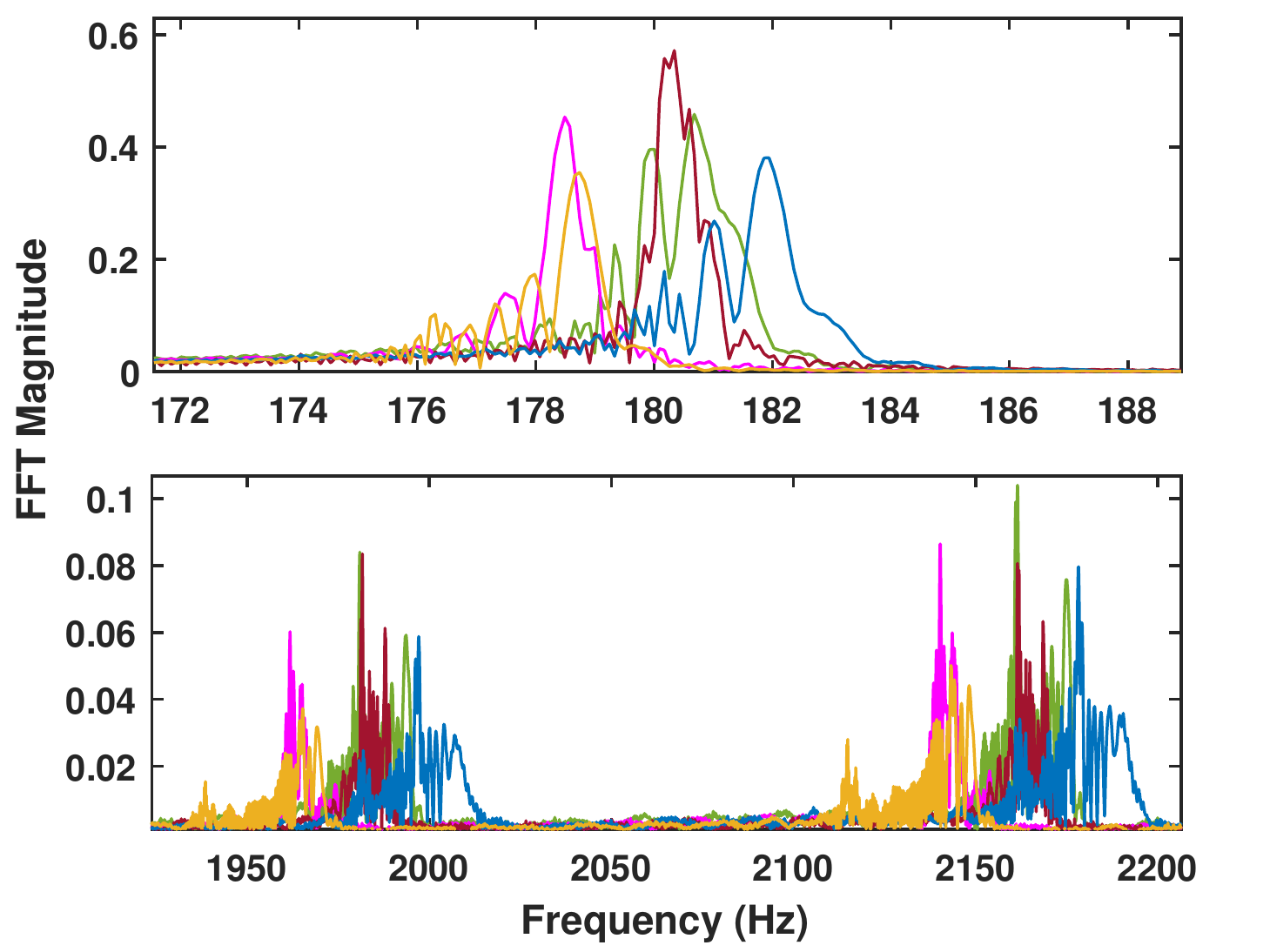}
		\label{fig:psdexp1}}
	\hspace{-0.23in}
	\subfigure[Home Living Room Sofa]{
		{\includegraphics[width=0.74\columnwidth]{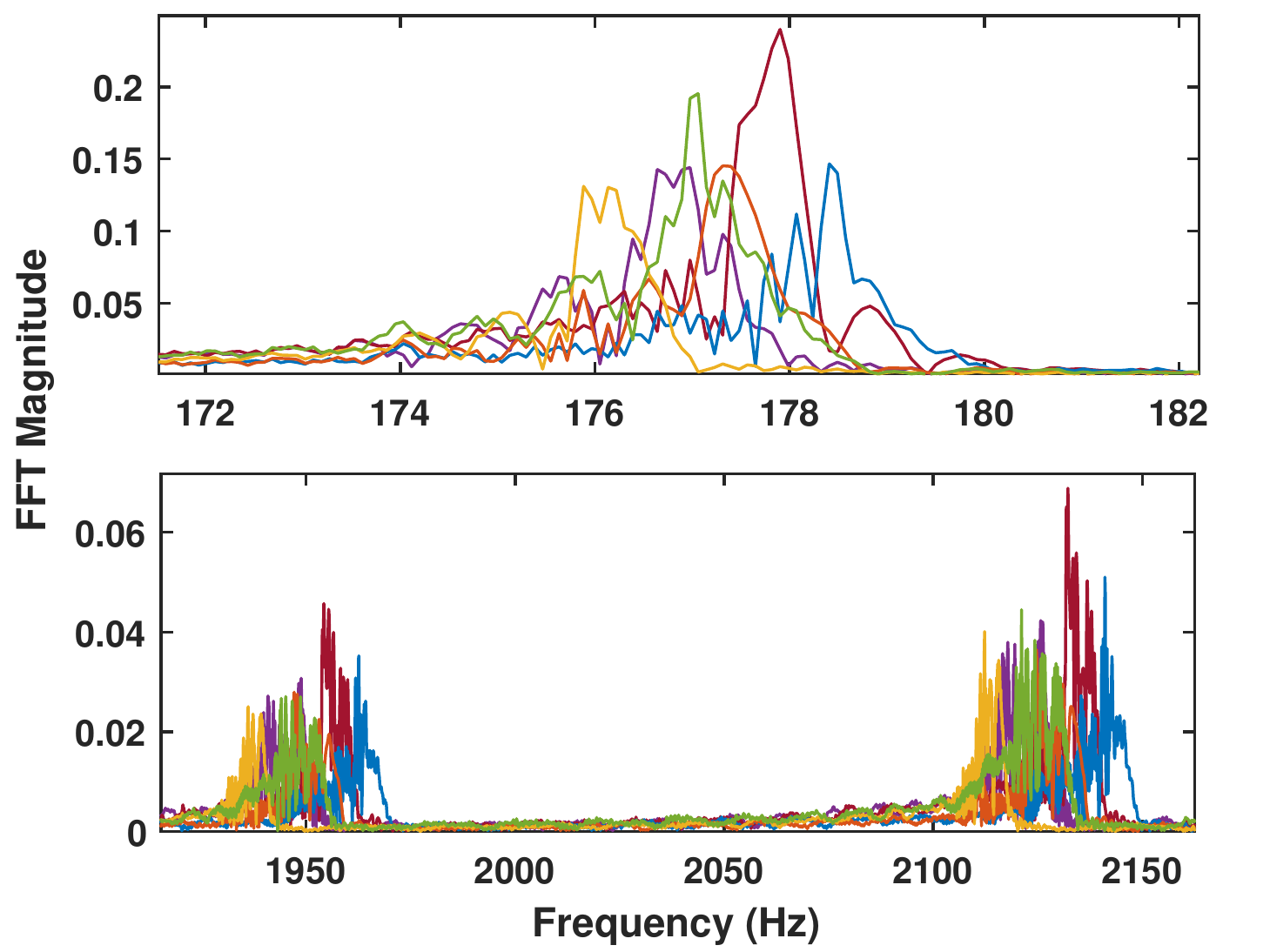}}
		\label{fig:psdexp2}}
	
	\vspace{-0.14in}
	
	\subfigure[Home Bed]{
		\includegraphics[width=0.75\columnwidth]{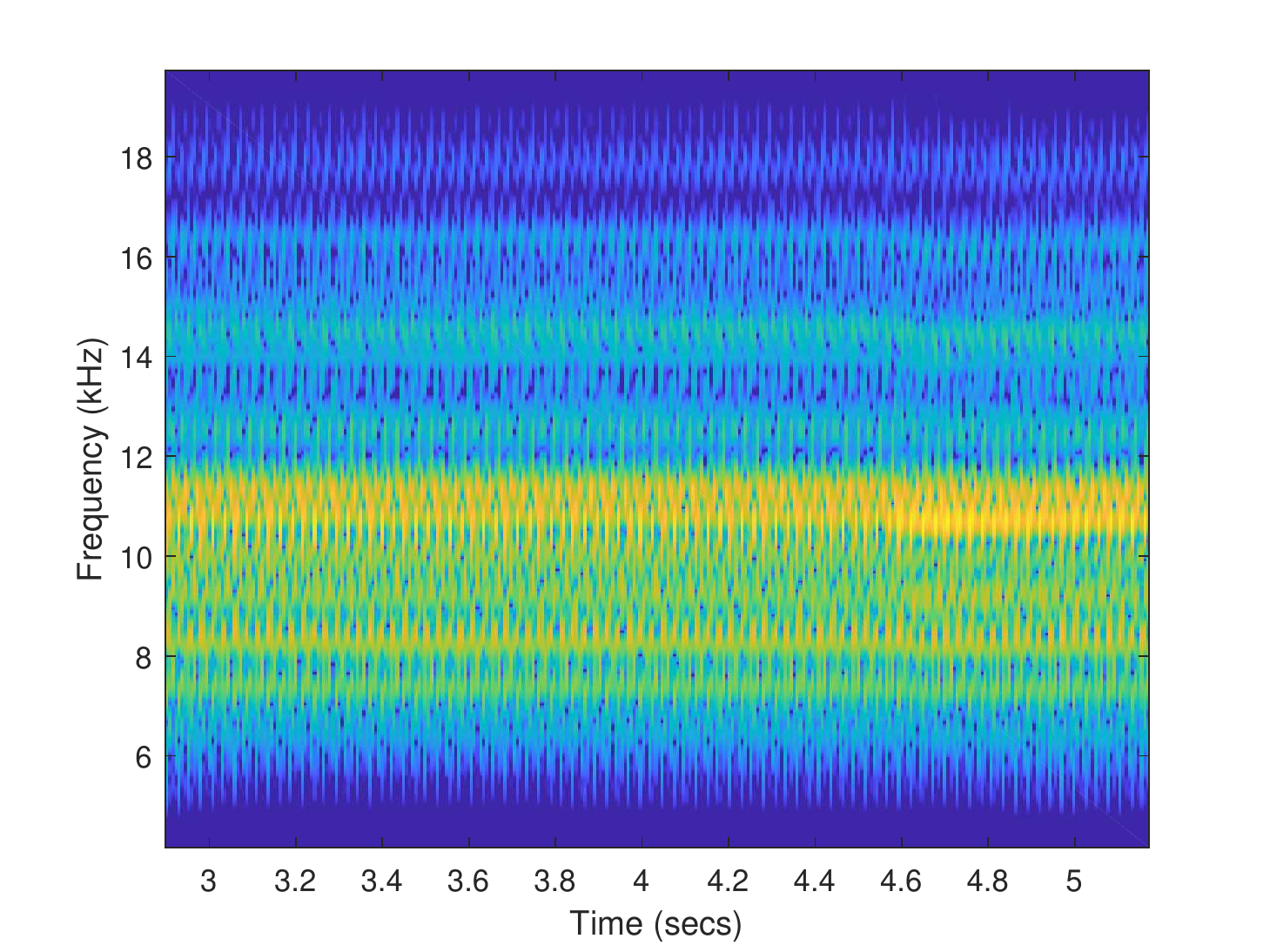}
		\label{fig:enveopestftexp1}}
	\hspace{-0.22in}
	\subfigure[Home Living Room Sofa]{
		{\includegraphics[width=0.75\columnwidth]{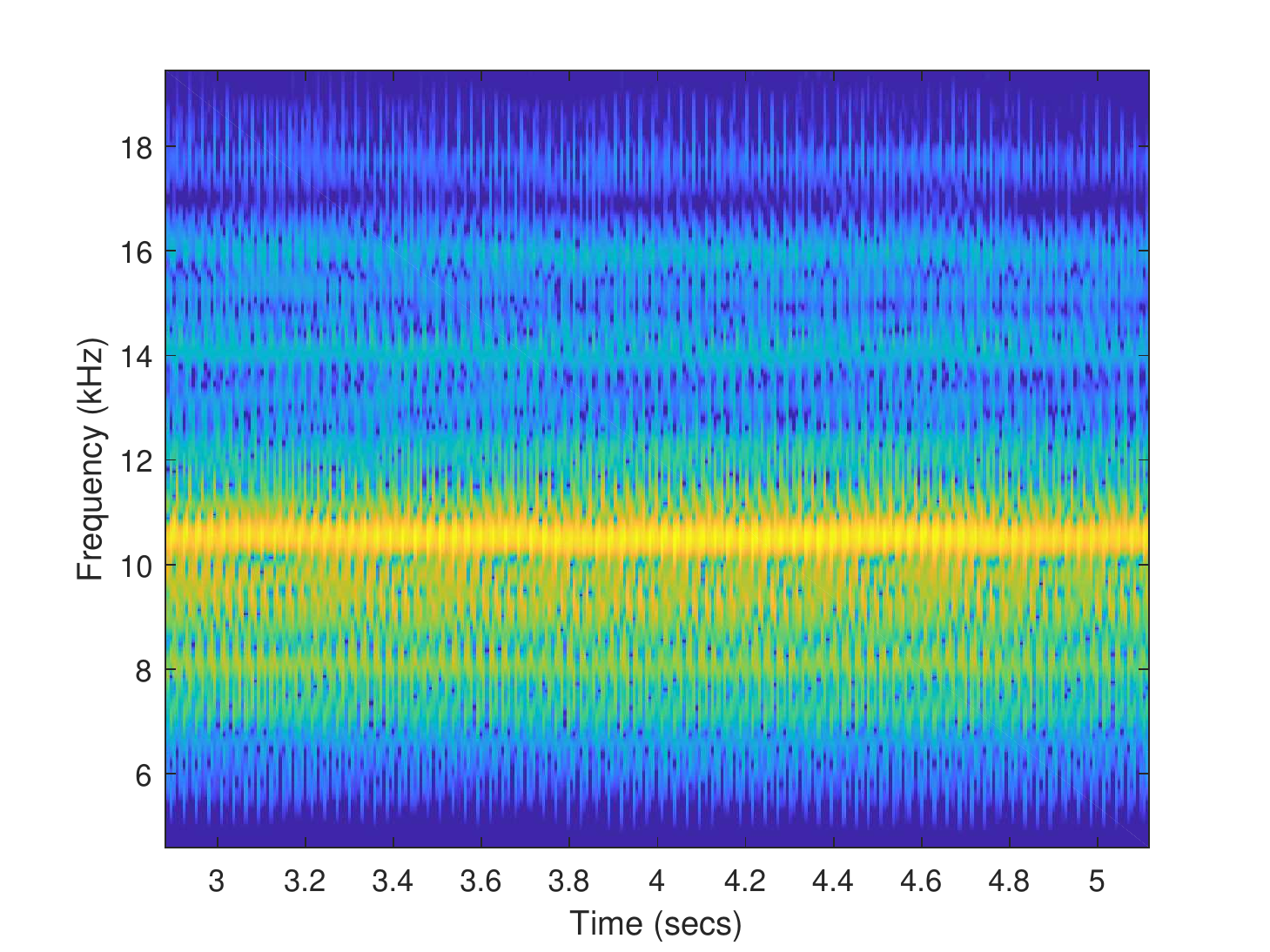}}
		\label{fig:envelopestftexp2}}
	\label{fig:psdexps}
	\vspace{-0.15in}
	\caption{Impact of hardware imperfection on features extracted by traditional techniques PSD and STFT}
	\vspace{-0.15in}
\end{figure*}

\postsub
To understand the challenge posed by smartphone hardware imperfections, we use PSD and Short Time Fourier Transform (STFT).
%
Figures \ref{fig:psdexp1} and \ref{fig:psdexp2} show the PSD of the recorded unprocessed sound signals for multiple different experiments, which we performed by placing a smartphone at the same location on a sofa and a bed, respectively.
Each figure shows the PSD coefficients over two different frequency ranges.
We can observe that PSDs for both the sofa and the bed are not consistent for repetitive experiments, even when the smartphone's location and the environmental scenario while performing the experiments remained unchanged.
This occurs because the smartphone, its vibrator motor, and the rest of its hardware vibrate at slightly different frequencies in each different experiment.
This behavior is random and uncontrollable, and therefore, is bound to cause intra-class (\ie within samples of the same class) variations, which will lead to classification errors.
Moreover, due to this inconsistency, it often happens that the variations due to vibration on two different surfaces occur in similar set of frequencies (as shown by some samples in Figs. \ref{fig:psdexp1} and \ref{fig:psdexp2}), which further makes the use of such frequency domain features difficult as they do not always represent different surfaces uniquely.
%

%
We hypothesize that even if there are irregularities in the repetition frequency of such vibration patterns, the patterns themselves must be very similar.
Figs. \ref{fig:enveopestftexp1} and \ref{fig:envelopestftexp2} show the STFT of the unprocessed sound signals from one experiment corresponding to each of the two surfaces.
%
%
Interestingly, we observe that some patterns repeating along time; however, the time period of their repetition is not consistent due to the aforementioned hardware irregularities.
As we discussed in \S \ref{sec:workingprinciples}, smartphone vibrations are generated because of the to and fro motion of a mass inside its vibration motor.
The vibration motor tries to move the smartphone (and the hardware inside) at its own vibration frequency (which is often irregular due hardware imperfections).
However, the smartphone's motion is constrained due to its own weight/structure and the absorption properties of the surface that it is placed on.
This whole process during vibration gives rise to peculiar pressure waves, which can be sensed by the built-in microphone.
Moreover, as the mass inside the motor repeatedly moves to and fro, it will give rise to similar pressure waves in every cycle of its ``irregular'' vibration period, which will reflect in time-series of the sound signals.
These intuitions form the basis of our time-series based analysis of the surfaces' vibration response (\S \ref{sec:timeseries}). 

\begin{figure*}[htbp]
	\centering
	\captionsetup{justification=centering}    
	\captionsetup[subfigure]{aboveskip=-1pt,belowskip=-1pt}
	\subfigure[Home Bed]{
		\includegraphics[width=1.4\columnwidth]{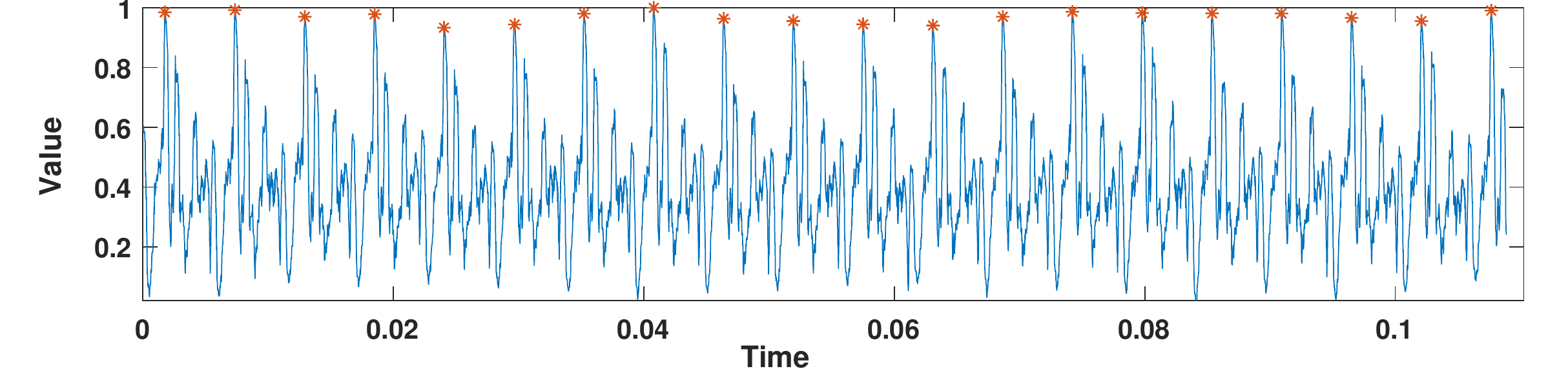}
		\label{fig:peaksexp1}}
	
	\vspace{-0.14in}
	
	\subfigure[Home Living Room Sofa]{
		{\includegraphics[width=1.4\columnwidth]{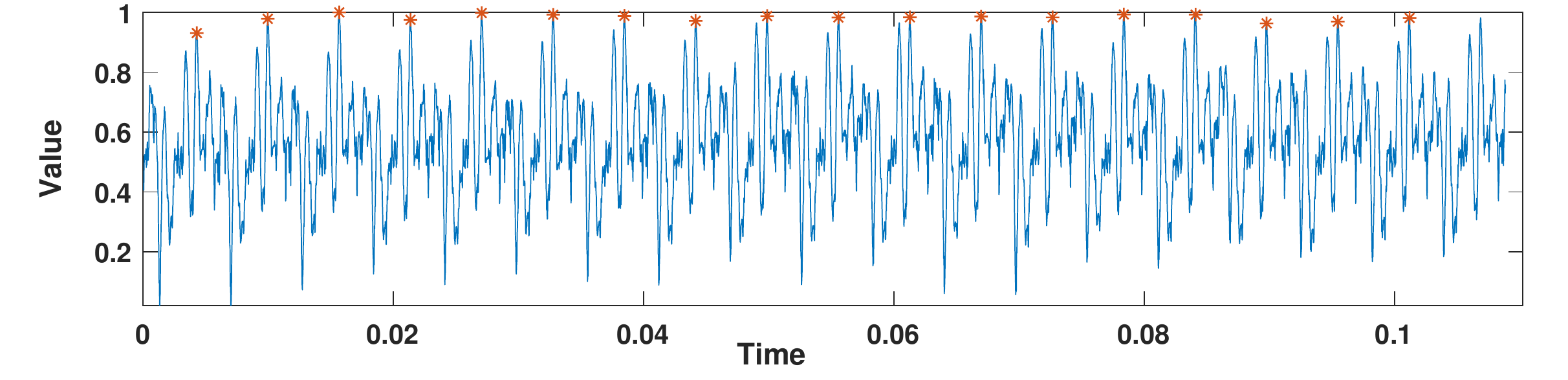}}
		\label{fig:peaksexp2}}
	\label{fig:peaksexps}
	\vspace{-0.15in}
	\caption{Repetitive patterns appearing in the processed sound signals corresponding to vibration}
	\vspace{-0.15in}
\end{figure*}

In this paper, we propose to address the issues due to smartphone hardware imperfections by extracting time-series based vibration signatures from the processed sound signals. 
Figures \ref{fig:peaksexp1} and \ref{fig:peaksexp2} show time-series of the sound signals (\ie first order difference of the RMS envelope) corresponding to one experiment from each of the two different surfaces (\ie Bed and Sofa), whose PSDs are shown in \ref{fig:psdexp1} and \ref{fig:psdexp2} and whose STFTs are shown in \ref{fig:enveopestftexp1} and \ref{fig:envelopestftexp2} , respectively.
The two time-series correspond to a window of 4800 sound samples (\ie $\sim$0.1088 seconds for $F_s =$ 44.1 kHz).
We can easily observe distinguishing patterns repeating in both time-series, which repeat approximately with the frequency of the vibrating mass in the smartphone's vibration motor.
We also observe that the patterns in both scenarios are consistent across time, and that the patterns in one scenario are different from the ones in other scenario.

\presec\postsec\subsection{Extraction of Vibration Signature} \label{sec:timeseries}\postsec
To robustly differentiate surfaces, we need to extract vibration patterns from time-series of the processed sound signals and then use those patterns to obtain consistent vibration signatures.
However, we face multiple challenges.
%
%
%
%
%
The first challenge is that intermittent short-term noises in real-life scenarios are uncontrollable, and therefore, can affect any part of the time-series.
VibroTag needs to extract the vibration patterns that are representative of the whole time-series, \ie the vibration patterns extracted from one segment of the time-series (\ie a time window) should repeat in other segments, and therefore, truly represent the surface's vibration response.
A naive approach is to extract all vibration patterns in the recorded signals and then combine them (\eg by taking their average), which is computationally expensive.
To address this challenge, we take a \textit{randomized} approach, where we first divide the whole time-series of the sound signal into equally sized time windows, and then randomly select multiple of those time windows to extract vibration patterns from.
Each window is of size $S$, where $S = $ 4800 sound samples in our current implementation of VibroTag.
Moreover, the windows are selected without replacement, \ie once selected, they are not selected again.
VibroTag keeps randomly selecting new time windows until $M$ vibration patterns are extracted ($M$=100 in our implementation).
\textit{In real-life scenarios, the number of iterations required to extract $M$ vibration patterns of a surface can be used to tell the user whether their environment is too noisy to extract a valid vibration signature or not.}
For example, the average (taken over 20 different samples) number of iterations it took for convergence when loud music (\ie high variable noise) was played on a laptop in the background was $\sim$1351, for medium noise/volume level the number was $\sim$495, whereas for no variable noise scenarios it was $\sim$136.
In the cases where our algorithm cannot find enough vibration patterns and runs out of possible time windows to search for patterns, it will not converge. 
However, we did not experience any such scenarios during our testing.


The second challenge is to extract the vibration patterns from different randomly selected time windows by localizing their place of occurrence in those time windows.
However, because of the inconsistencies in the vibration behavior of smartphone due to hardware imperfections, we cannot know the frequency of repetition of the vibration patterns, which makes it harder to localize the place of occurrence of such patterns.
To address this challenge, we develop a \textit{peak detection} based algorithm to extract vibration patterns.
Our algorithm is based on the observation that every vibration pattern has a peak value that occurs consistently at around the same part of every vibration cycle (as evident in Figs. \ref{fig:peaksexp1} and \ref{fig:peaksexp2}).
Based on this algorithm, VibroTag determines the locations of multiple such peaks in each of the randomly selected time windows.
Afterwards, VibroTag uses the consecutive peaks detected in each window to extract the vibration patterns between those peaks.
%

\presub
\subsubsection{Extraction of Vibration Patterns}
\postsub
There are two key challenges in developing our peak detection based vibration pattern extraction algorithm.
First, based on our experiments, we observe that the scale of variations due to vibration in the time-series of different randomly selected windows varies, even when the same smartphone is placed at the same location of the same surface, which happens due to hardware imperfections based inconsistencies in the vibration process. 
Moreover, different surfaces and different smartphones exhibit different scale of variations due to vibration.
This makes the parametrization of our peak detection algorithm difficult to generalize.
To address this challenge, VibroTag performs \textit{max-min normalization} on the time-series corresponding to every selected window before feeding it to the peak detection algorithm.
This step ensures that the parametrization of our algorithm can be easily generalized to different time windows and to different smartphones and surfaces.

\begin{figure}[htbp]
	\centering
	\captionsetup{justification=centering}    
	\captionsetup[subfigure]{aboveskip=20pt,belowskip=20pt}
	\subfigure[Home Bed]{
		\includegraphics[width=0.72\columnwidth]{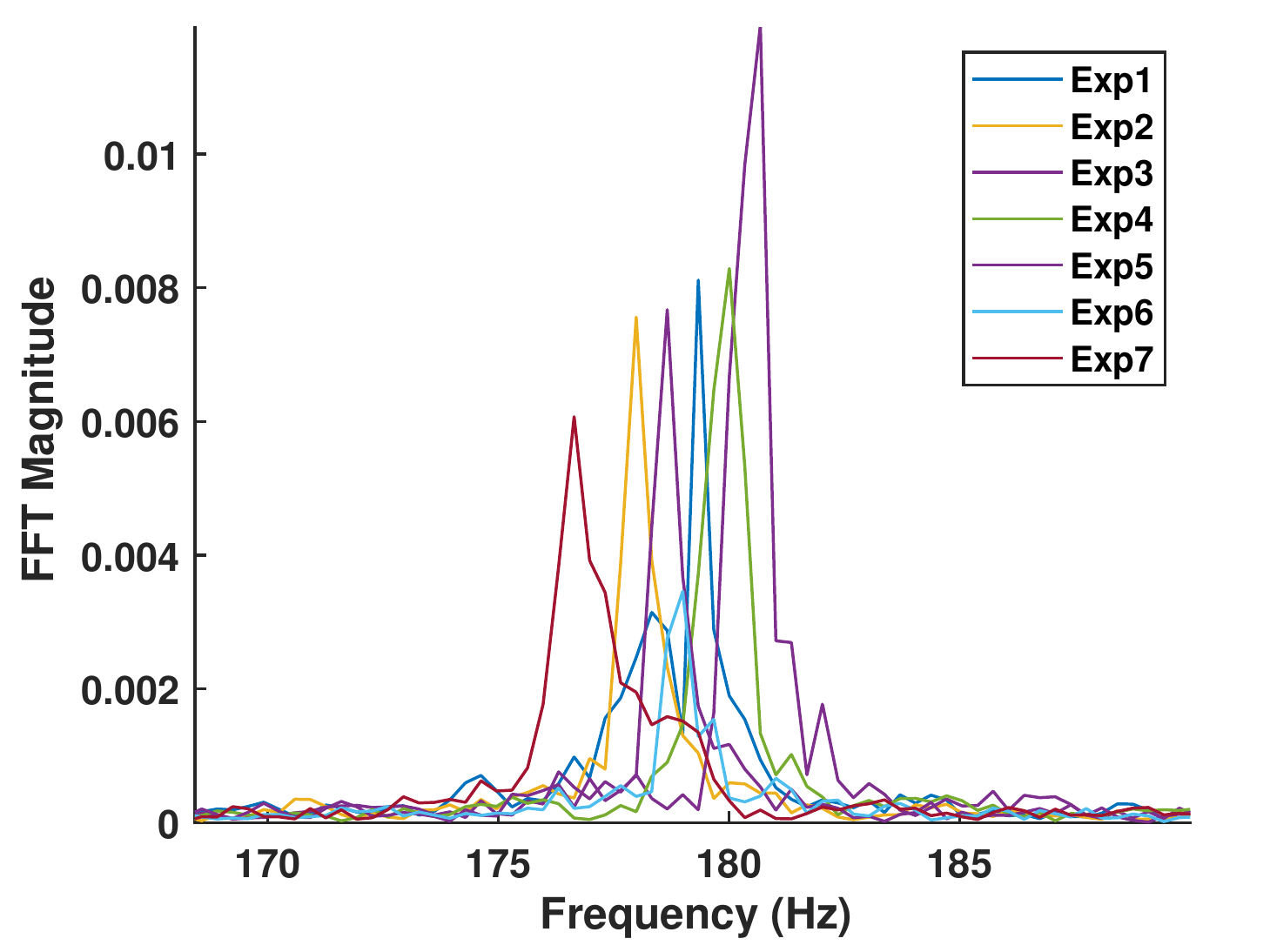}
		\label{fig:envelopermspsd1}}
	\hspace{-0.21in}
	\subfigure[Home Living Room Sofa]{
		{\includegraphics[width=0.72\columnwidth]{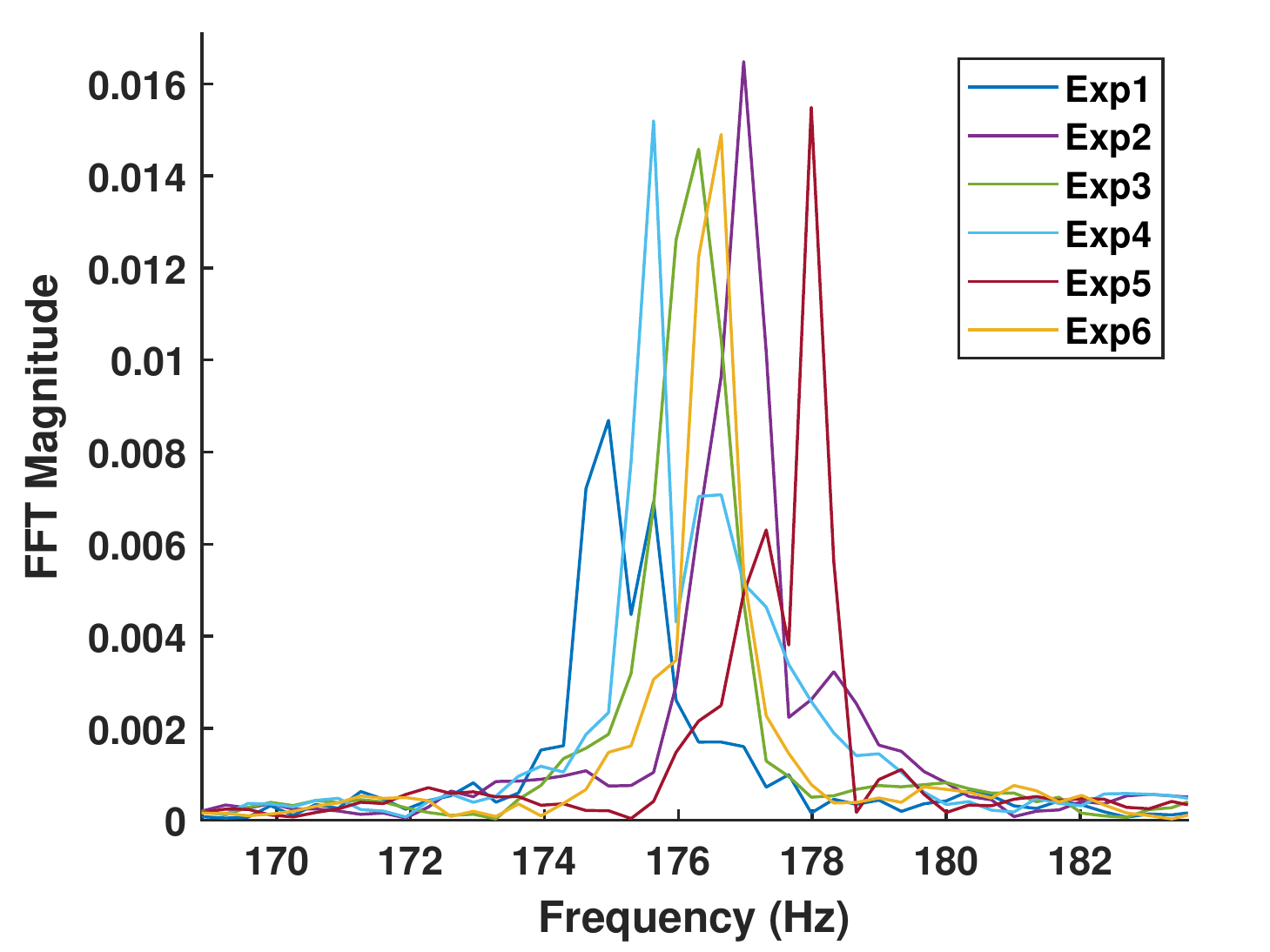}}
		\label{fig:envelopermspsd2}}
	\label{fig:envelopermspsd}
	\vspace{-0.15in}
	\caption{PSD of the sound signals in time windows corresponding to the scenarios in Figs. \ref{fig:peaksexp1}-\ref{fig:peaksexp2}}
	\vspace{-0.1in}
\end{figure}

The second challenge is to robustly determine the locations of peaks corresponding to different vibration patterns present in a time window.
To address this, VibroTag's peak detection algorithm determines the location of such peaks based on three key parameters, namely \textit{minimum peak prominence} (MINPRO), \textit{minimum peak distance} (MINDIST), and \textit{minimum peak strength} (MINSTR).
The prominence of a peak measures how much the peak stands out, due to its height and location, relative to other peaks around it. 
We tune MINPRO such that we only detect those peaks which have a relative importance of at least MINPRO.
We tune MINDIST according to the fact that maximum repetition rate of patterns is approximately $\hat{f}_{o} + \delta f$, such that the redundant peaks are discarded, where $\hat{f}_{o}$ is an approximate number for the the frequency of repetition ($f_o$) of vibration patterns.
As we discussed before, the frequency of repetition of vibration patterns in the processed sound signal is irregular.
In order to determine $\hat{f}_{o}$, VibroTag calculates PSD of the time-series in the selected window.
Figures \ref{fig:envelopermspsd1} and \ref{fig:envelopermspsd2} show example PSD's corresponding to one of the randomly selected time windows from seven different experiments performed on the Bed and the Sofa, respectively.
VibroTag determines the peak frequency from the PSD, which corresponds to the approximate repetition frequency (\ie $\hat{f}_{o}$) of the vibration patterns present in the window.
The term $\delta f$ represents the second standard deviation of the variation in vibration frequency around $\hat{f}_{o}$, which we estimate for every smartphone based on multiple different experiments.
$\delta f$ is required as some vibration patterns can repeat earlier than $1/\hat{f}_{o}$.

To further sift out redundant peaks, we only choose peaks of value greater than MINSTR times the median value of the peaks detected in the window.
In our current implementation, we chose MINPRO = 0.65, MINDIST = $\frac{1}{f_o + \delta f}$ seconds, $\delta f = 6.5$,  and MINSTR = 0.5.
We perform this parametrization only during the design time, which generalizes well for multiple different surfaces and smartphones (\ie Nexus 4 and OnePlus 2).
Our algorithm does not require any end-user calibration effort.
VibroTag uses the consecutive peaks detected in each randomly selected window to extract multiple vibration patterns between those peaks.

\presub
\subsubsection{Construction of Vibration Signature}
\begin{figure}[htbp]
	\centering
	\captionsetup{justification=centering}    
	\captionsetup[subfigure]{aboveskip=-1pt,belowskip=-1pt}
	\subfigure[Bed]{
		\includegraphics[width=0.67\columnwidth,height=0.46\columnwidth]{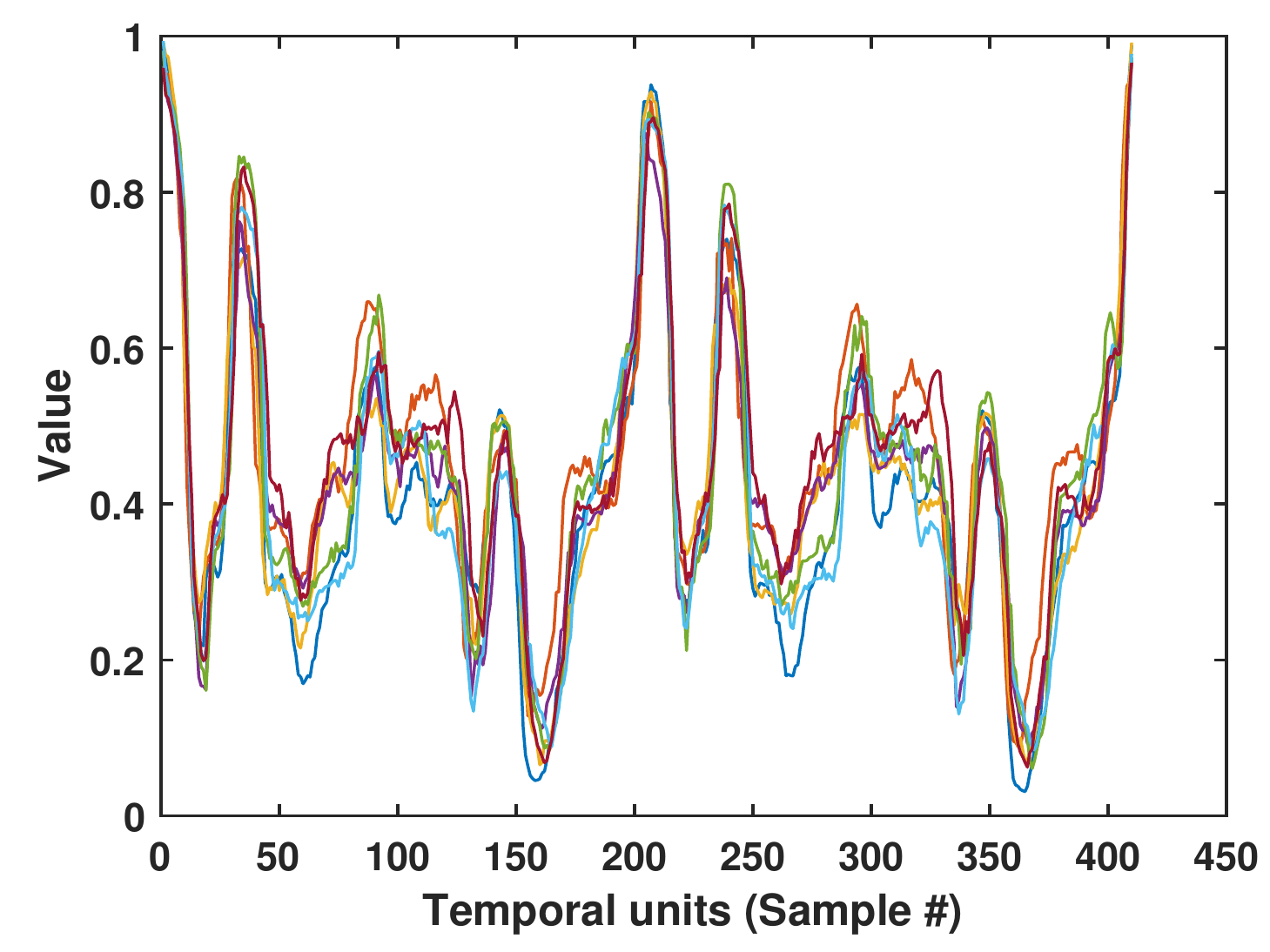}
		\label{fig:finalfeaturesexp1}}
	\hspace{-0.15in}
	\subfigure[Sofa]{
		{\includegraphics[width=0.67\columnwidth,height=0.46\columnwidth]{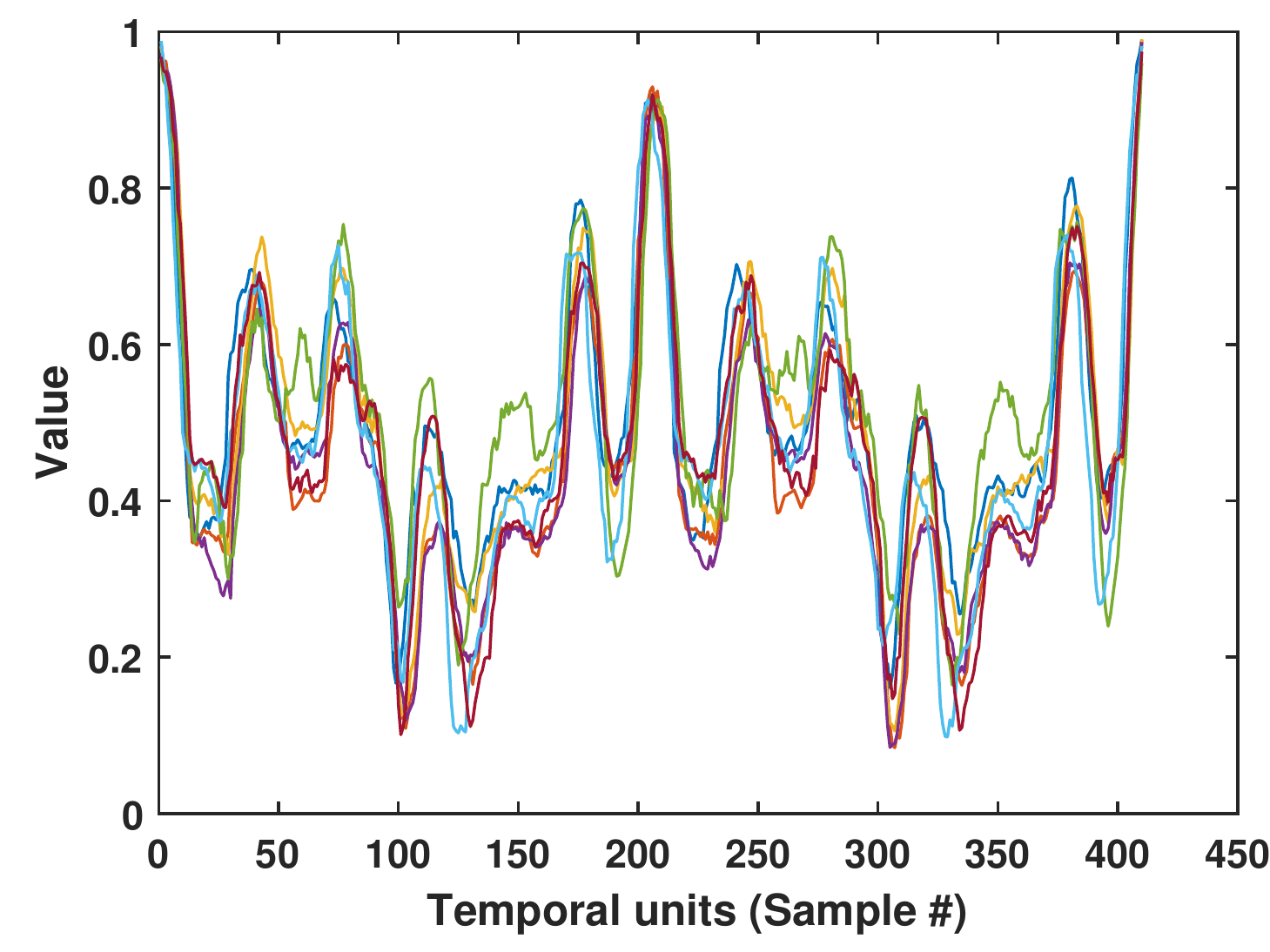}}
		\label{fig:finalfeaturesexp2}}
	\label{fig:finalfeaturesexps}
	\vspace{-0.15in}
	\caption{Extracted time-series features (Low Noise)}
	\vspace{-0.12in}
\end{figure}

\postsub
To construct a single consistent vibration signature, VibroTag first collects at least a total of $M=100$ patterns extracted from the randomly selected time windows. 
Once an enough number of vibration patterns are extracted from different time windows, VibroTag combines all those patterns using median (i.e. 50th percentile) to get a single vibration signature.
%
%
The median operation helps VibroTag remove short-term noisy variations in different vibration patterns, and therefore, allows it to extract a single robust vibration signature of the surface.
Figures \ref{fig:finalfeaturesexp1} and \ref{fig:finalfeaturesexp2} show the example signatures extracted for the Bed and the Sofa related experiments, where we can observe that the vibration signatures of each surface are consistent and almost identical.
Moreover, the extracted signatures uniquely represent their respective surfaces.
\begin{figure}[htbp]
	\centering
	\captionsetup{justification=centering}
	\hspace{-0.15in}
	\subfigure[Cafeteria Table 1]{
		{\includegraphics[width=0.702\columnwidth,height=0.484\columnwidth]{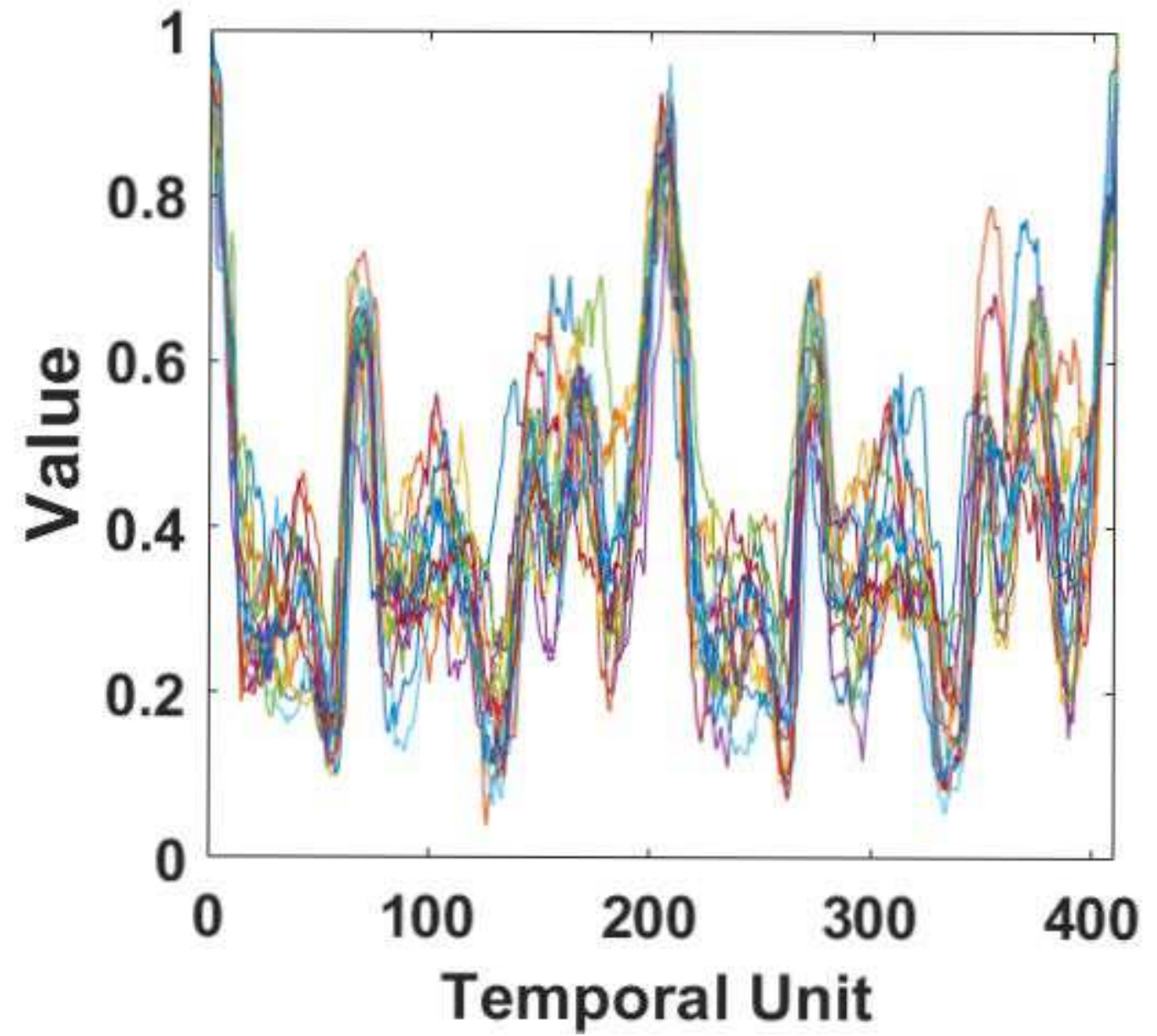}}
		\label{fig:cafetable1}}
	\hspace{-0.15in}
	\subfigure[Cafeteria Table 2]{
		{\includegraphics[width=0.702\columnwidth,height=0.484\columnwidth]{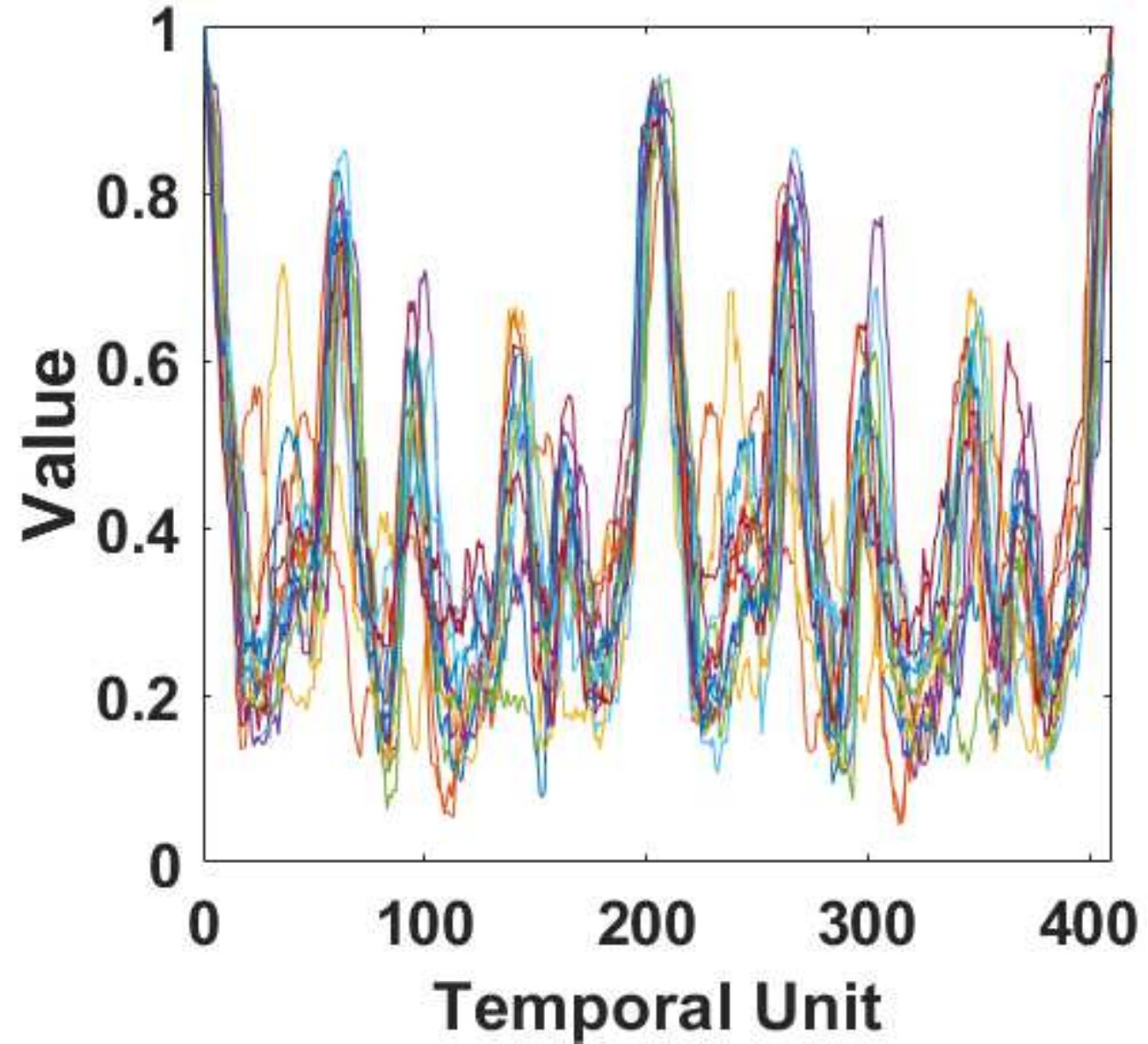}}
		\label{fig:cafetable2}}
	\hspace{-0.1in}
	\vspace{-0.15in}
	\caption{Extracted time-series features (High Noise)}
	\centering
	\vspace{-0.12in}
	\label{fig:cafetablesnoise}
\end{figure}

Figure \ref{fig:cafetablesnoise} shows the vibration signatures extracted for two similar tables during lunch time in a cafeteria on a university campus (\ie a highly noisy environment). 
We can see that although some vibration signatures that VibroTag extracted are inconsistent, yet it was able to extract several consistent vibration signatures even in such a highly noisy environment.

%% file: KamranTMC/classification.tex
\presec \postsec\section{Classification  \& Recognition} \postsec
We use the shapes of the extracted waveforms as features because the shapes retain both time and frequency domain information of the waveforms and are thus more suited for use in classification.
After obtaining the time-series based vibration signatures, VibroTag uses them to build training models for classification.
As VibroTag needs to compare vibration signatures obtained for different surfaces, we need a comparison metric that provides an effective measure of the similarity between vibration signatures of two surfaces.
To achieve this, VibroTag uses the technique of Dynamic Time Warping (DTW) that calculates the distance between waveforms by performing optimal alignment between them. 
Using DTW distance as the comparison metric between vibration signatures, VibroTag trains a k-nearest neighbour (kNN) classifier using those signatures. 

\begin{figure}[htbp]
	\centering
	\captionsetup{justification=centering}
	\subfigure[ViborTag]{
		{\includegraphics[width=0.6\columnwidth]{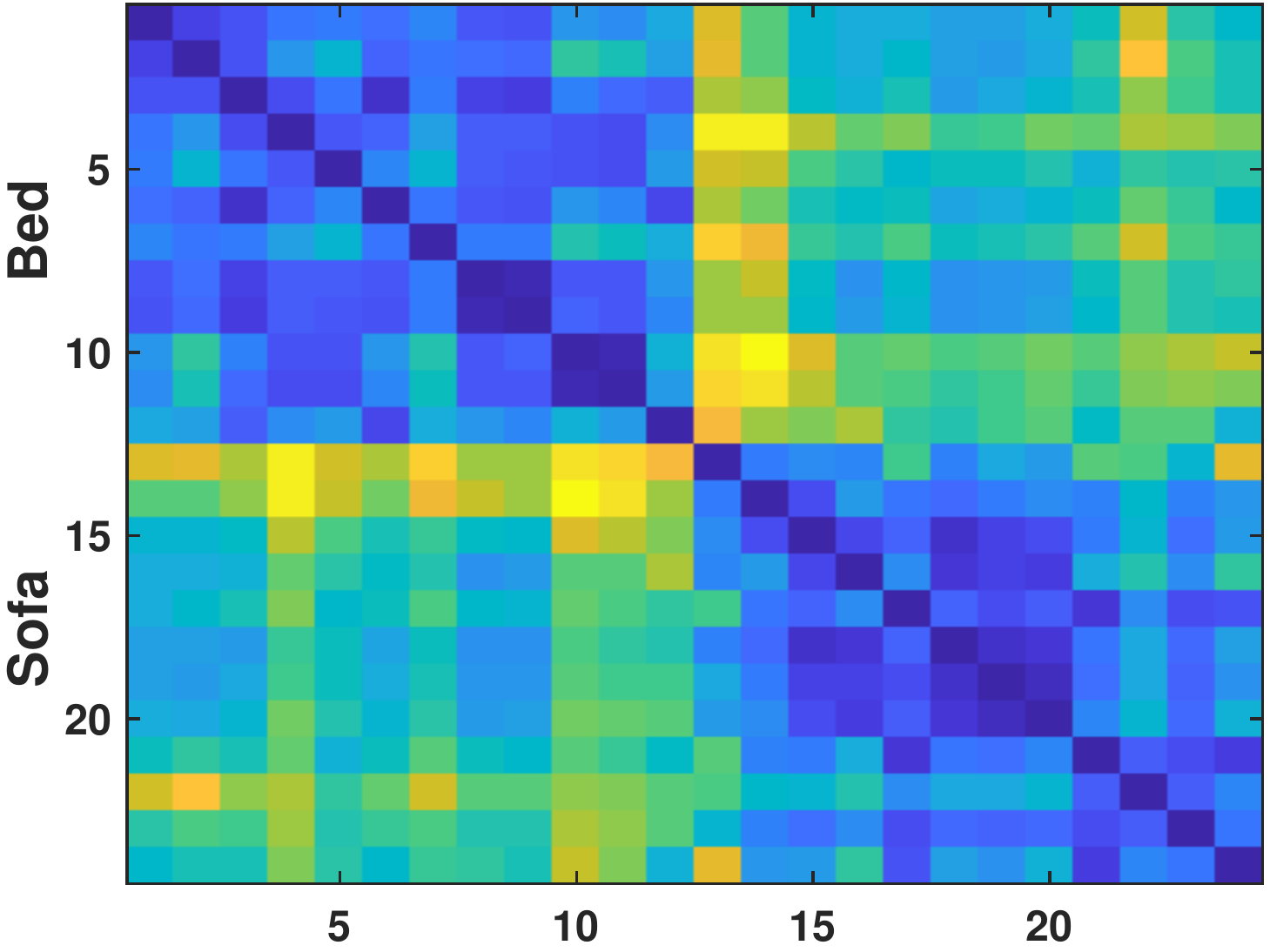}}
		\label{fig:vibrotagdistance}}
	\subfigure[IMU features \cite{cho2012vibration}]{
		{\includegraphics[width=0.6\columnwidth]{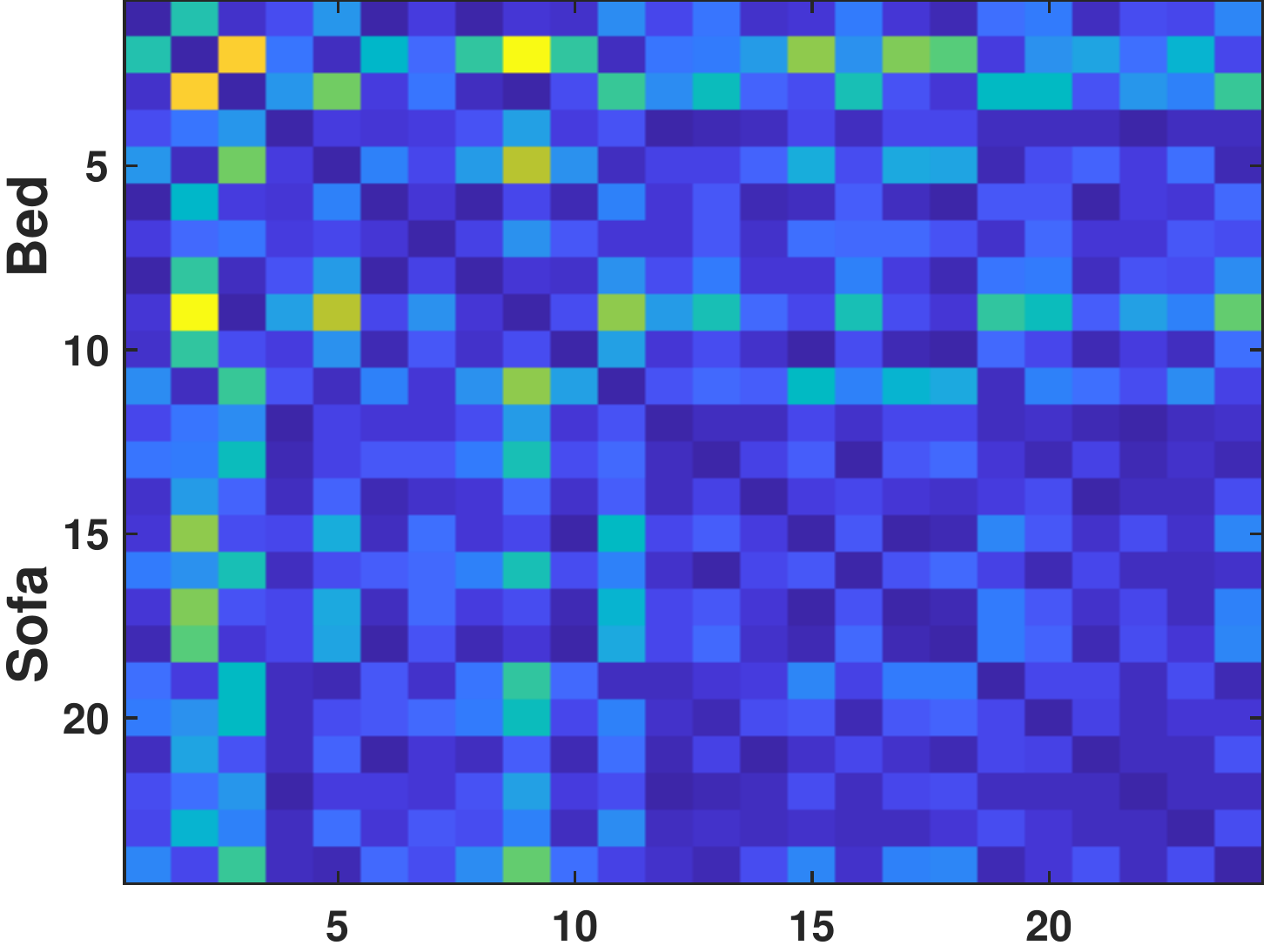}}
		\label{fig:imudistance}}
	\vspace{-0.12in}
	\caption{Colormaps of distance between features of (a) VibroTag (DTW) \& (b) IMU scheme (Euclidean).}
	\vspace{-0.12in}
	\label{fig:restricted-homelivingroomsofa-homebed-restricteddtw}
\end{figure}

DTW is a dynamic programming based solution for obtaining the minimum distance alignment between any two waveforms.
DTW can handle waveforms of different lengths and allows a non-linear mapping of one waveform to another by minimizing the distance between the two waveforms.
In contrast to Euclidean distance, DTW gives us the intuitive distance between two waveforms by determining the minimum distance warping path between them even if they are distorted or shifted versions of each other. 
DTW distance is the Euclidean distance of the optimal warping path between two waveforms calculated under boundary conditions and local path constraints.
In our experiments, DTW distance proves to be effective for comparing two vibration signatures of
different surfaces.
Figure \ref{fig:vibrotagdistance} shows the colormap of DTW distance between the vibration signatures extracted by VibroTag from the experiments corresponding to the bed and the sofa (12 signatures each). 
The average DTW distance among signatures of the bed is $\sim$2.3 and that for the sofa is $\sim$3.1.
However, the average DTW distance between the vibration signatures of the two surfaces was 16.59.
Figure \ref{fig:imudistance} shows the color map of Euclidean distance between features obtained using the IMUs based scheme proposed in \cite{cho2012vibration}.
We can see that IMU based features cannot successfully differentiate between the two surfaces due to high similarity, whereas VibroTag's signatures are significantly better at differentiating the two seemingly similar surfaces.

VibroTag requires training data for the surfaces to be recognized.
Afterwards, it trains a kNN classifier using the vibration signatures corresponding to those surfaces. 
To recognize a surface, VibroTag feeds the detected vibration signature of that surface to the trained kNN classifier. 
The kNN classifier searches for the majority class label among $k$ nearest neighbors of the corresponding vibration signature using the DTW distance metric.
VibroTag declares the majority class label obtained from the kNN classifier as label of the tested surface.
In the current implementation of VibroTag, we chose $k$ = 5 so that the classification process averages more voters in each prediction, which makes our classifier more resilient to outliers.

%% file: KamranTMC/implementionevaluation.tex
\presec \section{Implementation \& Evaluation} \postsec

\subsection{Implementation Details} \postsec
\begin{figure*}[htbp]
	\centering
	\captionsetup{justification=centering}
	
	\subfigure[ViborTag App interface]{
		{\includegraphics[width=0.608\columnwidth]{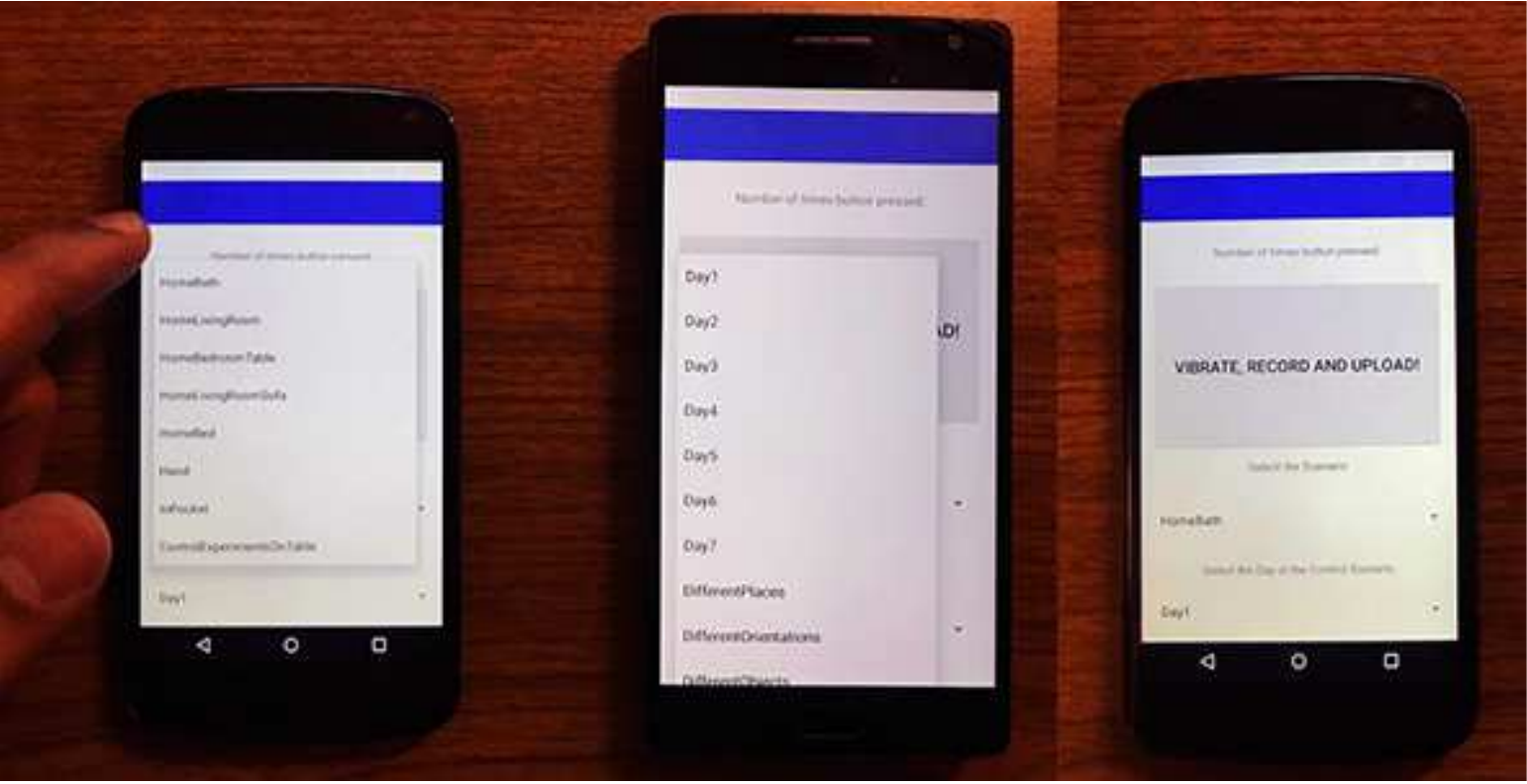}}
		\label{fig:vibrotaginterface}}
	\subfigure[Office]{
		{\includegraphics[width=0.408\columnwidth]{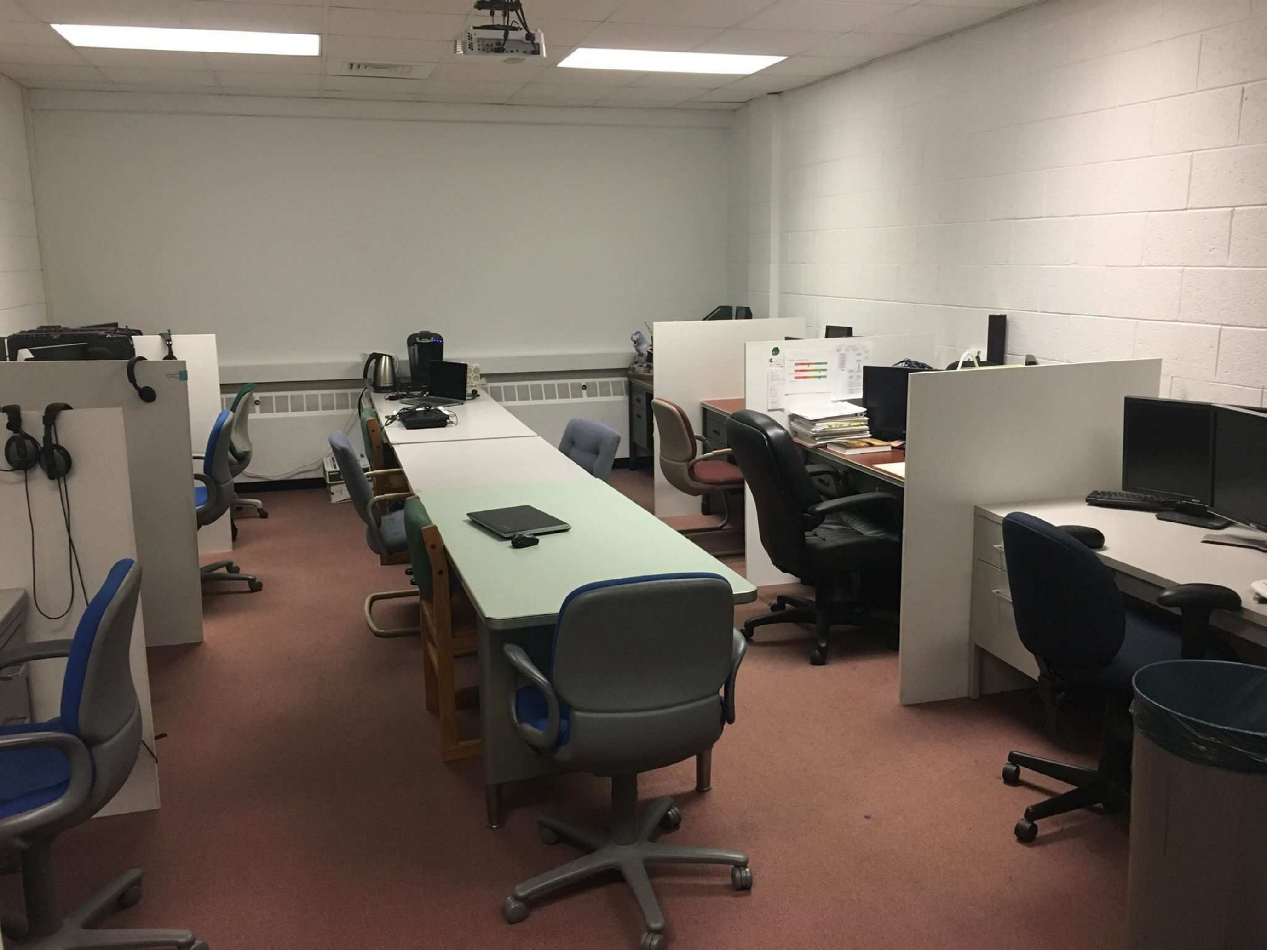}}
		\label{fig:officepic}}
	\subfigure[Office locations]{
		{\includegraphics[width=0.522\columnwidth]{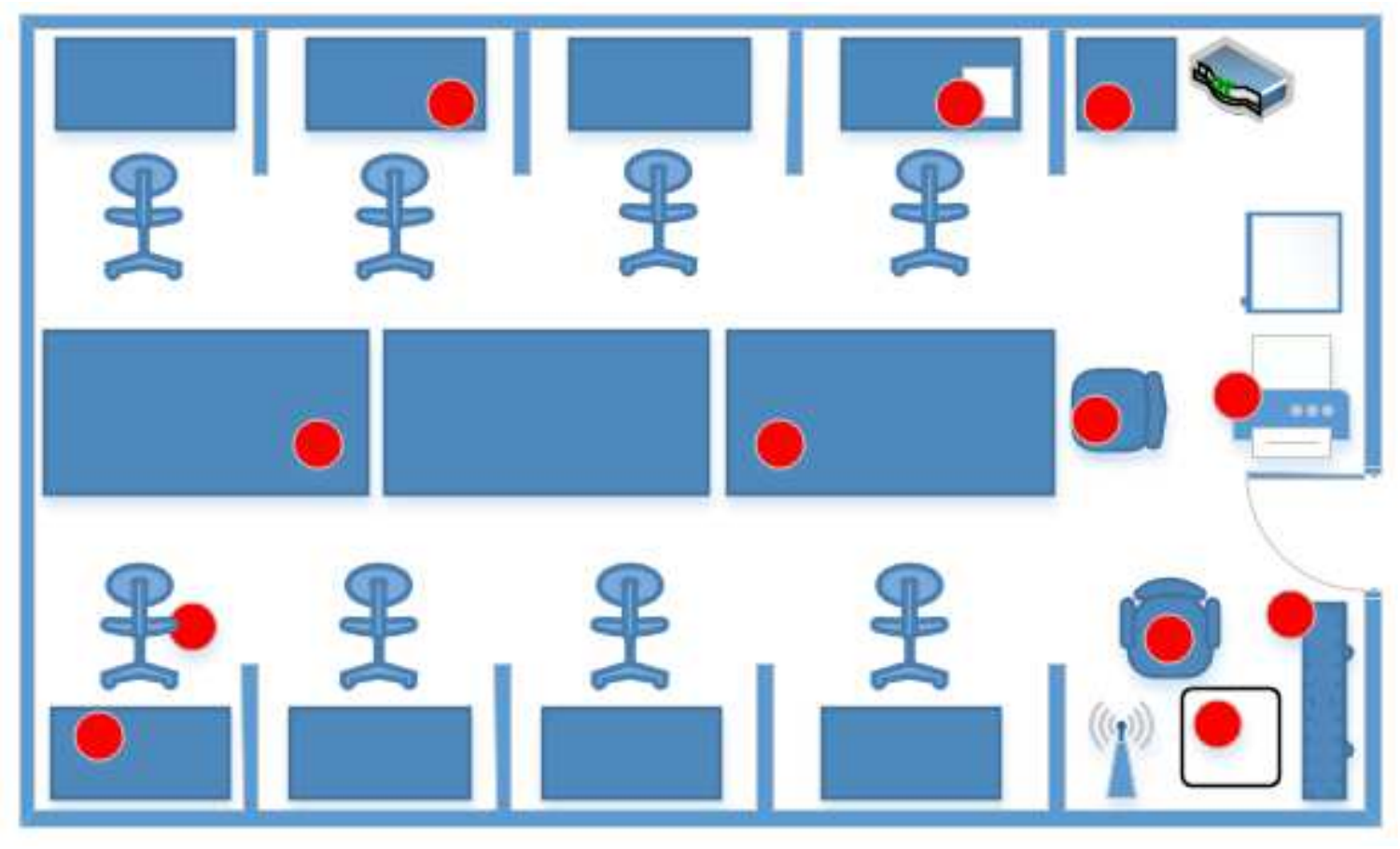}}
		\label{fig:officeenv}}
	
	\hspace{0.1cm}
	
	\subfigure[Apartment]{
		{\includegraphics[width=0.3375\columnwidth]{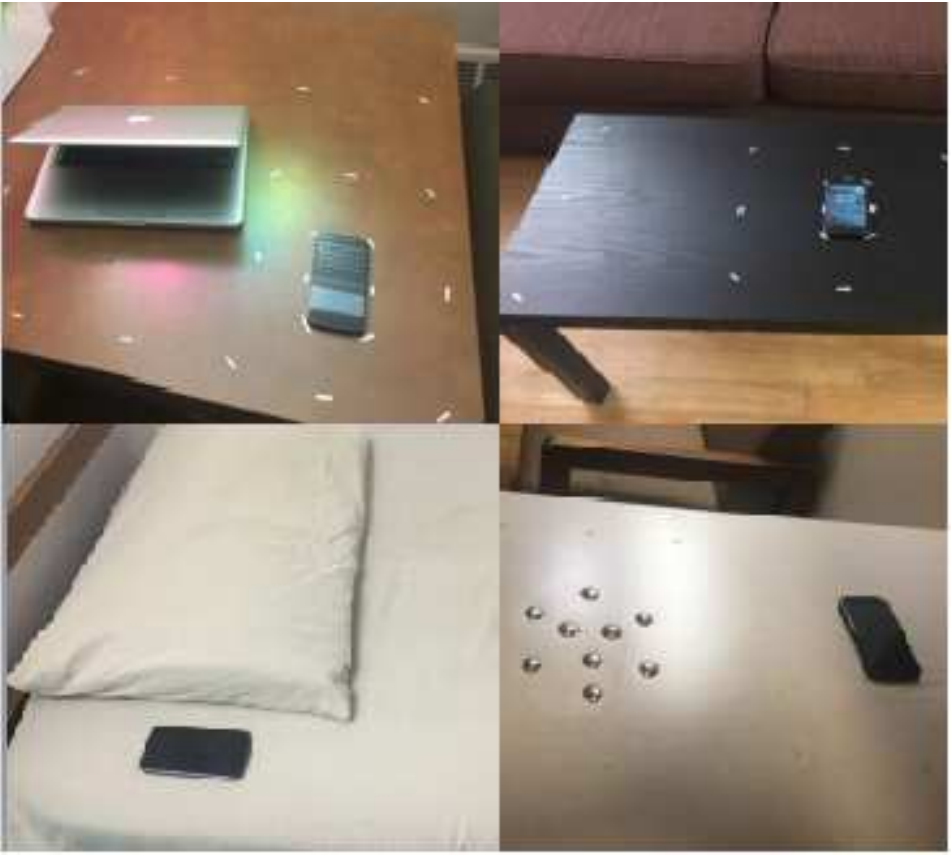}}
		\label{fig:homepics}}
	\subfigure[Apartment locations]{
		{\includegraphics[width=0.5475\columnwidth]{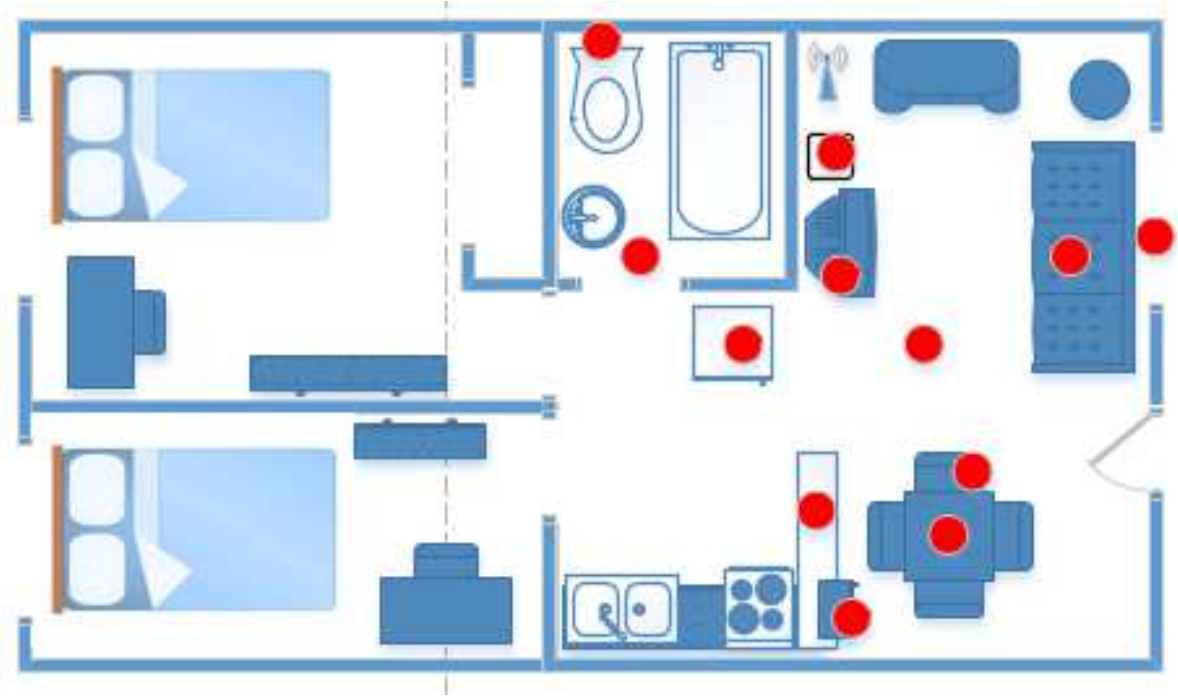}}
		\label{fig:homeenv}}
	\vspace{-0.15in}	
	\caption{VibroTag Setup (a) VibroTag App (b) office environment (c) example of data collection locations in office (d) example of surfaces used for data collection (e) example of data collection locations in apartment}
	\label{fig:experimentalsetupfigs}
	\vspace{-0.12in}
\end{figure*}
We developed an Android application for generating vibrations and sampling sound signals simultaneously (Fig. \ref{fig:vibrotaginterface} shows VibroTag's UI).
Our application can record sound in a separate high priority asynchronous thread which helps minimize sampling related irregularities.
We use a 16 bit PCM encoding on Mono channel with a sampling rate of 44,100Hz for sound recording.
We also record the data from the smartphone's IMU sensors (\ie accelerometer and gyroscope) in a separate high priority thread.
We use this data to implement the state-of-the-art IMUs based vibration sensing approach for single COTS smartphones proposed in \cite{cho2012vibration}, and then compare its surface recognition accuracy with VibroTag.
Each data instance constitutes $\sim$3 seconds of sound and IMU data, during which the vibration motor keeps vibrating.
%
%
%
Our application controls a smartphone's vibrator motor only in terms of turning it ON or OFF, and therefore, does not change the amplitude or pattern of the vibrations.
This allows the smartphone to vibrate at its default vibration settings, which helps keep data samples collected at the same surface/location consistent.
Moreover, it makes VibroTag applicable to smartphones which do not provide any amplitude control over their vibration motors.
We evaluated VibroTag using two smartphones, \ie Google Nexus 4 and OnePlus 2. 
%
%
%

%
%
\presub \subsection{Evaluation Setup} \postsub
%
%
%
We evaluated VibroTag's performance by conducting extensive experiments in two different type of environments, \ie office and apartment.
We selected these environments because they represent real-world use case scenarios where a user interacts with different objects and surfaces regularly.
We collected data from 4 different volunteers, three with Nexus 4 and one with OnePlus 2, whom we name User-1 (Nexus), User-2 (Nexus), User-3 (OnePlus 2), User-4 (Nexus), respectively.
All volunteers were university students who lived in different apartment complexes.
No restrictions were imposed on the movement or work conditions of people residing/working in the apartments/office.
For example, when we collected data in the office environment, other people in the office were working and chatting as they do on a normal working day.  
Similarly, data collection did not cause any interference in the daily activities (cooking, eating, watching TV, cleaning, etc.) in the volunteers' apartment mates. 
Therefore, our evaluation of VibroTag takes into account realistic environments where noise from human activities is present most of the time. 
We used metrics such as confusion matrices, True-Positive-Rates (TPRs) and False-Positive-Rates (FPRs) to evaluate VibroTag's classification performance.
We also compare VibroTag's performance with the IMUs based approach proposed in \cite{cho2012vibration}.

%
\presec \postsec \subsection{VibroTag's Sensitivity} \postsec
We define VibroTag's sensitivity as its ability to differentiate between different positions and orientations of the smartphone placed on the same location/surface.
For example, a user can place his smartphone on his office table in several possible positions and different orientations.
VibroTag's sensitivity is an important metric since we claim that a user can place his smartphone on a surface with reasonable flexibility, without having to worry about centimeter level differences in its position and orientation (unlike \eg EchoTag \cite{tung2015echotag}).
This claim will not be satisfied if VibroTag is too sensitive.

\begin{figure*}[htbp]
	\centering
	\captionsetup{justification=centering}
	\subfigure[High-User-1]{
		{\includegraphics[width=0.5598\columnwidth]{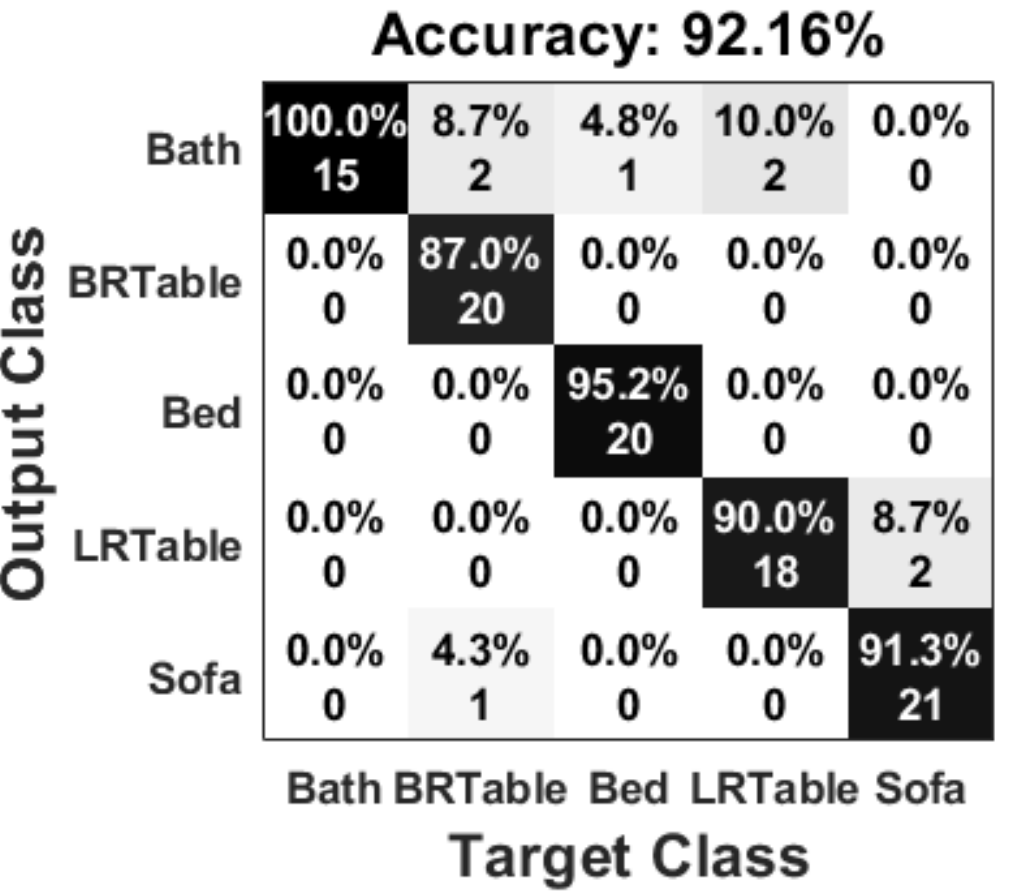}}
		\label{fig:user1-high}}
	\hspace{-0.13in}
	\subfigure[Moderate-User-1]{
		{\includegraphics[width=0.5267\columnwidth]{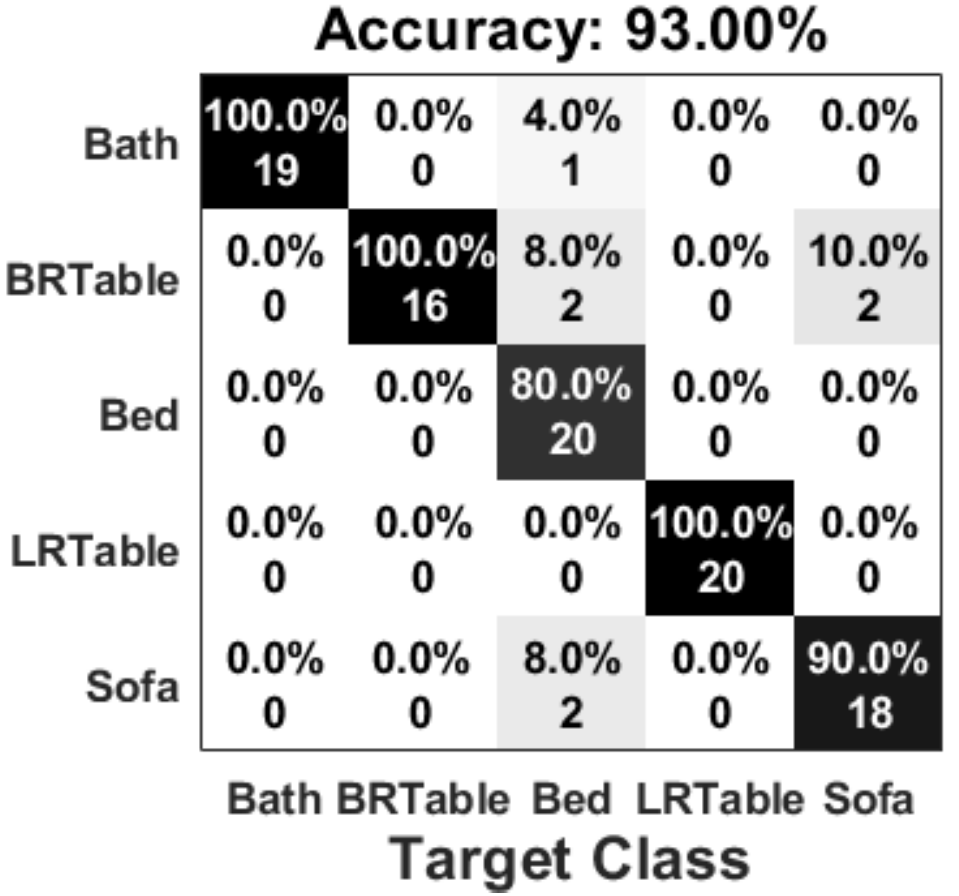}}
		\label{fig:user1-medium}}
	\hspace{-0.13in}
	\subfigure[Least-User-1]{
		{\includegraphics[width=0.5345\columnwidth]{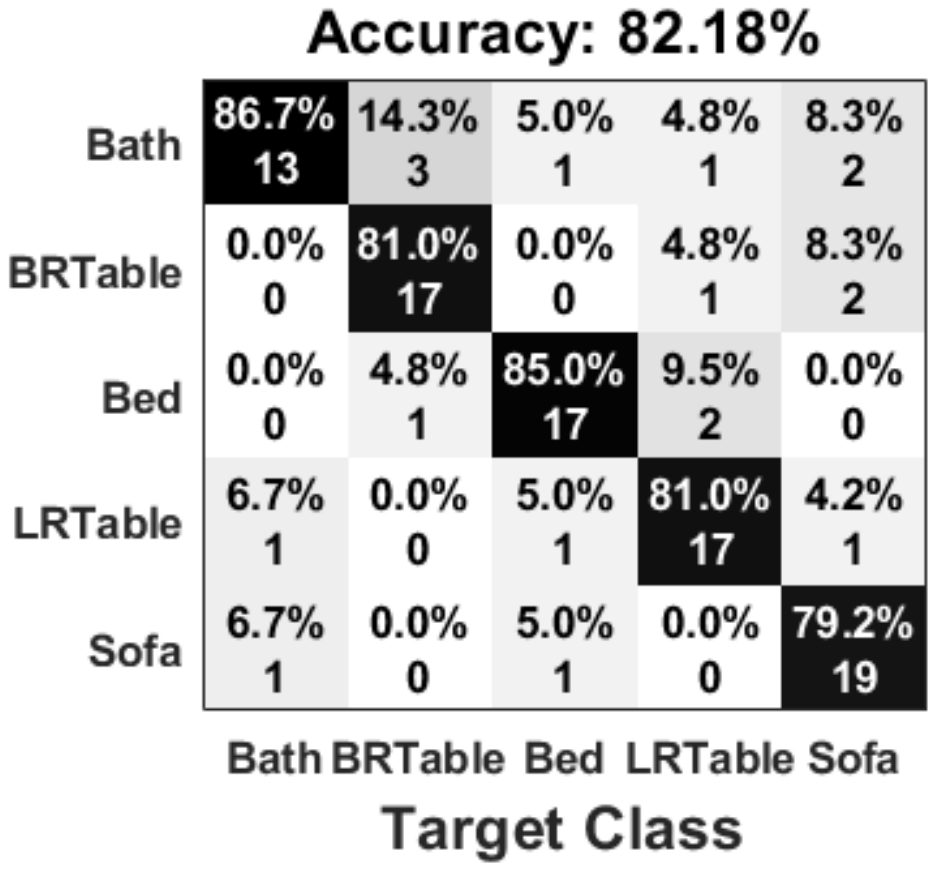}}
		\label{fig:user1-least}}
	
	\hspace{0.13in}
	
	\subfigure[High-User-2]{
		{\includegraphics[width=0.53255\columnwidth]{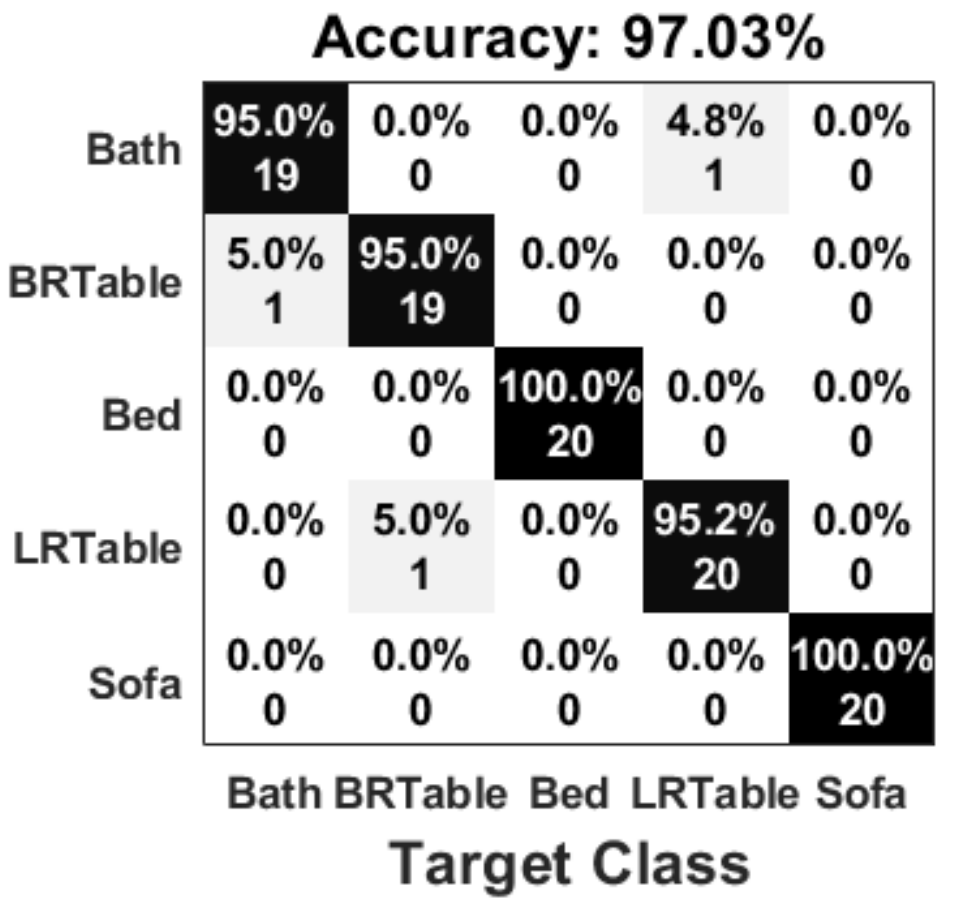}}
		\label{fig:user2-high}}
	\hspace{-0.13in}
	\subfigure[Moderate-User-2]{
		{\includegraphics[width=0.5265\columnwidth]{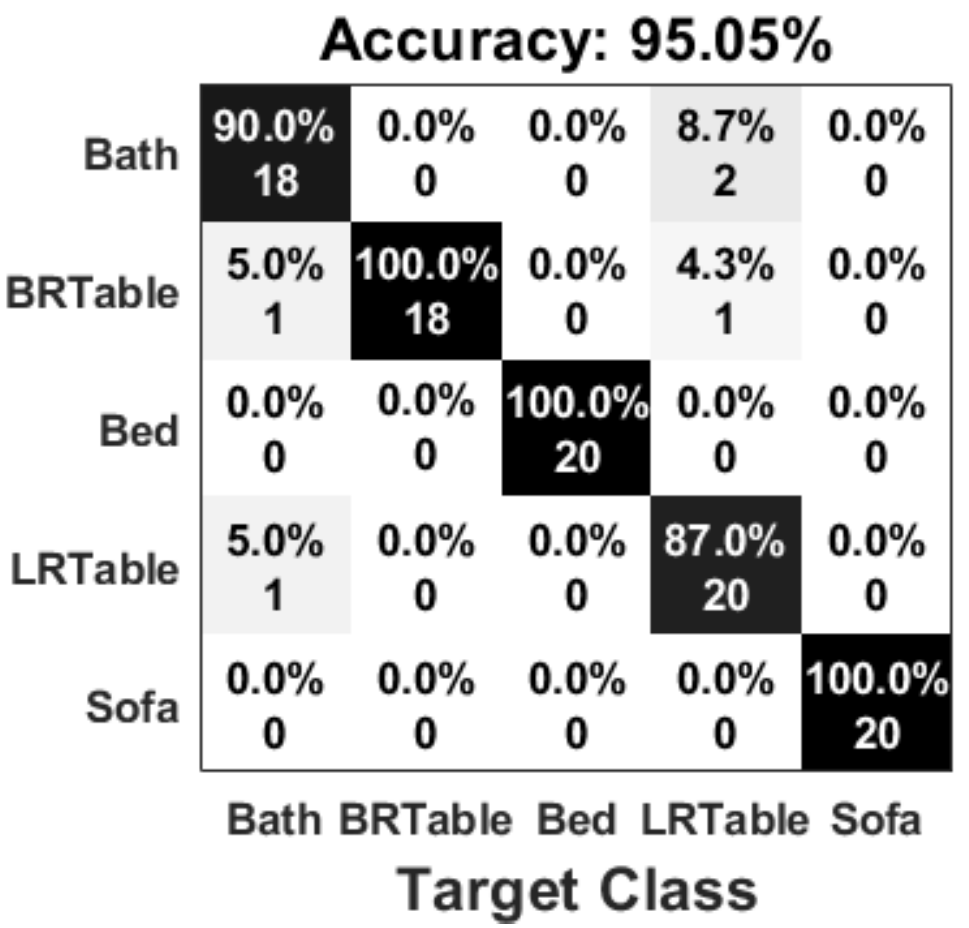}}
		\label{fig:user2-medium}}
	\hspace{-0.13in}
	\subfigure[Least-User-2]{
		{\includegraphics[width=0.564\columnwidth]{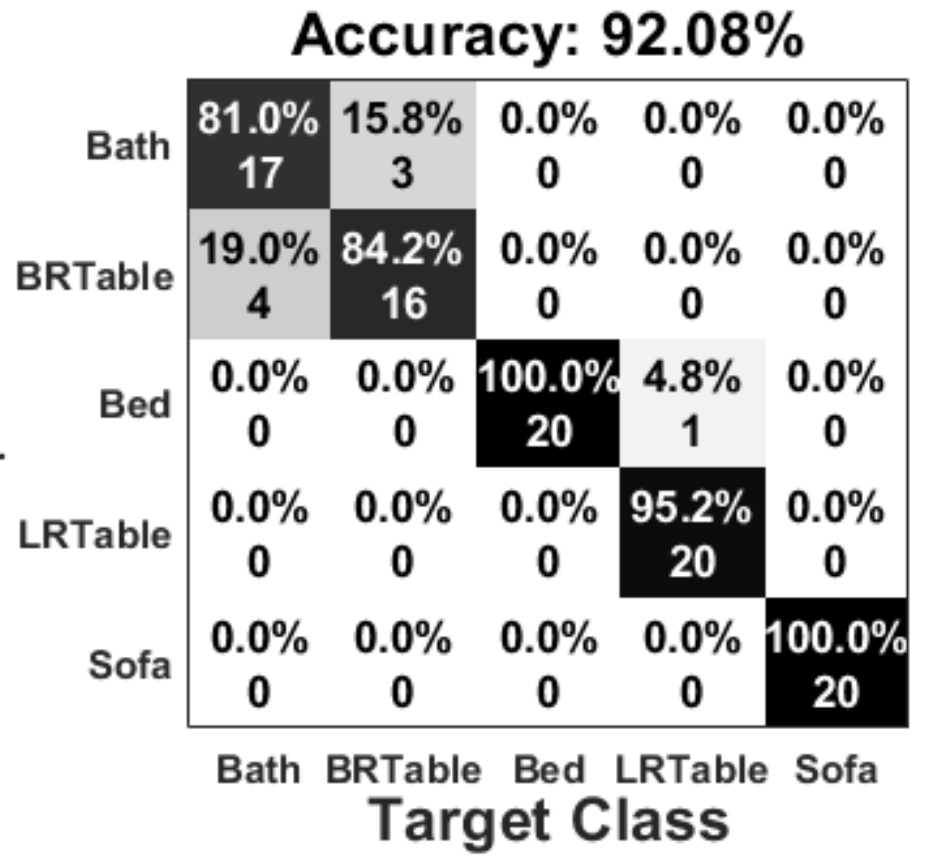}}
		\label{fig:user2-least}}
	\hspace{-0.13in}
	\vspace{-0.12in}
	\caption{Confusion matrices for experiments performed by User-1 and User-2 to determine VibroTag's sensitivity}
	\centering
	\vspace{-0.12in}
	\label{fig:sensitivity}
\end{figure*}

To understand VibroTag's sensitivity, each volunteer collected data in his restroom (on the toilet tank), on his bed, bedroom table, living room table and living room sofa. 
Users collected 25 to 30 samples from each surface for three different smartphone placement scenarios \ie (1) \textit{least restricted}, (2) \textit{moderately restricted} and (3) \textit{highly restricted}. 
Each scenario corresponded to three rectangular regions of different sizes.
We marked the highly restricted region to be approximately within a few inches of the same dimensions as that of the smartphone, the moderately restricted region to be $\sim$4 times larger than that of the highly restricted region, and the least restricted region to be about $\sim$3 times larger in size compared to moderately restricted region.
for example, Figs. \ref{fig:sofa} and \ref{fig:homework} show the marked regions for a sofa and a worktable, respectively, in an apartment.
For the experiments related to the least restricted region, volunteers were allowed to place their smartphone even beyond the third zone, as long as their smartphone was placed on the same surface.
%
%
%
%
%
%
%
%
%
%
Figure \ref{fig:sensitivity} shows confusion matrix plots for the tested 5 classes, (namely Bed, Living Room Table, Living Room Sofa, Restroom Ledge and Kitchen Counter), for both User-1 using Nexus 4 (Figs. \ref{fig:user1-high}-\ref{fig:user1-least}) and User-2 using OnePlus 2 (Figs. \ref{fig:user2-high}-\ref{fig:user2-least}).
For each scenario, confusion matrices are plotted using results from 2-fold cross-validation. 
We observe that for both users, highly restricting the device placement results in highest average prediction accuracy, (i.e. 92.16\% and 97.03\% respectively) which gradually decreases as restriction on smartphone's position and orientation changes from high to least. 
From the confusion plots (Fig. \ref{fig:sensitivity}), we observe that the accuracies corresponding to User-2 are higher that User-1's.
This may be because either OnePlus 2 is able to extract better quality vibration signatures than Nexus 4, or because User-2's environment and the tested surfaces therein were different from User-1's (\eg User-1's bedroom table might have some light objects (\eg keys) placed close to the smartphone that created noise when responding to the vibration).
We discuss such impact of surrounding objects on VibroTag in \S \ref{sec:accuracy}.

\begin{figure}[htbp]
	\centering
	\captionsetup{justification=centering}
	\subfigure[k-folds on User-1 data]{
		{\includegraphics[width=0.6\columnwidth]{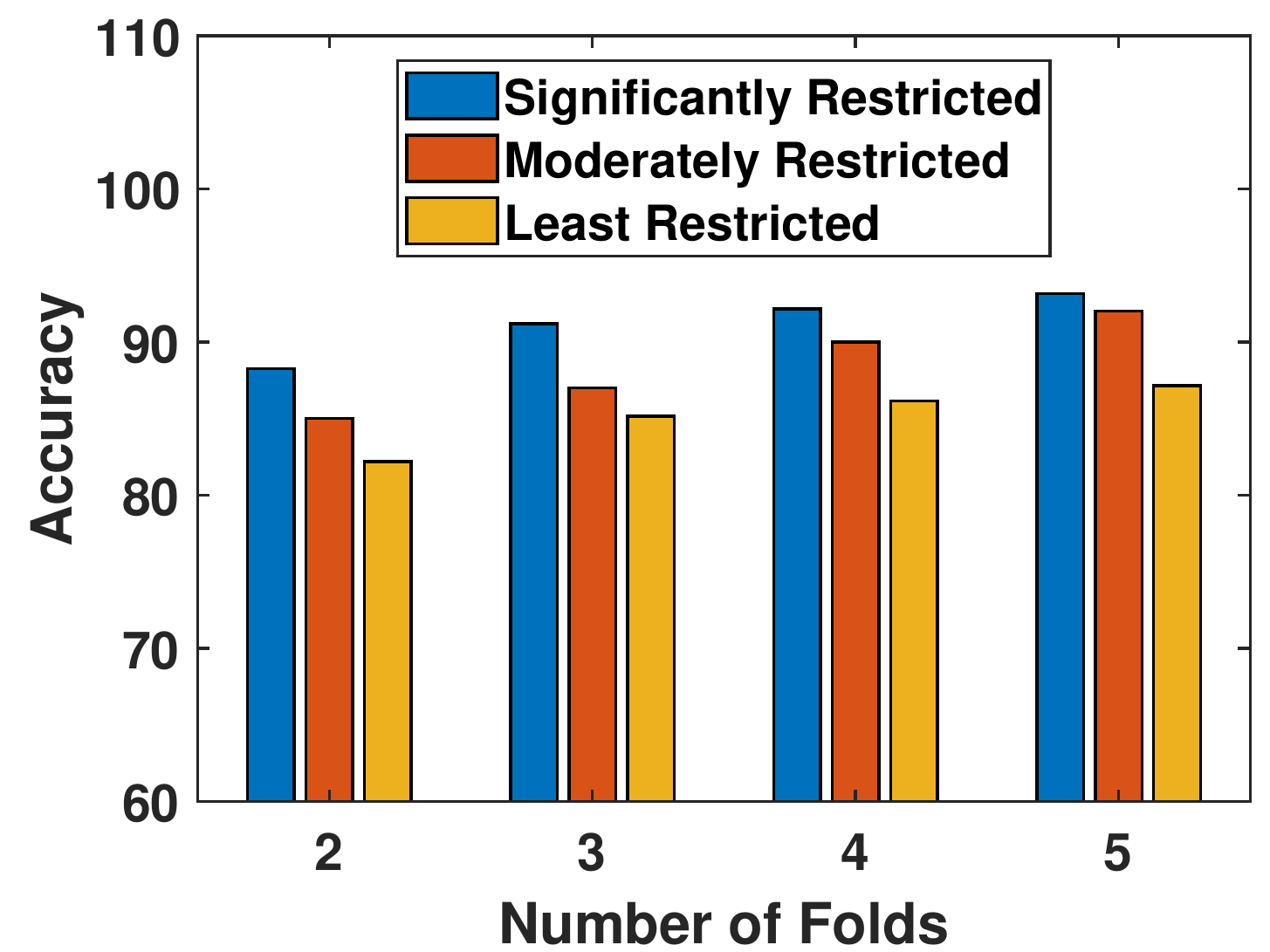}}
		\label{fig:foldssalman}}
	\hspace{-0.1in}
	\subfigure[k-folds on User-2 data]{
		{\includegraphics[width=0.6\columnwidth]{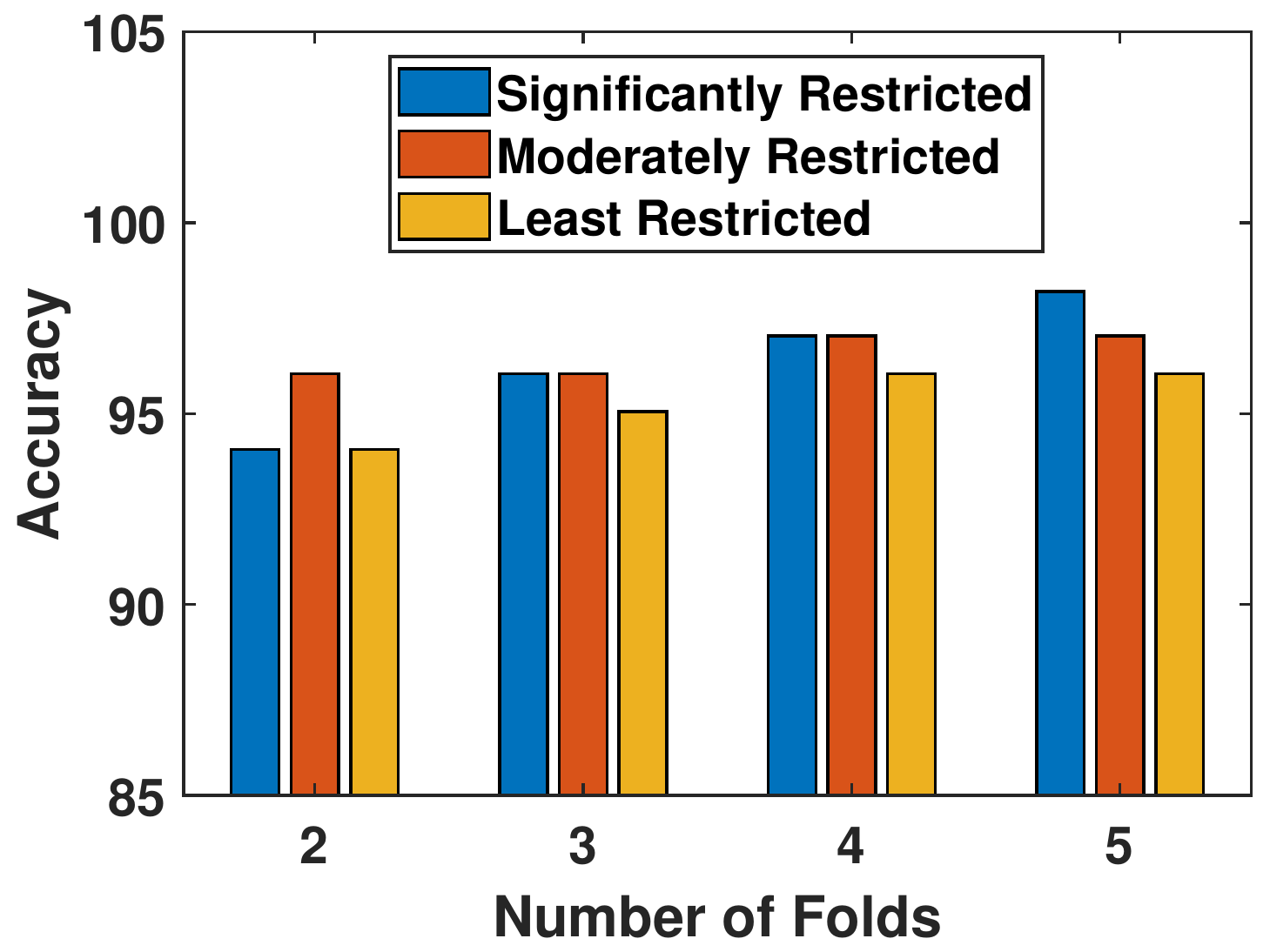}}
		\label{fig:foldskamran}}
	\vspace{-0.12in}
	\caption{Average accuracy with increasing number of training samples (VibroTag's sensitivity experiments)}
	\vspace{-0.25in}
	\label{fig:sensitivityfolds}
\end{figure}
Our results show that average accuracy corresponding to User-1's moderately restricted scenario are higher than highly restricted scenario (Figs. \ref{fig:user1-high}-\ref{fig:user1-least}), which may be attributed to more noisy samples obtained during highly restricted scenario. 
%
%
%
Figures \ref{fig:foldssalman} and \ref{fig:foldskamran} show average accuracy of all 5 classes for all 3 restriction scenarios obtained with 2-fold, 3-fold, 4-fold and 5-fold (\ie increasing percentage data used for training from 50\% to 80\%) cross-validation classification experiments.
We observe that VibroTag performs well for all 3 restriction scenarios even when only 50\% data is used for training and remaining for testing, and accuracies for even least restricted scenarios reach as high as 87\% for User-1 and 95\% for User-2 when percentage of training data is 80\%.
\begin{figure*}
	\captionsetup[subfigure]{justification=centering}
	\centering
	\subfigure[Confusion matrix for all 5 days data]{
		{\includegraphics[width=0.8\columnwidth]{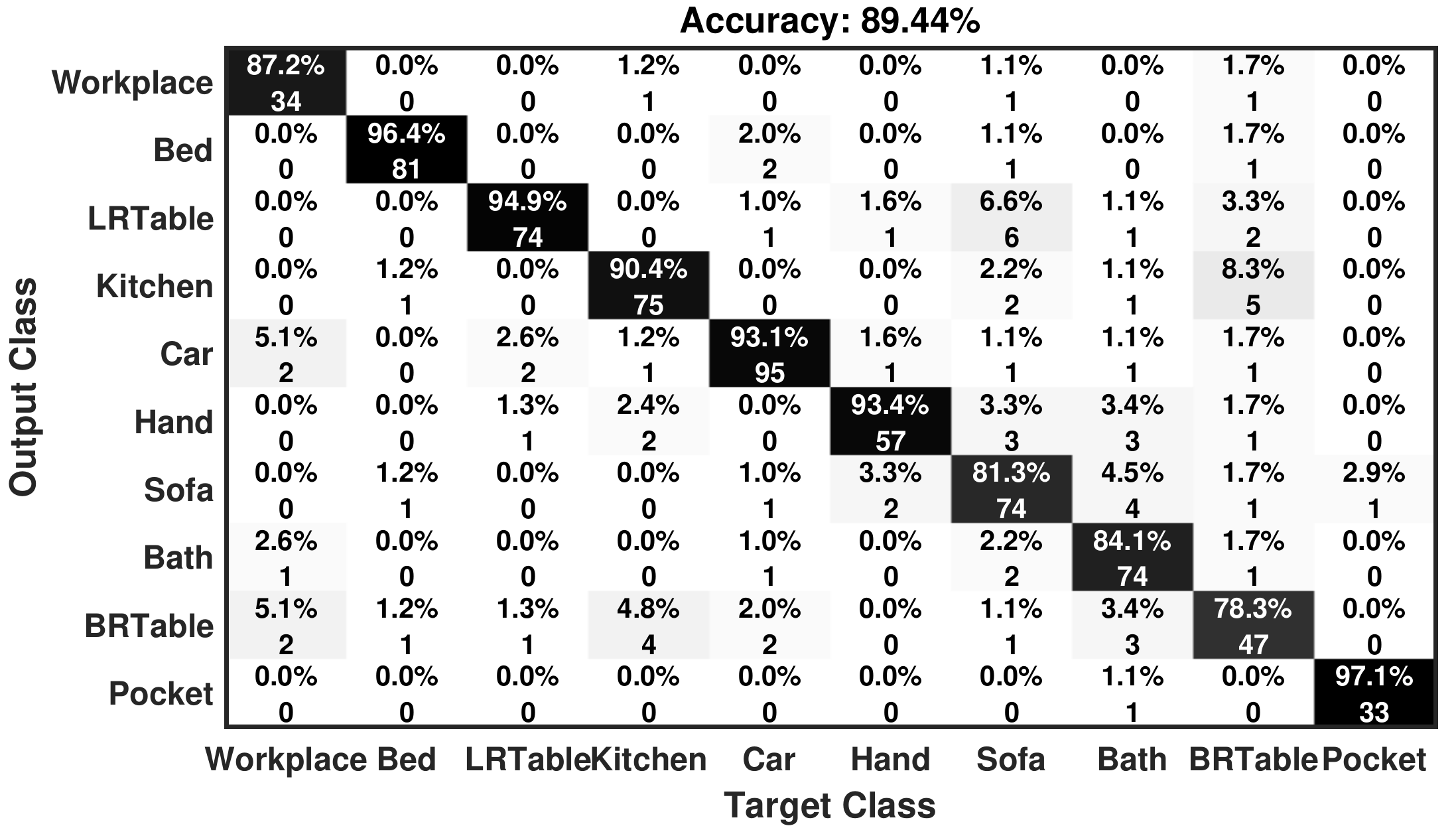}}
		\label{fig:user1-days-confusion}}
	\subfigure[Individual and all 5 days]{
		{\includegraphics[width=0.6\columnwidth]{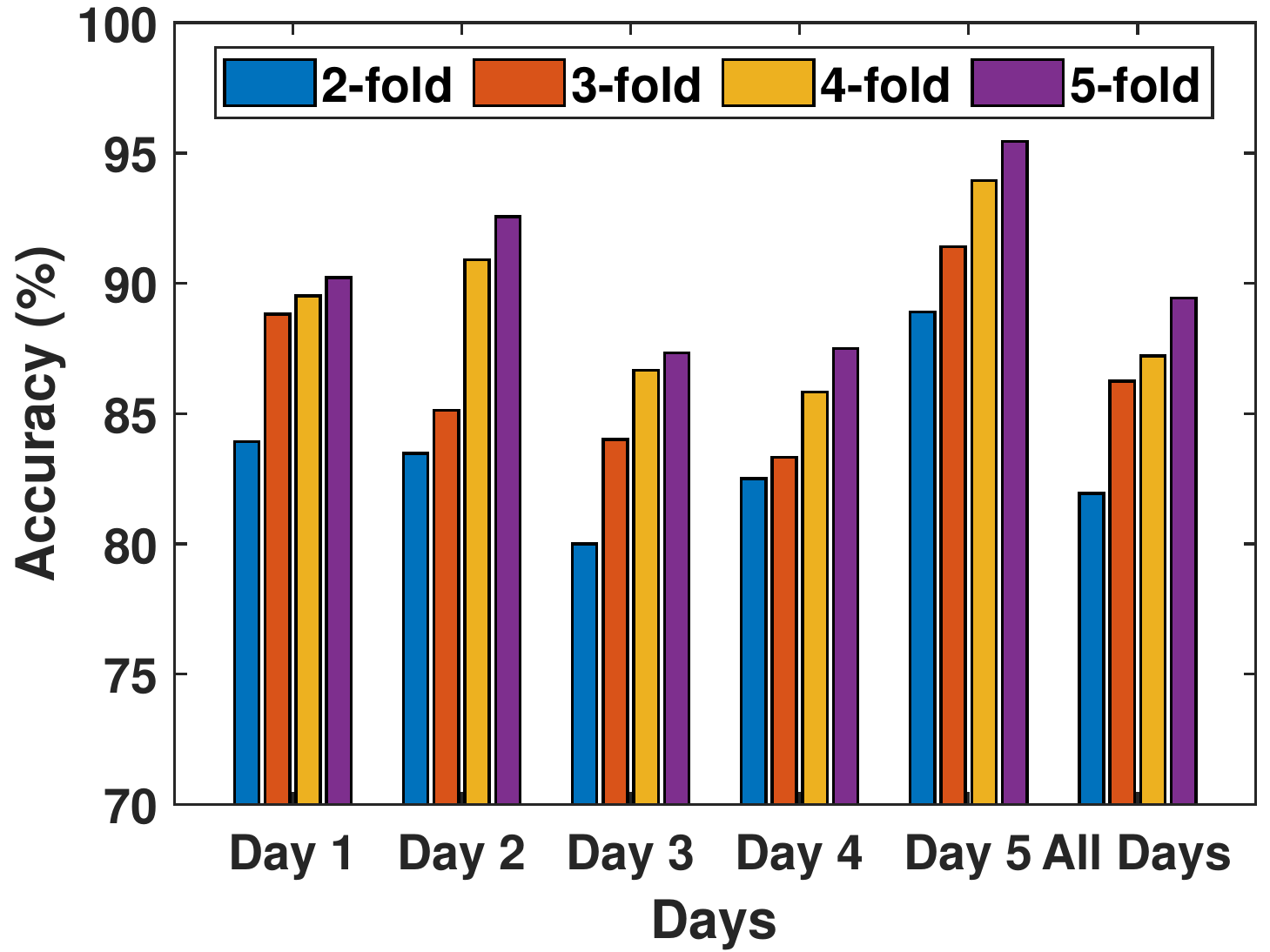}}
		\label{fig:user1-days}}
	\hspace{-0.1in}
	\subfigure[Consecutive days accuracy]{
		{\includegraphics[width=0.566\columnwidth]{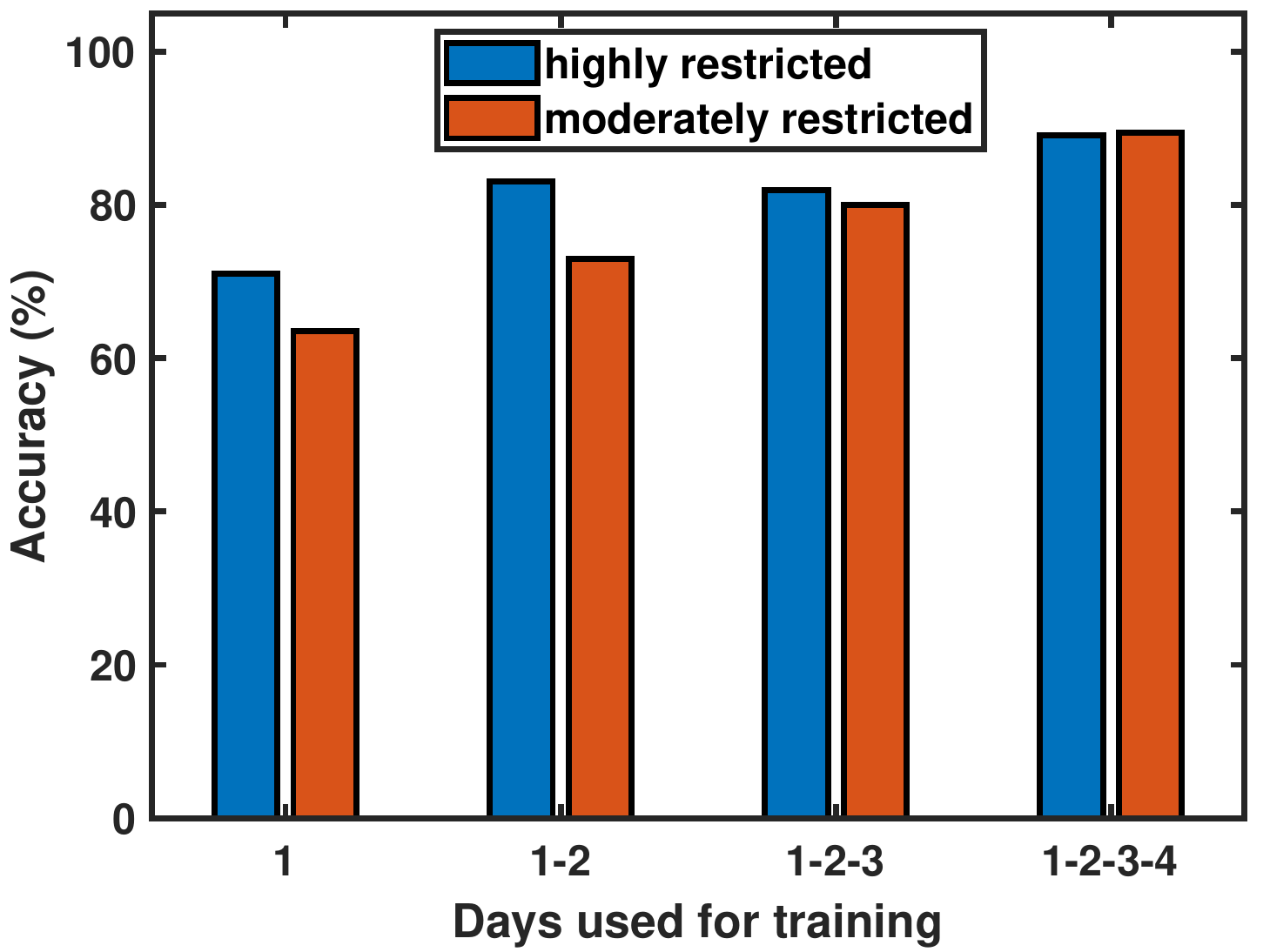}}
		\label{fig:consecutive_days_kamran}}
	\vspace{-0.12in}
	\caption{(a) Average 4-fold cross-validation accuracies over all classes (User-1) (moderately restricted experiments), (b) Confusion matrix after cross-validation, (c) Training on data from previous days, testing on subsequent days.}
	\label{fig:accuracyoverdays1}
	\vspace{-0.1in}
\end{figure*}
	
%
\presec \subsection{VibroTag's Accuracy} \label{sec:accuracy} 
\subsubsection{Location Recognition Accuracy}\postsub

\textit{VibroTag achieves an average 4-fold accuracy of 86.55\% when identifying different locations, whereas the IMU based approach achieves only 49.25\%.}
Table \ref{tab:allsurfacesaccuracy} shows average 2-fold and 4-fold classification accuracies obtained for 24 different locations/surfaces by User-1. 
For these experiment, User-1 collected 30-35 samples from each of those locations in a moderately restricted manner.
Our results show that VibroTag achieves an average (4-fold) recognition accuracy of 86.55\%, which is 37\% higher than the average accuracy achieved by the latest IMUs based approach \cite{cho2012vibration}, which achieves only 49.25\%.
Moreover, the lowest accuracy achieved by VibroTag is 70.33\%, whereas the IMUs based method's accuracy goes as low as 16.38\%.
This shows that the features extracted using VibroTag can successfully differentiate between different locations/surfaces, even when the surfaces are made of very similar material (\eg wood chair vs wood table, or metal drawer vs metal shelve).

\begin{table*}[htbp] 
	\caption{Average accuracy of recognizing different surfaces in office and apartment scenarios}
	\vspace{-0.1in}
	\renewcommand{\arraystretch}{1.6}
	\centering
	\resizebox{2.05\columnwidth}{!}{
		\begin{tabular}{|c|m{1.3cm}|c|m{0.9cm}|m{1.2cm}|m{1.1cm}|m{1.3cm}|m{1.2cm}|m{1cm}|m{1.2cm}|m{1.6cm}|m{1.9cm}|m{1cm}|m{1.3cm}|m{1.61cm}|m{1.35cm}|m{1.65cm}|m{1cm}|m{1.4cm}|m{1cm}|m{1.3cm}|m{1.65cm}|m{1.25cm}|m{1.35cm}|c|c|m{1cm}|c|}
			\cline{2-25}
			\multicolumn{1}{c}{\textbf{}}  & \multicolumn{11}{|c|}{\textbf{\large Office environment}}  & \multicolumn{13}{c|}{\textbf{\large Apartment environment}}  \\ 		
			\cline{2-25}
			\multicolumn{1}{c|}{\textbf{}}& \bf Pages Bundle  &  \bf Printer  & \bf Foam Chair &\bf Metal Drawer   & \bf Carpet & \bf Wooden Chair & \bf Metal Shelve & \bf Mouse Pad & \bf Leather Chair & \bf Center Desk (Wood) & \bf Cardboard Box & \bf XBox& \bf Work Desk (Wood)&\bf Window Ledge (Marble)&  \bf Bathtub Ledge &\bf Living Table (Wood)&\bf Glass Table& \bf Kitchen Counter & \bf Fridge & \bf Wooden Floor & \bf Bedroom Table (Wood)& \bf TV Table (Wood) & \bf \bf Living Room Sofa & \bf Microwave\\  	\cline{2-25}
			\multicolumn{1}{c|}{\textbf{}}& \multicolumn{24}{c|}{\bf \large VibroTag's Accuracy} \\ 	\cline{1-25}
			\bf 2-Fold &\bf\large77.60&\bf\large97.72&\bf\large68.50&\bf\large83.31&\bf\large86.55&\bf\large88.31&\bf\large89.09&\bf78.53&\bf\large91.76&\bf\large79.02&\bf\large95.79&\bf\large73.46&\bf\large80.62&\bf\large92.86&\bf\large93.19&\bf\large81.9&\bf\large79.03&\bf\large87.36&\bf\large87.65&\bf\large84.31&\bf\large82.29&\bf\large76.39&\bf\large72.94&\bf\large91.06 \\ 	\hline
			\bf 4-Fold &\bf\large80.87&\bf\large99.28&\bf\large70.33&\bf\large86.77&\bf\large91.15&\bf\large88.69&\bf\large90.84&\bf\large81.25&\bf\large93.19&\bf\large81.53&\bf\large95.52&\bf\large75.96&\bf\large83.49&\bf\large96.16&\bf\large94.06&\bf\large82.84&\bf\large82.43&\bf\large91.17&\bf\large90.09&\bf\large88.26&\bf\large88.26&\bf\large77.61&\bf\large76.99&\bf\large92.56 \\ \hline
			\multicolumn{1}{c|}{\textbf{}}& \multicolumn{24}{c|}{\bf \large State-of-the-Art IMUs Based Method's Accuracy \cite{cho2012vibration}} \\ 	\cline{1-25}
			\bf 2-Fold &\bf\large68.64&\bf\large87&\bf\large30.35&\bf\large68.3&\bf\large35.13&\bf\large94.94&\bf\large89.96&\bf\large64.05&\bf\large75.45&\bf\large27.07&\bf\large34.29&\bf\large19.55&\bf\large49.41&\bf\large65.08&\bf\large38.78&\bf\large25.18&\bf\large16.89&\bf\large19.42&\bf\large56.60&\bf\large32.98&\bf\large51.10&\bf\large30.07&\bf\large29.31&\bf\large64.98 \\ 	\hline
			\bf 4-Fold &\bf\large75.63&\bf\large90.39&\bf\large31.23&\bf\large70.02&\bf\large37.48&\bf\large95.91&\bf\large91.52&\bf\large66.31&\bf\large78.23&\bf\large29.69&\bf\large40.21&\bf\large21.56&\bf\large52.92&\bf\large66.95&\bf\large40.26&\bf\large25.92&\bf\large17.04&\bf\large20.23&\bf\large63.40&\bf\large33.98&\bf\large52.35&\bf\large32.53&\bf\large32.27&\bf\large65.27 \\ \hline
	\end{tabular}}
	\vspace{-0.06in}
	\label{tab:allsurfacesaccuracy}
\end{table*}

\presub\subsubsection{Location Recognition Accuracy over Days}\postsub

\begin{figure*}[htbp]
	\centering
	\captionsetup{justification=centering}
	\includegraphics[width=1.3\columnwidth]{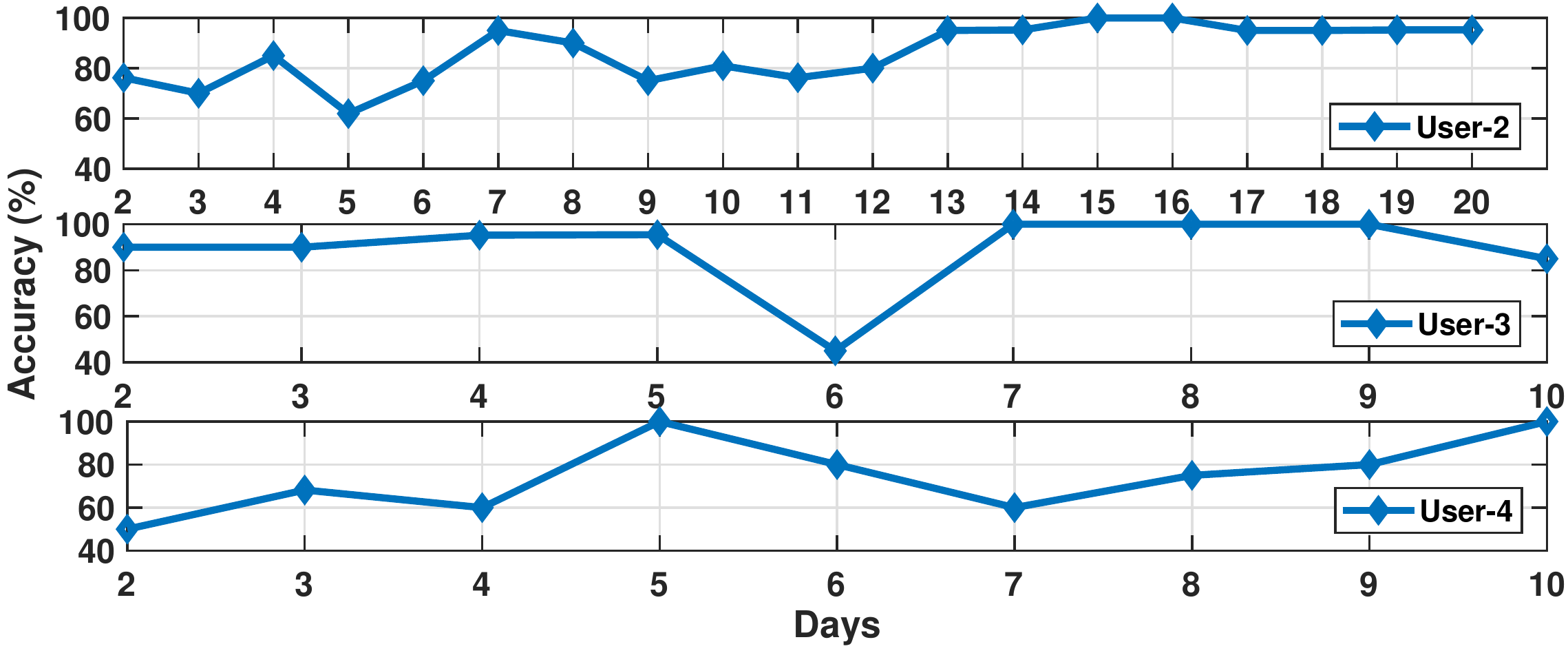}
	\vspace{-0.12in}
	\caption{Accuracies on consecutive days}
	\label{fig:consecutiveDaysAccuraciesUsers2to4}
	\vspace{-0.12in}
\end{figure*}

\textit{VibroTag can maintain an average accuracy of up to 85\% using training samples obtained for 3-4 days only.}
VibroTag's accuracy can change over days due to several different reasons, for example, changes in environmental noise and/or changes in position of other items placed on a surface (\eg light objects such as keys, etc.).
Next, we explore how VibroTag's accuracies change over days and how much training VibroTag requires to maintain high accuracies when testing on data from a new day.

Figure \ref{fig:user1-days} shows average cross-validation accuracy over all classes for data obtained from User-1 on 5 different days. 
The figure also shows cross-validation accuracy and confusion matrix obtained when data from all 5 days was combined.
For these experiments, User-1 collected 6-20 samples from 10 different locations (\ie  Workplace (office table), Bed, Living Room Table, Kitchen (on marble counter), Car (small compartment in front of the gear stick), Hand, Living Room Sofa, Restroom (on toilet tank), Bedroom Table and Pocket) every day.
For this set of experiments, data was collected for both highly and moderately restricted smartphone placement scenarios.
Our results show that VibroTag achieves at least 79\% (and at most 87\%) accuracy everyday when using only 50\% of a day's data for training.
Figure \ref{fig:consecutive_days_kamran} shows how combining data from previous days improves accuracy for data collected on the subsequent days from User-1.
%
%
We observe that VibroTag can achieve accuracy of more than 80\% on the unknown samples on day 5 for both restriction scenarios.
Moreover, we observe that the accuracy of moderately restricted smartphone placement scenarios approaches highly restricted scenarios.
%

To understand how VibroTag's accuracy changes over days across multiple users, we collected 5 samples from Users 2, 3, and 4 for 4 different locations (\ie Kitchen (on marble counter), Living Room Sofa, Restroom (on toilet tank), Bedroom Table and Living Room Table) for 20, 10, and 10 consecutive days, respectively. 
Fig. \ref{fig:consecutiveDaysAccuraciesUsers2to4} shows how the classification accuracy changes for different users, where VibroTag is trained using the data from previous days and test on the subsequent days.
We observe that the accuracy generally increases with days for User-2, however, there is a major dip for User-3 on day 5 and for User-4 between days 5-7, which can be attributed to major changes in the surrounding environment in terms of noise and/or addition/removal of different objects (such as keys or a pen) on the surface.

\presec\postsec \subsubsection{Impact of Surrounding Objects} \label{sec:noiserobustnesseval} 
\begin{figure}[htbp]
	\centering
	\captionsetup{justification=centering}
	\subfigure[setup]{
		{\includegraphics[width=0.23\columnwidth]{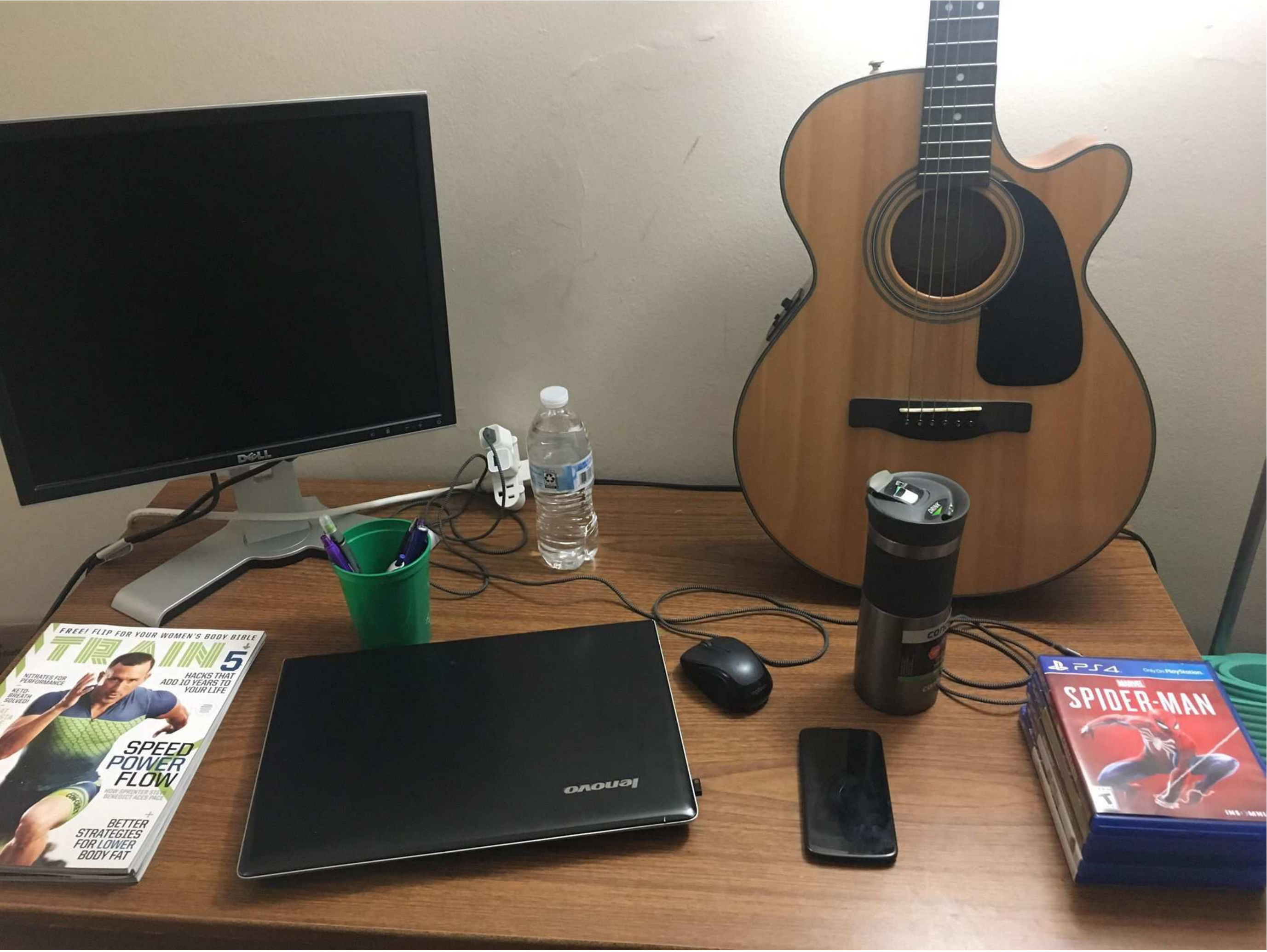}}
		\label{fig:normal}}
	\hspace{-0.1in}
	\subfigure[no guitar]{
		{\includegraphics[width=0.23\columnwidth]{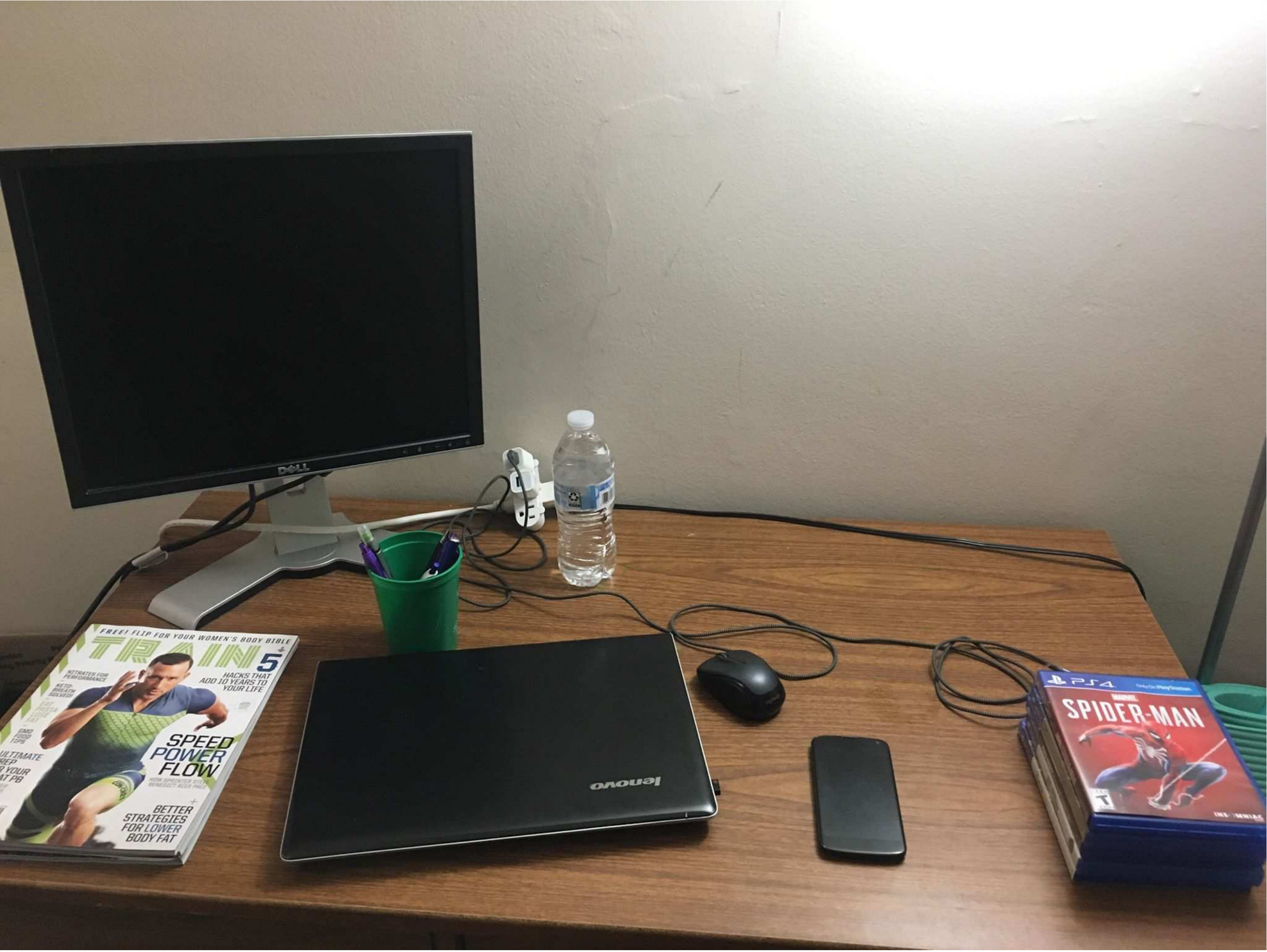}}
		\label{fig:removeguitar}}
	\hspace{-0.1in}
	\subfigure[no laptop]{
		{\includegraphics[width=0.23\columnwidth]{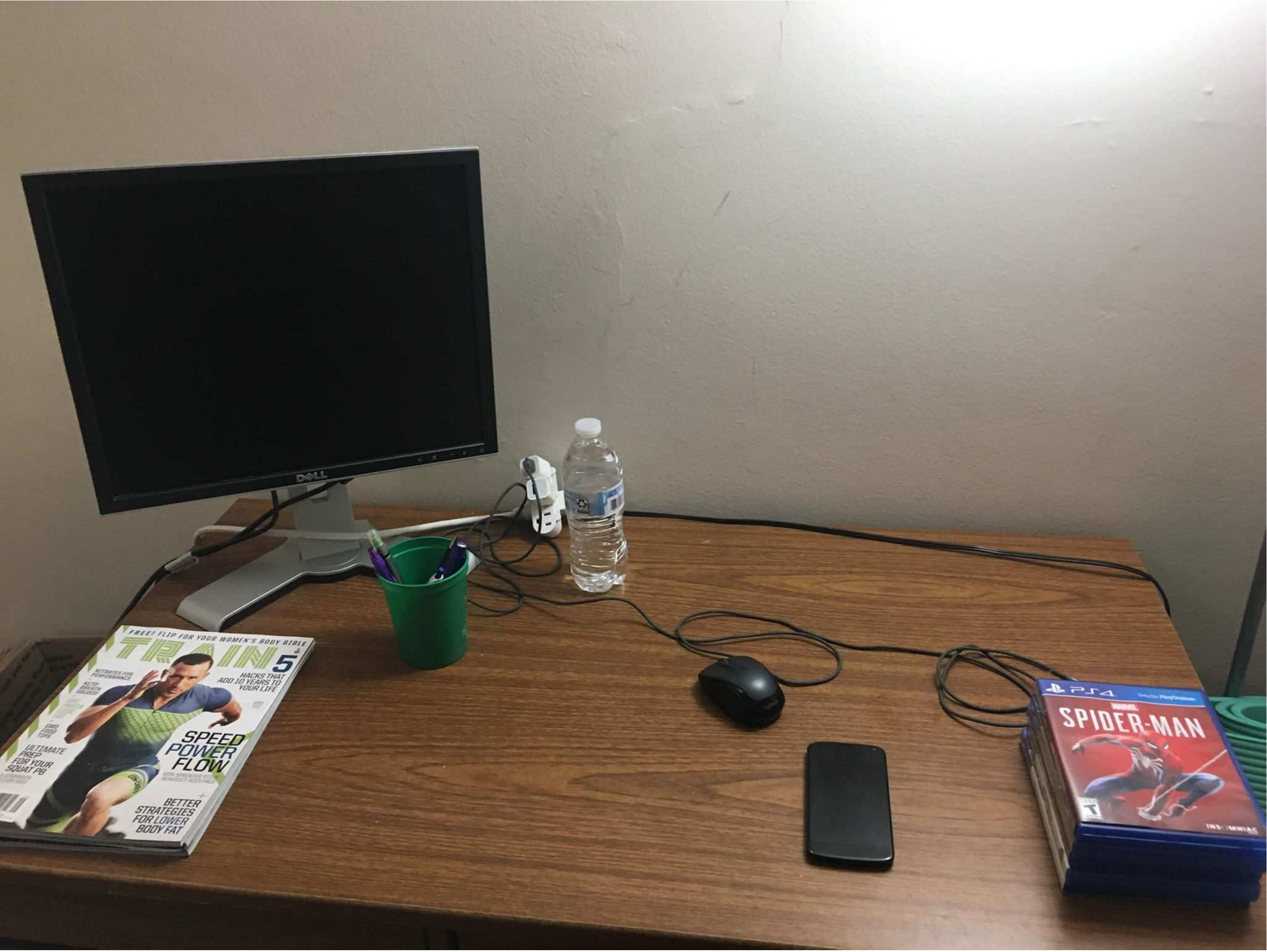}}
		\label{fig:removelaptop}}
	\hspace{-0.1in}
	\subfigure[no LCD]{
		{\includegraphics[width=0.23\columnwidth]{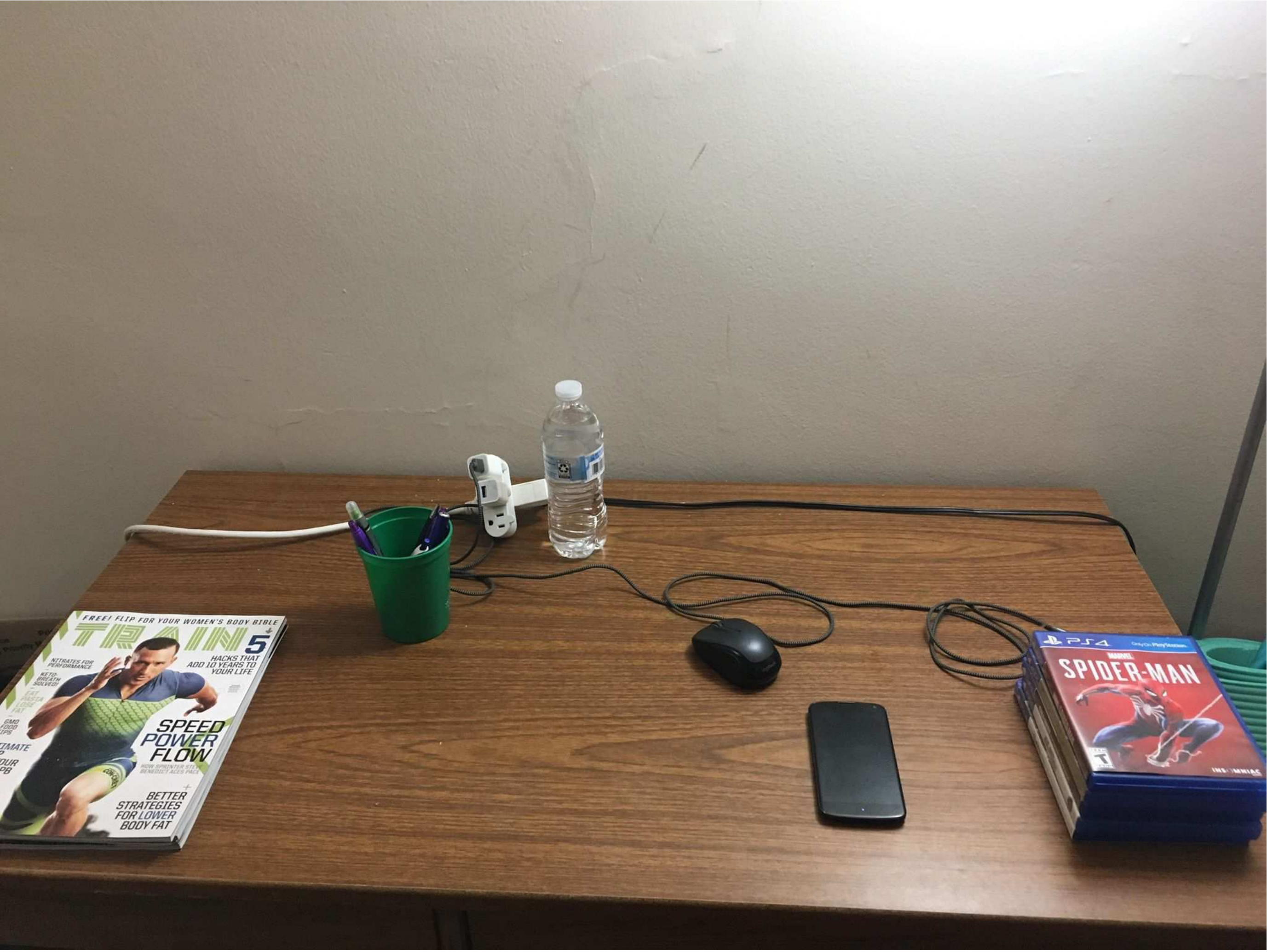}}
		\label{fig:removemonitor}}
	\vspace{-0.12in}	
	\caption{Removing things from bedroom table}
	\vspace{-0.12in}
	\label{fig:removingthingssalman}
\end{figure}

\begin{figure}[htbp]
	\centering
	\captionsetup{justification=centering}
	\subfigure[setup]{
		{\includegraphics[width=0.17\columnwidth]{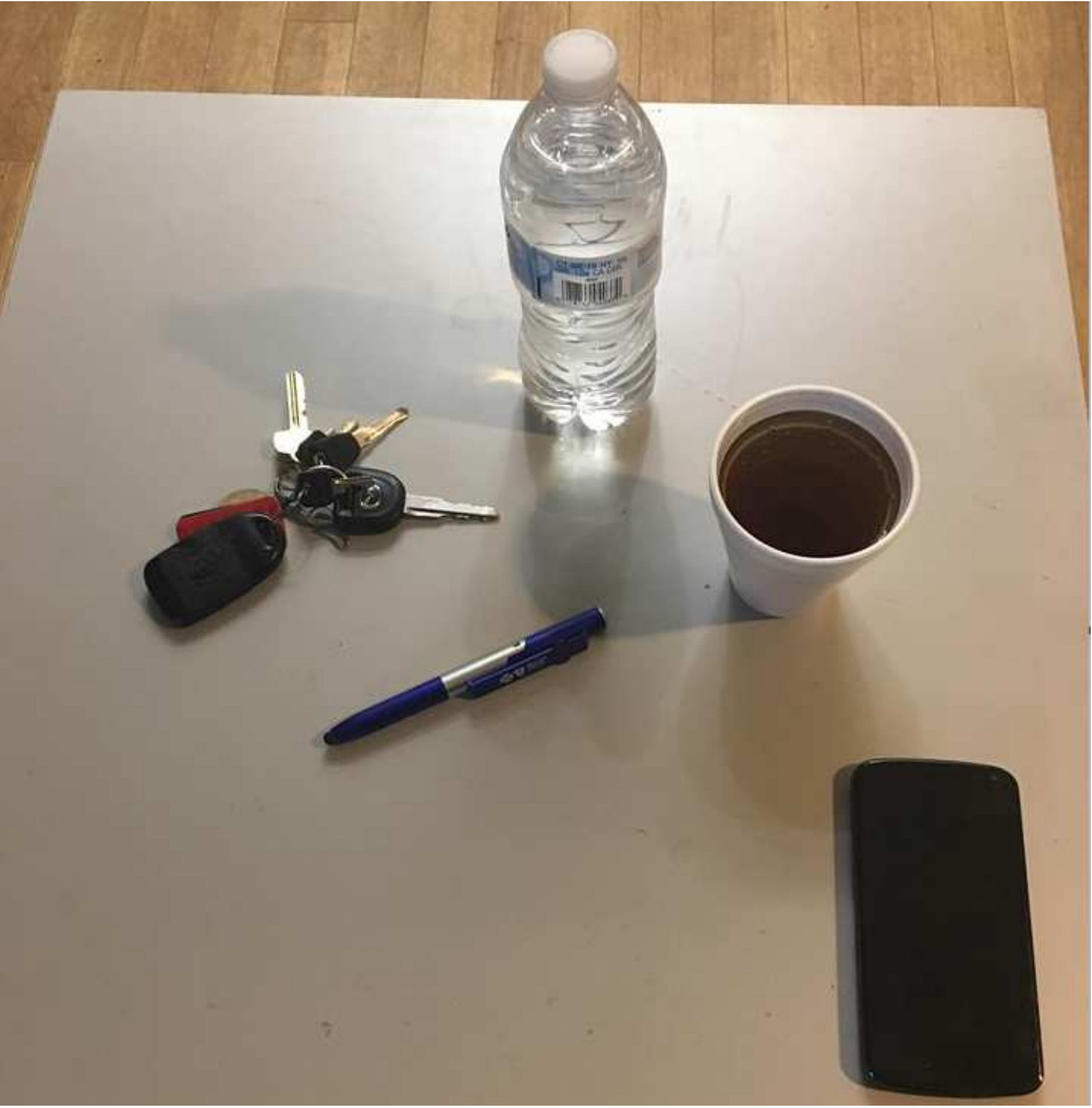}}
		\label{fig:things}}
	\hspace{-0.1in}
	\subfigure[3 inches]{
		{\includegraphics[width=0.23\columnwidth]{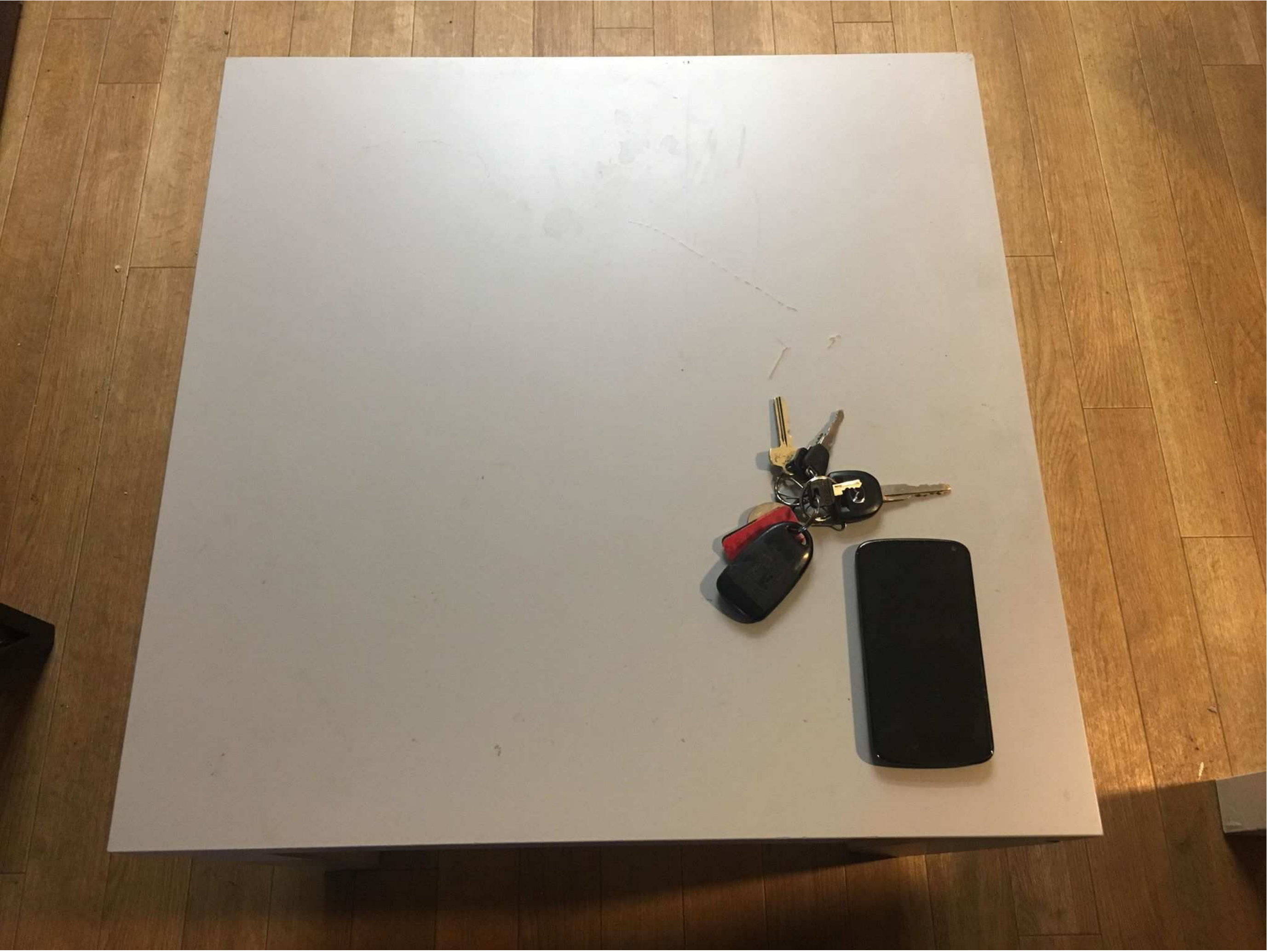}}
		\label{fig:close1}}
	\hspace{-0.1in}
	\subfigure[9 inches]{
		{\includegraphics[width=0.23\columnwidth]{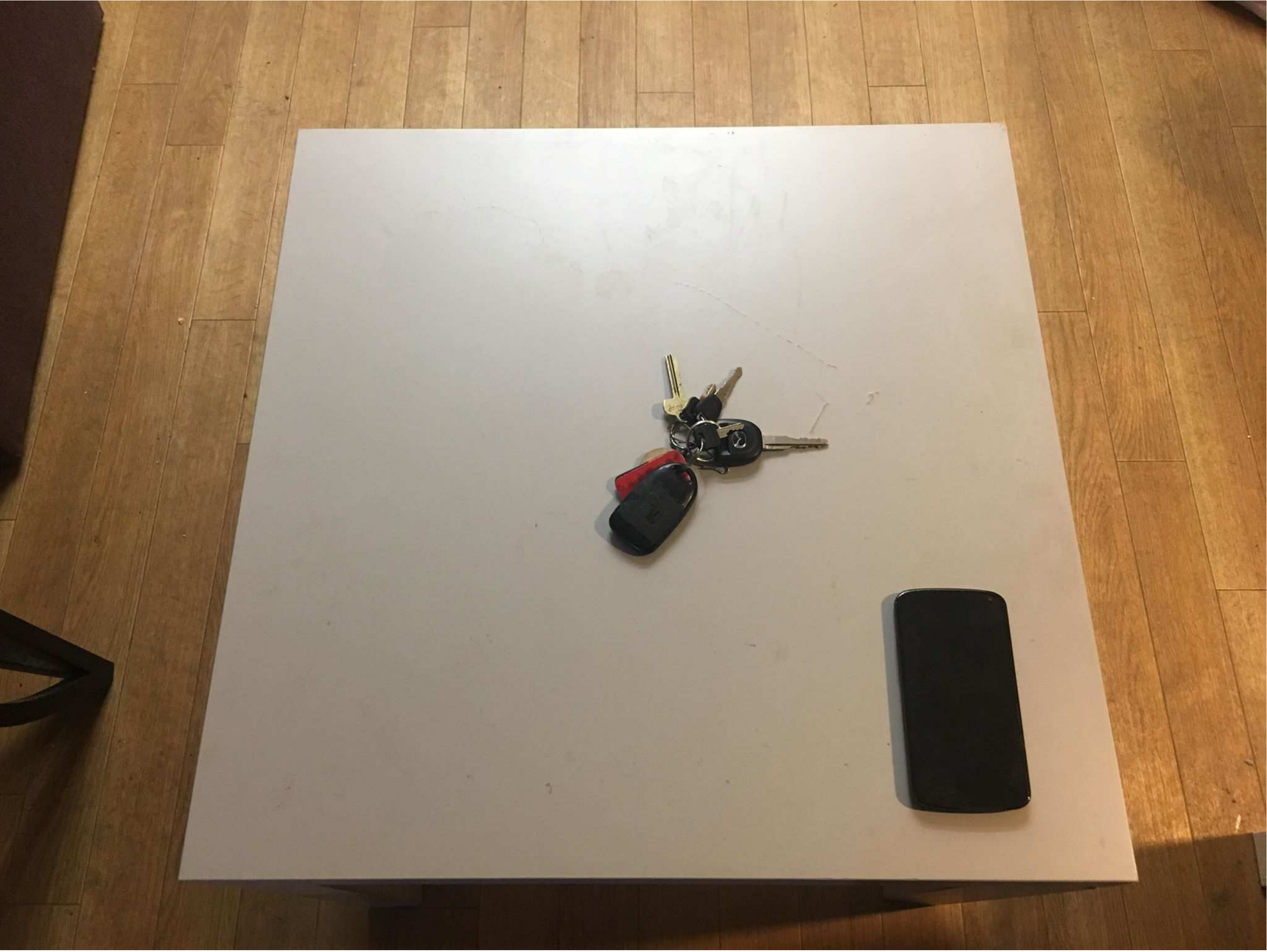}}
		\label{fig:close2}}
	\hspace{-0.1in}
	\subfigure[12 inches]{
		{\includegraphics[width=0.23\columnwidth]{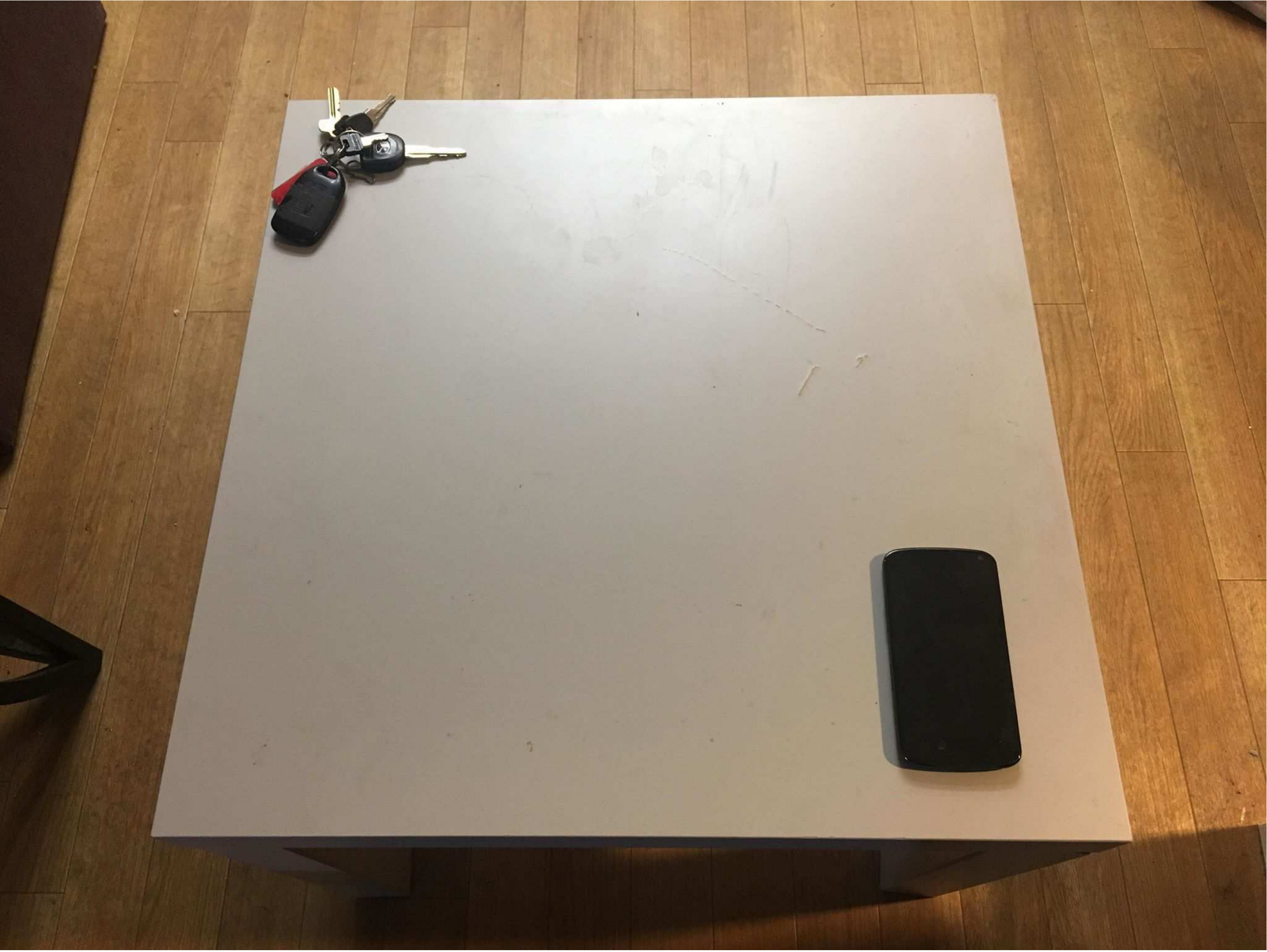}}
		\label{fig:close3}}
	\vspace{-0.12in}
	\caption{Bringing light objects closer to smartphone}
	\label{fig:bringcloser}
	\vspace{-0.12in}
\end{figure}

\begin{figure}[htbp]
	\centering
	\captionsetup{justification=centering}
	\subfigure[Living Room Table]{
		{\includegraphics[width=0.6\columnwidth]{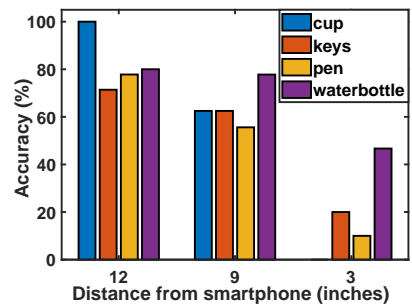}}
		\label{fig:closenessoftthingsclassify}}
	\hspace{-0.1in}
	\subfigure[Bedroom Table]{
		{\includegraphics[width=0.6\columnwidth]{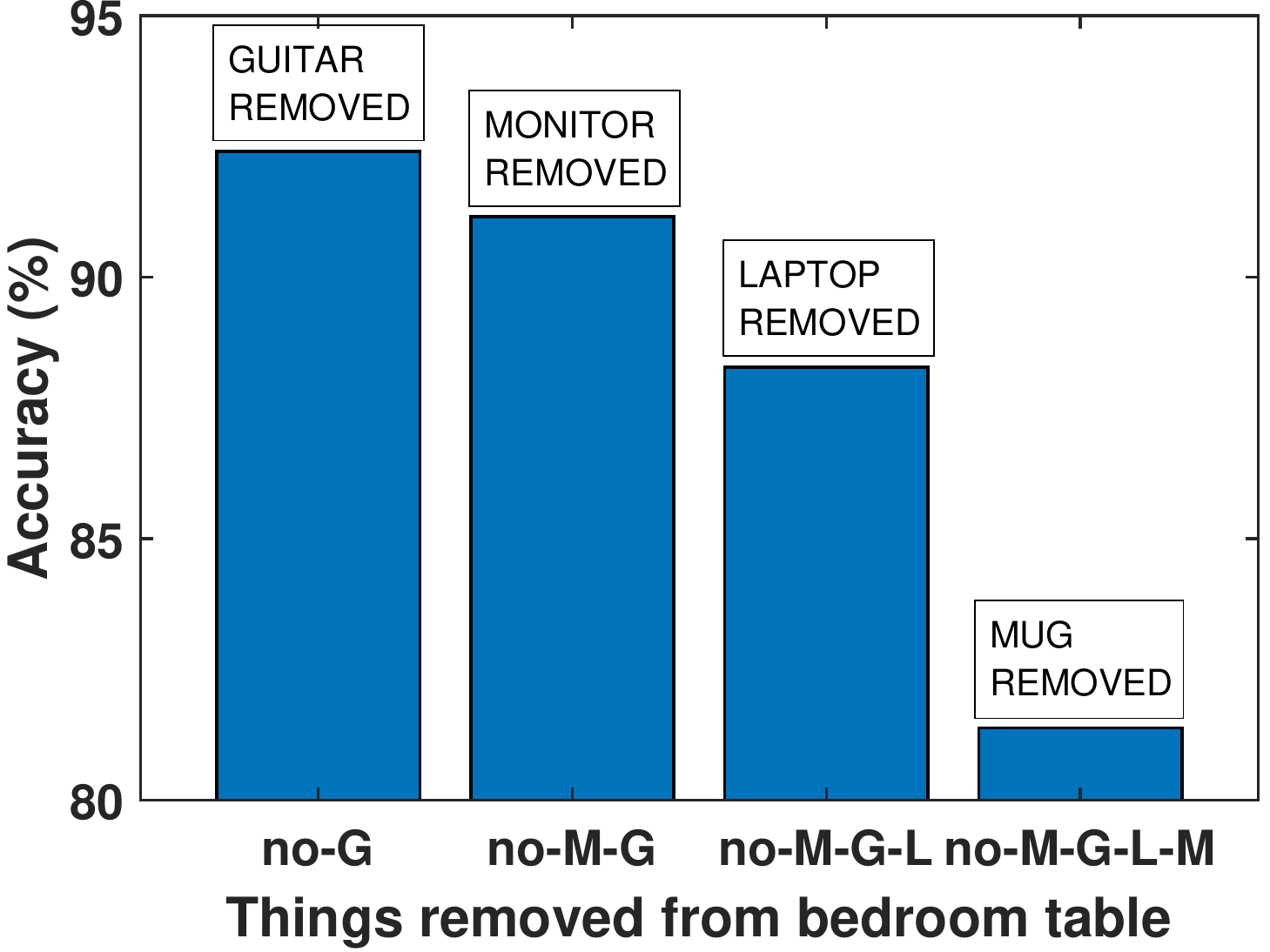}}
		\label{fig:removingobjectsbedroom}}
	\hspace{-0.1in}
	\vspace{-0.12in}
	\caption{Effect of (a) moving objects closer and of (b) removing objects on classification}
	\centering
	\vspace{-0.12in}
	\label{fig:removingobjects}
\end{figure}

\postsec
To understand the impact of surrounding objects on VibroTag's accuracy, we performed two different sets of experiments.
In the first set of experiments, we collected data on a participant's bedroom table before and after removing 4 different heavier objects (\ie a guitar, an LCD, a laptop and a mug) from the table one by one.
Figure \ref{fig:removingthingssalman} shows the setup for these experiments, where \ref{fig:normal} corresponds to the scenario where all objects were on the table, and \ref{fig:removemonitor} corresponds to the scenario where all three objects were removed.
In the second set of experiments, we collected data on the participant's living room table by bringing 4 different lighter objects (\ie a cup, a set of keys, a pen, and a water bottle) closer to the smartphone.
Fig. \ref{fig:bringcloser} shows the setup, where \ref{fig:things} shows the different objects used in these experiments, and \ref{fig:close1} - \ref{fig:close3} shows a set of keys being brought closer to the smartphone.
Fig. \ref{fig:removingobjects} shows results for the aforementioned sets of experiments.

We observe from Fig. \ref{fig:closenessoftthingsclassify} that as the lighter objects come closer to the smartphone (\ie down from 12 inches to 3 inches closer), the classification accuracy of the table decreases significantly.
For example, when the pen is within 3 inches of the phone, the accuracy goes as low as 9\%.
From Fig. \ref{fig:removingobjectsbedroom}, we observe that the impact of heavier objects on VibroTag's accuracy is not as significant as the lighter ones, which happens because the energy transfered by smartphone's vibration is not enough to make those objects vibrate significantly.
However, we observe that the classification accuracy still drops more than 10\% as we slowly remove the objects that were previously placed on the table.
This is because each of those objects has its own vibration response which was contributing to the overall vibration signature of the bedroom table.
So, when those objects are removed one by one, their response is subsequently omitted in the new vibration signature of the surface, which leads to loss in the classification performance.

\presub\subsubsection{Impact of Upper Cut-Off Frequency}
\begin{figure}[htbp]
	\vspace{-0.07in}
	\centering
	\captionsetup{justification=centering}
	\includegraphics[width=0.8\columnwidth]{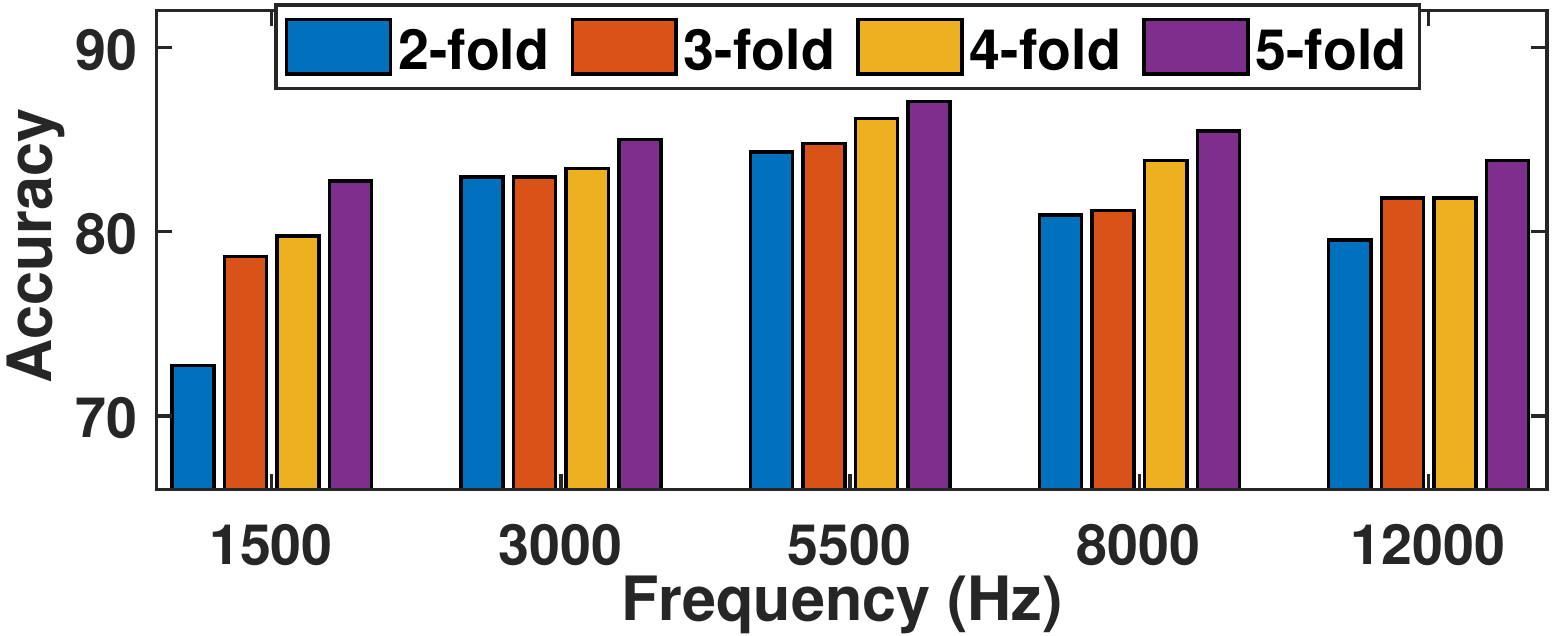}
	\vspace{-0.1in}
	\caption{Cross-validation accuracies for different band-pass filter upper cut-off frequencies (User-3)}
	\label{fig:user1-frequency}
	\vspace{-0.13in}
\end{figure}

\postsub
\textit{VibroTag achieves best accuracy when frequencies above 5500Hz are filtered out from the recorded sound signals.}
A smartphone's microphone can usually capture sounds in the frequency range of 20Hz - 20kHz.
However, the smartphone's vibration usually causes variations in lower frequencies, and therefore, filtering out higher frequencies can reduce impact of background noise and any unwanted noisy variations.
To understand which frequencies can be filtered out to achieve best accuracy in VibroTag, we employ a Butterworth band-pass filter, and determine the average accuracy for 5 different upper cut-off frequencies of the filter.
Figure \ref{fig:user1-frequency} shows how User-3's (OnePlus 2) multi-fold cross-validation accuracies vary as upper cut-off frequency increase from 1500 to 12000.
We observe that VibroTag achieves best accuracy at cut-off frequency of 5500Hz.
As the upper cut-off frequency increases, it allows higher frequency noisy variations in the vibration signatures, which leads to lower classification accuracies.
We observed similar results for other users as well.
Therefore, all accuracies reported in this paper correspond to 5500Hz cut-off frequency.
Note that other smartphones may exhibit better accuracies for cut-off frequencies which are slightly different from 5500Hz, however, this is an aspect which is out of the scope of this paper.

%% file: KamranTMC/usabilitystudy.tex
\presec
\subsection{Usability Study}
\begin{figure}[htbp]
	\centering
	\captionsetup{justification=centering}
	\includegraphics[width=0.7\columnwidth]{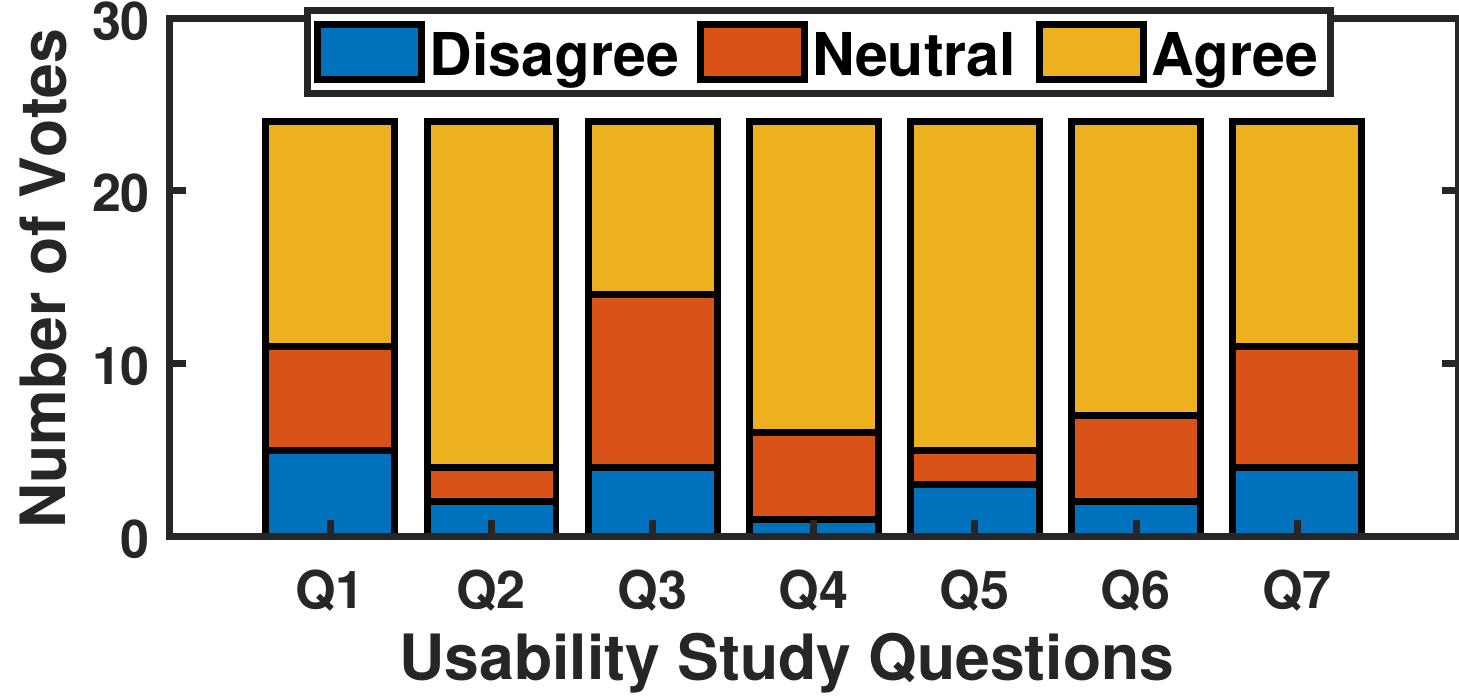}
	\vspace{-0.1in}
	\caption{Vote distribution of 7 VibroTag's usability questions asked from 24 participants}
	\label{fig:survey_combined}
	\vspace{-0.12in}
\end{figure}

\postsec
%
%
%
%
We carried out a usability study and asked 24 participants (20 male, 4 female), recruited at university, about the flexibility and usability of the VibroTag application in daily life. 
The participants comprised of students and university employees of ages 19 to 35. They were first briefed about the working of VibroTag and then its target applications such as symbolic localization.
The volunteers were shown the VibroTag application interface as pictured in Figure \ref{fig:vibrotaginterface} and given a demo of the acoustic trace collection. 
They were also briefed on how the smartphone can be placed on a surface with 3 different levels of restriction flexibility. 
At the end, they were given a set of usability questions given below and summarized in Figure \ref{fig:survey_combined}:
%
\textbf{Q1.} Are you comfortable using smartphone for location recognition?
\textbf{Q2.} Can VibroTag help you save time by setting reminders? 
\textbf{Q3.} Are you comfortable with VibroTags' use of vibration? 
\textbf{Q4.} Is it easy to place smartphone on preferred locations for learning? 
\textbf{Q5.} Can VibroTag help you in setting smart notifications linked to locations? 
\textbf{Q6.} Is VibroTag useful in activating other smart applications?  
\textbf{Q7.} Do you find VibroTag application valuable and fun to use?
%
%
Our study indicates reasonably high agreement on the usefulness of VibroTag based smart notifications and reminders.

%% file: KamranTMC/conclusion.tex
\presec
\section{Conclusion}
\postsec
In this work, we make the following contributions.
First, we propose the first fine-grained vibration based sensing scheme, that can recognize different surfaces using the vibration mechanism and microphone of a single COTS smartphone.
The intuition is that the smartphone's vibration causes the whole smartphone structure and the hardware inside it to vibrate in a peculiar pattern, which depends upon the absorption properties of the surface that the smartphone is placed on.
These vibrations produce peculiar sound waves that we detect using the smartphone's microphone.
Second, we propose a time-series based signal processing technique to extract fine-grained vibration signatures that are robust to hardware irregularities and background environmental noises.
%
%
We implemented VibroTag on two different Android phones and evaluated in multiple different environments.
Our results show that VibroTag achieves an average surface recognition accuracy of 86.55\%, which is 37\% higher than the average accuracy of only 49.25\% achieved by the state-of-the-art IMUs based schemes.

%% file: KamranTMC/acknowledgements.tex
\presec
\section{Acknowledgements}
\postsec
We would like to thank Salman Ali for his significant help with experiments and conducting the usability study.